\documentclass[paper]{JHEP} 
\usepackage{cite,epsfig,graphicx}

\def\APA{{\tt APACIC++}}
\def\AME{{\tt AMEGIC++}\ }

\newcommand{\nnb}{\nonumber}
\newcommand{\bc}{\begin{center}}
\newcommand{\ec}{\end{center}}
\newcommand{\bea}{\begin{eqnarray}}
\newcommand{\eea}{\end{eqnarray}}
\newcommand{\barl}{\begin{array}{rl}}
\newcommand{\ba}{\begin{array}}
\newcommand{\ea}{\end{array}}

\newcommand{\slaa}{\!\!\!/}

\title{AMEGIC++ 1.0\\
A Matrix Element Generator In C++}

\author{F.\ Krauss$^a$, R.\ Kuhn$^{b,c}$ and G.\ Soff$^c$\\
$^a$Cavendish Laboratory, University of Cambridge, Cambridge CB3 0HE,
U.K.\\[1mm]
$^b$Max Planck Institut f\"ur Physik Komplexer Systeme, 01187 Dresden,
Germany\\[1mm]
$^c$Institut f\"ur Theoretische Physik, TU Dresden, 01062 Dresden,
Germany\\[1mm]
E-mail: \email{krauss@hep.phy.cam.ac.uk,}\email{kuhn@theory.phy.tu-dresden.de}
}
\abstract{The new matrix element generator \AME is introduced,
dedicated to describe multi--particle production in high energy
particle collisions. It automatically generates helicity amplitudes
for the processes under consideration and constructs suitable,
efficient integration channels for the multi--channel phase space
integration. The corresponding expressions for the amplitudes and the
integrators are stored in library files to be linked to the main
program.}   

\keywords{
Standard Model; QCD; Electroweak Theory; LEP Physics;
High Energy Electron--Positron Collisions; Multi--Jet Production; 
Matrix Element Generation; Helicity Amplitudes; Phase Space Integration;
Multi--channel Method; Monte Carlo}
\preprint{Cavendish--HEP--11/01}

\begin{document}
\newpage
\section*{Program Summary}
{\it Title of the program :} \AME\\[0.5cm]
{\it Program obtainable from :} authors upon request\\[0.5cm]
{\it Licensing provisions :} none\\[0.5cm]
{\it Operating systems under which the program has been tested :} UNIX,
LINUX, VMS\\[0.5cm]
{\it Programming language :} C++\\[0.5cm]
{\it Separate documentation available :} no\\[0.5cm]
{\it Nature of the physical problem: }\\
The theoretical description of multi particle production, even at the
tree--level, suffers from two problems :
\begin{enumerate}
\item The rapidly increasing number of amplitudes forbids the
      traditional method of summing and squaring individual Feynman
      amplitudes by means of the completeness relations for spinors
      and polarization vectors. Instead, the helicity method is
      employed, translating the amplitudes into complex
      numbers. Still, the helicity amplitudes for a large number of
      diagrams have to be constructed which itself is a formidable
      task.
\item The complex structure of the high--dimensional phase space
      imperatively requires using Monte Carlo methods for its
      evaluation. Here, efficiency is of paramount importance, and one
      has to employ non--flat phase space measures which must be
      optimized for the process under consideration and its specific 
      singularity structure in phase space.\\
\end{enumerate}
{\it Method of solution:} \\
Automatic generation of helicity amplitudes related to Feynman
diagrams describing the process at the tree--level. Translating of the
amplitudes into character strings and storing in libraries for increased
efficiency during evaluation, i.e. phase space integration. Integration
by means of multichannel methods with specific channels which are
constructed from the Feynman diagrams.\\   

\tableofcontents
\newpage
\section{Introduction}

In the last decades, the observation of the production and subsequent
decay of particles with rising masses \cite{WZtop} has lead to more
and more far reaching qualitative and increasingly accurate
quantitative knowledge about physics on subnuclear scales. As a
consequence of this experimental success, the Standard Model \cite{SM} 
nowadays is widely accepted as the best model for the scales under
current consideration. However, many particle physicists believe that
the Standard Model is only the effective realization of a theory valid
up to much higher scales with even higher energetic particles
participating. Consequently, for the experimental verification or
falsification of such ideas, colliders with even higher energy release
are currently being planned or constructed \cite{Colliders}.  
From a theoretical point of view the increasing energy release does
not only allow the creation of these presumably unstable new
particles, in fact, with rising energies the number of particles or
jets produced increases drastically as well. Therefore, to analyse both the
signal and the background related to processes, where new heavy
particles are produced and decay into high energetic secondaries,
the description of multi--particle states is mandatory.  

However, such a description, even in the lowest accessible
perturbative approximation is far from being trivial. Basically, two
major problems emerge with an increasing number of final state
particles :
\begin{enumerate}
\item The number of amplitudes related to a process with $n$ outgoing
      particles rises approximately like a factorial in $n$. Thus,
      the traditional method of summing and squaring the amplitudes by
      use of the completeness relations leads to an exploding number
      of terms to be evaluated. 
\item The phase space populated by the final state particles is
      multidimensional, enforcing the use of Monte Carlo methods for
      the integration. However, non--trivial experimental cuts on the
      phase space in connection with possibly wildly fluctuating
      integrands due to nearly on--shell intermediate particles make
      sophisticated mapping and sampling methods indispensable.
\end{enumerate}

\noindent
These two problems necessitate the usage of computer programs. 
Examples are {\tt CompHep} \cite{CompHep}, 
{\tt FeynArts/FeynCalc} \cite{FAFC}, {\tt Grace} \cite{Grace}, 
{\tt MADGRAPH} \cite{MADGRAPH}, and {\tt O'Mega/WHIZARD} \cite{OMEGA}.

\noindent
The new program \AME which will be presented in this publication,
proposes a solution to both problems discussed above. In its present
form, \AME automatically creates all Feynman diagrams related to 
user defined processes in the framework of the Standard Model. It
should be emphasized here that a generalization to a large variety of
other models is straightforward and currently in preparation. However,
the diagrams are then transformed into helicity amplitudes, yielding 
complex numbers for every combination of external momenta. These
amplitudes, corresponding to a specific process, are translated into
character strings and stored in a library. For the phase space integration,
the structure of the individual Feynman diagrams is analysed and
suitable integration channels are generated. Again, they are
translated into a library file. In this sense, \AME is a program to
generate highly efficient matrix element generators for a large variety
of processes. Of course, this allows for an easy implementation into 
Monte Carlo event generators like \APA\ (for a detailed description see
\cite{Kuhn:2001dk}) which has been performed successfully.  

The outline of this paper is as follows: First, the theoretical
background for both the evaluation of amplitudes and the efficient
Monte Carlo generation of phase space points is reviewed, see
Sec.~\ref{TB}. Special emphasis is given to the helicity method, 
Sec.~\ref{HA}, and on some of the features of this method as
implemented within {\tt AMEGIC++}. Additionally, the multi--channel
algorithm for an efficient integration is discussed in some detail and
both default channels available in \AME and building blocks for
process dependent channels are listed. At the end, some higher order
effects will be discussed in Sec.~\ref{HO}. 

Then, large parts of the algorithms involved are discussed, see 
Sec.~\ref{Algorithms}. They can roughly be divided into three parts,
namely into methods for the generation of diagrams presented in
Sec.~\ref{GenAmp}, for their translation into helicity amplitudes,
cf.\ Sec.~\ref{HelAmpGen}, and finally, into algorithms for the
construction of suitable integration channels, see Sec.~\ref{GenChan}.
 
In Sec.~\ref{Classes} the corresponding classes are listed and their
interplay is described. This description is roughly split into the
following parts:
\begin{itemize}
\item First, Sec.~\ref{Organization} deals with the overhead,
      i.e.\ classes to drive {\tt AMEGIC++}.
\item Then, in Sec.~\ref{Model}, the general set--up of models is
      discussed. 
\item Sec.~\ref{Amplitude} describes the classes responsible for the
      construction of Feynman diagrams, their translation into
      helicity amplitudes and their calculation.
\item In Sec.~\ref{Strings} the essential methods for the translation
      of the helicity amplitudes into character strings and their storage
      in corresponding library files will be highlighted.
\item Sec.~\ref{PhaseSpace} explains the construction of the phase
      space integrators, the channels, and how they are used during the
      integration.
\item In Sec.~\ref{ParaSwitch}, the parameters
      and switches to be set by the user are explained and how they
      are read in.
\item Finally, in Sec.~\ref{Help} some more general helper classes are
      listed.
\end{itemize}
At the end an installation guide is given in Sec.~\ref{Install} and a
summary, Sec.~\ref{Summary} concludes this paper. Note that a Test Run
Output can be found in Appendix~\ref{TRO} as well.

\clearpage\newpage
\section{\label{TB}Theoretical background}

Before we present the algorithms and their implementation in \AME in
some detail, we will give a brief review of the underlying calculational 
methods. More specifically, in Sec.~\ref{HA} we will dwell on the method
of helicity amplitudes as introduced in \cite{SPINORS1,SPINORS2} and
clarify our notation. Then, in Sec.~\ref{PI}, we will review the
multi--channel methods \cite{Weightopt} employed for the subsequent
phase space integration of the matrix element.  

\subsection{\label{HA}Helicity amplitudes}

The basic idea behind the method of helicity amplitudes is that for
every combination of helicities and momenta for the incoming and 
outgoing particles any Feynman amplitude (represented by propagators
and vertices for the internal lines and spinors and polarization
vectors for the external particles) is nothing else but a complex
number. Once every such number is computed, all amplitudes for 
a process under consideration can be easily summed and the result
can then be squared. Inclusion of colour factors etc. is
straightforward and reduces in this framework to the multiplication of
two vectors of complex numbers (for the amplitudes and their complex
conjugated) with a colour matrix in the middle. The question to be
answered in this section is how we can extract the numerical value of
the amplitudes as a  function of the external momenta and
helicities. In {\tt AMEGIC++}, we follow quite an old approach to
answer this question which has been proven to be adequate and
successful in a large number of specific processes \cite{SPINORS2}
\footnote{However, we want to point out that other approaches exist
\cite{Caravaglios:1995cd} which are based on the LSZ theorem and
allow the direct recursive construction of matrix elements without
taking the detour of Feynman amplitudes. These approaches are currently
under further investigation and discussion. Nevertheless, we are
convinced that in the light of the need for efficient mappings of the
phase space to specific channels reproducing the singularity structure
of the process under consideration, the construction of Feynman
amplitudes is still a sine qua non.}. 
This approach relies on decomposition of the amplitude
into scalar products of four--momenta and into spinor products of the
form $u(p_i,\lambda_i)\bar u(p_j,\lambda_j)$, where the $p_{i,j}$ and
$\lambda_{i,j}$ are momenta and helicities, respectively. Knowing the
four--momenta, the scalar products can be calculated easily,
and for fixed helicities, the spinor products can be computed as well.
As we will see, the result for such a spinor product is a function of
the components of the momenta involved. This allows for the translation 
of any Feynman amplitude into generic building blocks which can be 
evaluated numerically. In the following we will give a short introduction 
into the spinor techniques we use for this decomposition and we will 
list the building blocks employed in {\tt AMEGIC++}.

\subsubsection{Spinors}

As it is well known, appropriate spinors $u(p,\lambda)$ for fermions with
mass $m$, momentum $p$ and with definite helicity $\lambda$ obey
Dirac's equation of motion (E.o.M.) 
\bea\label{DiracEoM}
(p\slaa - m) u(p,\lambda) &=& 0\,,\;\;
(p\slaa + m) v(p,\lambda)  =  0\,.
\eea
Note that since $p\slaa$ is not necessarily hermitian, $m$ does not
need to be real, but in any case $p^2 = m^2$ is fulfilled. With
$p^2<0$, $m$ turns imaginary and one can identify the particle spinor
with the positive imaginary eigenvalue $m$ (${\rm
Im(m)}>0$). Additionally, the spinors should fulfill
\bea\label{Spins}
(1\mp\gamma^5s\slaa) u(p,\pm) = 0\,,\;\;
(1\mp\gamma^5s\slaa) v(p,\pm) = 0
\eea
with the polarization vector $s$ obeying $s\cdot p=0$ and $s^2=1$.

To construct spinors satisfying both Eqs.~(\ref{DiracEoM}) and 
(\ref{Spins}), even for imaginary masses, we start with massless chiral 
spinors $w$, i.e.\ spinors satisfying 
\bea\label{wlambda}
w(k_0,\lambda)\bar w(k_0,\lambda) = 
\frac{1+\lambda\gamma_5}{2}\,k\slaa_0\,,
\eea
for arbitrary light--like $k_0$. Starting from such a chiral spinor
$w(k_0,\lambda)$ with definite chirality, spinors of opposite
chirality $w(k_0,-\lambda)$ can be constructed via
\bea\label{k1}
w(k_0,\lambda) = \lambda k\slaa_1 w(k_0,-\lambda) 
\eea
with the vectors $k_{0,1}$ satisfying
\footnote{Note that we have some freedom in choosing the $k$--vectors
which provides a valuable tool for checking the calculation.}
\bea
k_0^2 = 0\;,\;\;
k_0\cdot k_1 = 0\;,\;\;
k_1^2 = -1\,.
\eea
Massive spinors can then be expressed as
\bea\label{u2w}
u(p,\lambda) &=& \frac{p\slaa + m}{\sqrt{2p\cdot k_0}}\,w(k_0,-\lambda)\,,\;\;
v(p,\lambda)  = \frac{p\slaa - m}{\sqrt{2p\cdot k_0}}\,w(k_0,-\lambda)
\eea
in terms of these chiral spinors. These relations hold also true for
$p^2<0$ and imaginary $m$. For the construction of conjugate spinors
we will abandon the proper definition
\bea
\bar u = u^\dagger\gamma^0\,,\;\;\bar v = v^\dagger\gamma^0\,,
\eea
since this definition does not lead to the E.o.M. 
\bea\label{AntiEoM}
\bar u(p,\lambda) (p\slaa - m) = 0\,,\;\;\bar v(p,\lambda) (p\slaa + m) = 0
\eea
for spinors with imaginary masses. Instead we use Eqs.~(\ref{AntiEoM})
and
\bea
\bar u(p,\pm)(1\mp\gamma^5s\slaa) = 0\,,\;\;
\bar v(p,\pm)(1\mp\gamma^5s\slaa) = 0
\eea
to define conjugate spinors. With this definition, and choosing the
normalization
\bea
\bar u(p,\lambda)u(p,\lambda) = 2m\,,\;\;
\bar v(p,\lambda)v(p,\lambda) = -2m\,,
\eea
the conjugate spinors are constructed as
\bea\label{baru2w}
\bar u(p,\lambda) &=& \bar w(k_0,-\lambda)\,
                      \frac{p\slaa + m}{\sqrt{2p\cdot k_0}}
                      \,,\;\;
\bar v(p,\lambda)  =  \bar w(k_0,-\lambda)\,
                      \frac{p\slaa - m}{\sqrt{2p\cdot k_0}}\,.
\eea
They fulfill the completeness relation
\bea\label{completeness}
1 = \sum\limits_\lambda
\frac{\bar u(p,\lambda)u(p,\lambda)-\bar v(p,\lambda)v(p,\lambda)}{2m}\,.
\eea
The last relation can now be used to decompose the numerators
of fermion propagators according to
\bea\label{FermionPropsEq}
\lefteqn{\bar u(p_1,\lambda_1)\,\chi_1\,(p\slaa_2+\mu_2)\,
         \chi_3\, u(p_3,\lambda_3)}\nnb\\
  &\to & \frac12\left[\sum\limits_{\lambda_2}
         \bar u(p_1,\lambda_1)\,\chi_1\, u(p_2,\lambda_2)\,
         \bar u(p_2,\lambda_2)\,\chi_3\, u(p_3,\lambda_3)\,
         \left(1+\frac{\mu_2}{m_2}\right)
         \right.\nnb\\
  &&     \left.\;\;\;\;\;\;\;+
         \bar u(p_1,\lambda_1)\,\chi_1\, v(p_2,\lambda_2)\,
         \bar v(p_2,\lambda_2)\,\chi_3\, u(p_3,\lambda_3)\,
         \left(1-\frac{\mu_2}{m_2}\right)
         \right]\,,
\eea
where $m_2^2=p_2^2$ and the $\chi$ denote the (scalar or vector)
couplings involved.

\subsubsection{Polarization vectors}

\begin{enumerate}
\item{Massless bosons :\\
The next task on our list is to translate the polarization vectors 
$\epsilon(p,\lambda)$ of massless external spin--1 bosons with
momentum $p$ and helicity $\lambda$ into suitable spinor products. 
Massive bosons will experience a different treatment, we will deal
with them later. However, for massless bosons, the polarization
vectors satisfy
\bea\label{PolProps}
\begin{array}{lcllcl}
\epsilon(p,\lambda)\cdot p &=& 0\;,\;\;&
\epsilon(p,\lambda)\cdot \epsilon(p,\lambda) &=& 0\;,\\
\epsilon^\mu(p,-\lambda) &=& \epsilon^{\mu*}(p,\lambda) \;,\;\;&
\epsilon(p,\lambda)\cdot \epsilon(p,-\lambda) &=& -1\,.
\end{array}
\eea
In the axial gauge the polarization sum reads
\bea\label{PolSumE}
\sum\limits_{\lambda = \pm}
\epsilon^\mu(p,\lambda)\epsilon^{\nu*}(p,\lambda) =
-g^{\mu\nu}+\frac{q^\mu p^\nu+q^\nu p^\mu}{p\cdot q}\,,
\eea
where $q^\mu$ is some arbitrary, light--like four vector not parallel
to $p^\mu$. It can be shown that the spinor objects
\bea\label{PolSpin}
\frac{1}{\sqrt{4p\cdot q}}\,\bar u(q,\lambda)\gamma^\mu u(p,\lambda)
\eea
satisfy the properties of Eqs.~(\ref{PolProps}) and (\ref{PolSumE}) and hence
form an acceptable choice for the polarization vectors of massless
external spin--1 bosons. From this identification, we obtain an
additional constraint for $q^\mu$, namely that it should not be
collinear to $k_0^\mu$. Note, however, that the freedom in choosing $q$ 
reflects the freedom in choosing a gauge vector in the axial
gauge. Since final results should not depend on such a choice,
we have another tool at hand to check our calculations.\\
}
\item{Massive bosons:\\ 
For massive bosons, the situation changes. First of all, there is
another, longitudinal polarization state which satisfies the same
properties as the transversal states, Eq.~(\ref{PolProps}). In contrast
to the massless case, however, the spin sum changes and becomes
\bea\label{PolSumEM}
\sum\limits_{\lambda = \pm,0}
\epsilon^\mu(p,\lambda)\epsilon^{\nu*}(p,\lambda) =
-g^{\mu\nu}+\frac{p^\mu p^\nu}{m^2}\,,
\eea
where $m$ is the mass of the boson. We introduce
\bea\label{PolSpinMass}
a^\mu = \bar u(r_2,-)\gamma^\mu u(r_1,-)
\eea
with light--like four vectors $r_1$ and $r_2$ satisfying
\bea
r_1^2=r_2^2 = 0\;\;\mbox{\rm and}\;\;
r_1^\mu+r_2^\mu = p^\mu\,.
\eea
Replacing the spin sum by an integration over the
solid angle of $r_1$ in the rest frame of $p$, we obtain
\bea\label{PolSpinMassNorm}
\int\limits_{4\pi}d\Omega \, a^\mu a^{\nu*} =
\frac{8\pi m^2}{3}\left(-g^{\mu\nu}+\frac{p^\mu p^\nu}{m^2}\right)
\eea
which is of the desired form. Therefore, identifying
\bea
\epsilon^\mu \Longrightarrow a^\mu\;\;\mbox{\rm and}\;\;
\sum\limits_{\lambda = \pm,0} \Longrightarrow 
\frac{3}{8\pi m^2}\int\limits_{4\pi}d\Omega  
\eea
we will arrive at the correct form for the unpolarized cross section.

In other words, the polarization vector of a massive vector boson can
be constructed in terms of spinors via a pseudo--decay of the boson
into two massless fermions with coupling constant 1. 
}
\end{enumerate}

\subsubsection{Building blocks}

Finally, we will present the basic spinor products which serve as
elementary building blocks for our amplitudes. We introduce the
following short--hand notations :
\begin{itemize}
\item projectors
\bea
P_{L,R} = \frac{1\mp\gamma_5}{2}\,;
\eea
\item $S$--functions
\bea\label{Sfunctions}
S(+;p_1,p_2) = 2c_L\frac{k_0\cdot p_1k_1\cdot p_2-k_0\cdot p_2k_1\cdot p_1+
               i\epsilon_{\mu\nu\rho\sigma}
               k_0^\mu k_1^\nu p_1^\rho p_2^\sigma}
              {\eta_1\eta_2}
\eea
with the symmetry properties
\bea
S(-;p_1,p_2) = S^*(+;p_2,p_1)\;\;\mbox{\rm and}\;\;
S(\pm;p_1,p_2) = -S(\pm;p_2,p_1)\,;
\eea
\item normalizations and mass-terms
\bea
\label{muneta}
\eta_i=\sqrt{2p_i\cdot k_0}\;\;\mbox{\rm and}\;\;
\mu_i = \pm\frac{m_i}{\eta_i}\,,
\eea
where the signs denote particles and anti-particles.\\
\end{itemize}

The following spinor expressions are then the basic building 
blocks for helicity amplitudes within \AME:
\begin{enumerate}
\item $\displaystyle Y(p_1,\lambda_1;p_2,\lambda_2;c_L,c_R) = 
       \bar u(p_1,\lambda_1)[c_LP_L+c_RP_R]u(p_2,\lambda_2)$\,;\\
\item $\displaystyle X(p_1,\lambda_1;p_2;p_3,\lambda_3;c_L,c_R) = 
       \bar u(p_1,\lambda_1)p\slaa_2[c_LP_L+c_RP_R]u(p_3,\lambda_3)$\,;\\
\item $\displaystyle Z(p_1,\lambda_1;p_2,\lambda_2;
         p_3,\lambda_3;p_4,\lambda_4;         
         c_L^{12},c_R^{12};c_L^{34},c_R^{34}) =$\\[0.1cm] 
       $\displaystyle \bar u(p_1,\lambda_1)\gamma^\mu
           [c_L^{12}P_L+c_R^{12}P_R]u(p_2,\lambda_2)
       \bar u(p_3,\lambda_3)\gamma_\mu
           [c_L^{34}P_L+c_R^{34}P_R]u(p_4,\lambda_4)$\,.\\
\end{enumerate}
Their explicit form can be found in Tabs.~\ref{Yfuncs}, \ref{Xfuncs},
and \ref{Zfuncs}. 
\begin{table}[h]
\bc
\begin{tabular}{|c|c||c|c|}
\hline
$\lambda_1\lambda_2$ & $Y(p_1,\lambda_1;p_2,\lambda_2;c_L,c_R)$ &
$\lambda_1\lambda_2$ & $Y(p_1,\lambda_1;p_2,\lambda_2;c_L,c_R)$ \\\hline
$++$ & $c_R\mu_1\eta_2 + c_L\mu_2\eta_1$ &
$+-$ & $c_LS(+;p_1,p_2)$\\\hline
\end{tabular}
\caption{\label{Yfuncs} $Y$--functions for different helicity
combinations $(\lambda_1,\,\lambda_2)$.
The remaining $Y$--functions can be obtained by exchanging
$L\leftrightarrow R$ and $+\leftrightarrow -$. }
\ec
\end{table}
\begin{table}[h]
\bc
\begin{tabular}{|c|c|}
\hline
$\lambda_1\lambda_3$ & $X(p_1,\lambda_1;p_2;p_3,\lambda_3;c_L,c_R)$\\\hline
$++$ & $\displaystyle \left(\mu_1\eta_2+\mu_2\eta_1\right)
        \left(\mu_2\eta_3c_R+\mu_3\eta_2c_L\right)
        +c_RS(+;p_1,p_2)S(-;p_2,p_3)$\\[0.1cm]
$+-$ & $c_L\left(\mu_1\eta_2+\mu_2\eta_1\right)S(+;p_2,p_3)
       +\left(c_L\mu_2\eta_3+c_R\mu_3\eta_2\right)S(+;p_1,p_2)$\\\hline
\end{tabular}
\caption{\label{Xfuncs} $X$-functions for different helicity
combinations $(\lambda_1,\,\lambda_3)$.
The remaining $X$--functions can be obtained by exchanging
$L\leftrightarrow R$ and $+\leftrightarrow -$. }
\ec
\end{table}
\begin{table}[h]
\bc
\begin{tabular}{|c|c|}
\hline
$\lambda_1\lambda_2\lambda_3\lambda_4$ 
   & $Z(p_1,\lambda_1;p_2,\lambda_2;
         p_3,\lambda_3;p_4,\lambda_4;         
         c_L^{12},c_R^{12};c_L^{34},c_R^{34})$\\\hline
$++++$ & $2\left[S(+;p_3,p_1)S(-;p_2,p_4)c_R^{12}c_R^{34}+
          \mu_1\mu_2\eta_3\eta_4c_L^{12}c_R^{34}+
          \mu_3\mu_4\eta_1\eta_2c_R^{12}c_L^{34}\right]$\\[0.1cm]
$+++-$ & $2\eta_2c_R^{12}\left[S(+;p_1,p_4)\mu_3c_L^{34}
                         +S(+;p_1,p_3)\mu_4c_R^{34}\right]$\\[0.1cm] 
$++-+$ & $2\eta_1c_R^{12}\left[S(-;p_3,p_2)\mu_4c_L^{34}
q                         +S(-;p_4,p_2)\mu_3c_R^{34}\right]$\\[0.1cm] 
$++--$ & $2\left[S(+;p_4,p_1)S(-;p_2,p_3)c_R^{12}c_L^{34}+
          \mu_1\mu_2\eta_3\eta_4c_L^{12}c_L^{34}+
          \mu_3\mu_4\eta_1\eta_2c_R^{12}c_R^{34}\right]$\\[0.1cm]
$+-++$ & $2\eta_4c_R^{34}\left[S(+;p_1,p_3)\mu_2c_R^{12}
                         +S(+;p_2,p_3)\mu_1c_L^{12}\right]$\\[0.1cm] 
$+-+-$ & $0$\\[0.1cm]
$+--+$ & $-2\left[ \mu_1\mu_4\eta_2\eta_3c_L^{12}c_L^{34}
                  +\mu_2\mu_3\eta_1\eta_4c_R^{12}c_R^{34}
                  -\mu_1\mu_3\eta_2\eta_4c_L^{12}c_R^{34}
                  -\mu_2\mu_4\eta_1\eta_3c_R^{12}c_L^{34}\right]$\\[0.1cm]
$+---$ & $2\eta_3c_L^{34}\left[S(+;p_4,p_2)\mu_1c_L^{12}
                         +S(+;p_1,p_4)\mu_2c_R^{12}\right]$\\\hline
\end{tabular}
\caption{\label{Zfuncs} $Z$--functions for different helicity
combinations $(\lambda_1,\,\lambda_2,\,\lambda_3,\,\lambda_4)$.
The remaining $Z$--functions can be obtained by exchanging
$L\leftrightarrow R$ and $+\leftrightarrow -$. }
\ec
\end{table}
However, let us note that, in general, products like
\bea
\bar u(p_1)\gamma_\mu u(p_2)
\bar u(p_3)\gamma_\nu u(p_4)\cdot\
\left(g^{\mu\nu}-\frac{P^\mu P^\nu}{M^2}\right)
\eea
(stemming from the exchange of a massive boson with mass $M$ and
momentum $P$ between the two fermion lines) cannot be represented 
in terms of one elementary $Z$--function per helicity
combination. Instead, some composite functions, consisting of $Z$--
and $X$--functions, become necessary. 

\begin{figure}[h]
\begin{center}
\includegraphics[height=9cm]{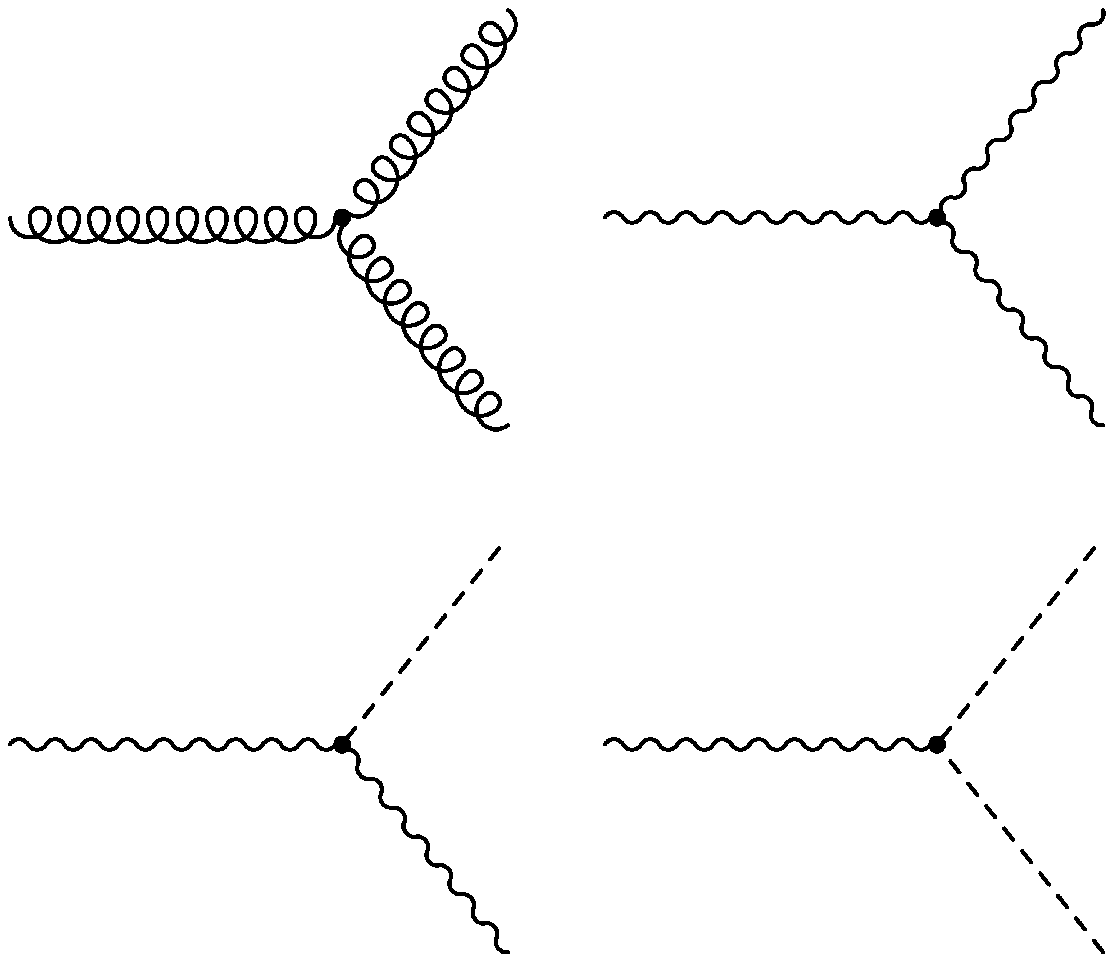}
\caption{\label{Figv3} Three boson vertices available within {\tt AMEGIC++}.
Note that all boson lines might be on-- or off--shell. Scalar and
vector bosons are depicted as dotted and wavy lines, respectively,
whereas a gluon is marked by a curly line.}
\end{center}
\end{figure}

\begin{figure}[h]
\begin{center}
\includegraphics[height=9cm]{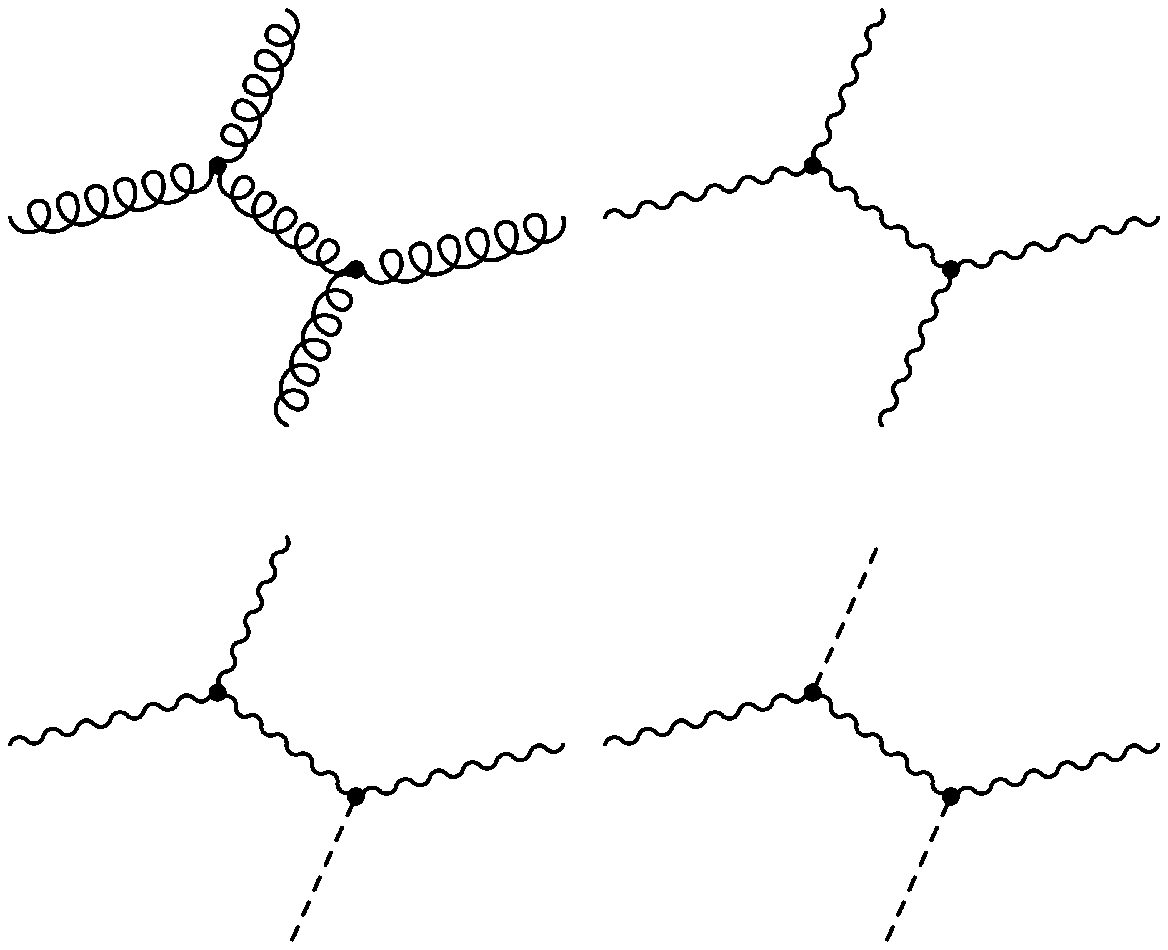}
\caption{\label{Figv4} Four boson vertices available within {\tt AMEGIC++}.
Note that all boson lines might be on-- or off--shell. Scalar and
vector bosons are depicted as dotted and wavy lines, respectively,
whereas a gluon is marked by a curly line.}
\end{center}
\end{figure}

\begin{figure}[h]
\begin{center}
\includegraphics[height=5cm]{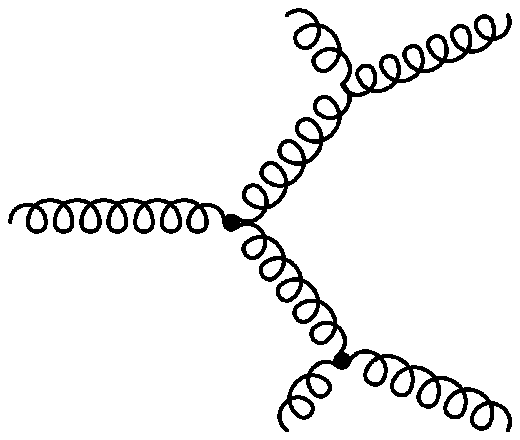}
\caption{\label{Figv5} Five gluon vertex available within {\tt AMEGIC++}.
Note that all boson lines might be on-- or off--shell.}
\end{center}
\end{figure}

\begin{figure}[h]
\begin{center}
\includegraphics[height=3cm]{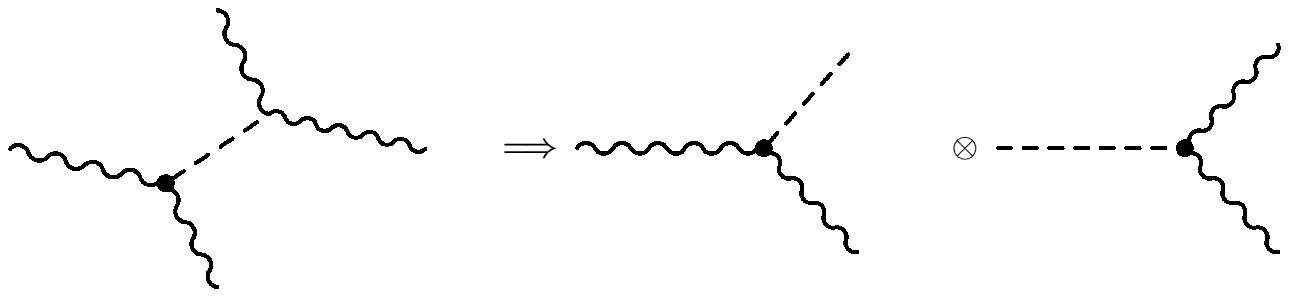}
\caption{\label{FigSv} Scalar propagators can be cut easily. Scalar and
vector bosons are depicted as dotted and wavy lines, respectively.}
\end{center}
\end{figure}

Similarly, the propagation of longitudinal modes in off--shell vector
bosons prevents us from employing polarization vectors to replace the
numerators of the propagators. Therefore, structures with three or
more bosons can not be ``cut open'' and suitable building blocks need to
be constructed for any such structure. Within {\tt AMEGIC++}, building blocks
for the structures depicted in Figs.~\ref{Figv3}, \ref{Figv4}, and \ref{Figv5}
are available at the moment, somewhat limiting the maximal number of vector
boson lines. 
Note that since vertices involving only scalar bosons are proportional
to scalars, there is no need to define specific three or four scalar
vertices. Similarly, scalar propagators can easily be cut open and
even more complicated structures can be decomposed into easier ones as
depicted in Fig.~\ref{FigSv}. 
Here, we feel, a further clarification is in order. As the observant 
reader might have noticed, we omitted vertices with four vector bosons in 
our pictorial presentation of the building blocks. Instead, we chose to 
disentangle the gauge structure of such vertices and map them piecewise 
on corresponding sequences of two three--boson vertices as depicted in the 
upper left diagram of Fig.~\ref{Figv4}.
%
\subsection{\label{PI}Phase space integration}

Now we are in the position to discuss the issue of integrating the
amplitudes. Due to the high dimensionality of the phase space ($3n-4$
for $n$ outgoing particles) and because of possible non--trivial cuts
on the phase space reflecting the experimental situation, we have to
abandon the idea of exact analytical integration. Instead we will
employ Monte Carlo methods, i.e.\ we will sample the matrix element,
here denoted with $f(\vec x)$, over the allowed phase space. Its
points $\vec x$ are then distributed randomly according to the
probability distribution $g(\vec x)$. Defining the weight
\bea
w(\vec x) = \frac{f(\vec x)}{g(\vec x)}
\eea
we can transform the true integral into a Monte Carlo estimate $E(f)$ 
via
\bea
\int_{\rm true} d\vec x\, f(\vec x) = 
\int_{\rm true} d\vec x\, g(\vec x)w(\vec x)
\Longrightarrow \left\langle w(\vec x)\right\rangle_g = E(f)\,.
\eea
To estimate the error of this estimate $E(f)$, we need the square root
of the variance $V(f)$ given by  
\bea\label{variance}
V(g) = \frac{1}{N}\left[
\left\langle w^2(\vec x)\right\rangle_g - 
\left\langle w(\vec x)\right\rangle_g^2\right]
= \frac{1}{N}\left[W(g)-E^2(f)\right]\,,
\eea
where we have identified 
\bea
W(g) = \left\langle w^2(\vec x)\right\rangle_g\,.
\eea
The name of the game is now to find a $g$ that efficiently minimizes 
$V(f)$. This might become quite a tricky business, since usually, the 
amplitude squared $f(\vec x)$ is wildly fluctuating due to resonant 
structures in propagators and -- in the Monte Carlo sense -- singular 
behavior at some edges of the allowed phase space. As illustrative 
examples, consider the phase space for the processes 
\bea
e^+e^-\to W^+W^-\to 4f\;\;\mbox{\rm and}\;\;
e^+e^-\to q\bar qg
\eea
which suffer from the resonant $W$--propagators or from the soft and
collinear divergencies related to the gluon emission, respectively.
To solve this problem of mapping the phase space efficiently to
suitable probability distributions, \AME employs the multi--channel
method described for instance in \cite{Weightopt}. The cornerstone of this
method is the knowledge about the singularity structure of individual
processes and the possibility of constructing an efficient mapping as a
sum of specific channels. Each of these channels is constructed to
cover the full phase space under consideration with special emphasis,
i.e.\ higher probability density, on one of the singular regions. In
the following we will briefly review this method and list the
ready--to--use standard channels in \AME plus the building blocks for
the constructed channels
\footnote{The building blocks used in \AME were used 
already in \cite{EXCALIBUR}. Note that similar approaches
were discussed in the literature, for example see
\cite{Ohl:1999jn} and \cite{Papadopoulos:2001tt}.}.

\subsubsection{Multi--channel method}

As indicated above, the basic idea of the multi--channel method is to
introduce different mappings of the phase space, i.e.\ different
distributions of the $\vec x$.  Each of these channels $g_i(\vec x)$
with $i=1, \dots, n$ is nonnegative over the full phase space and
normalized to unity,
\bea
\int d\vec x\, g_i(\vec x) = 1\,.
\eea
The various $g_i$ contribute according to so--called a priori weights
$\alpha_i \ge 0$ which again are normalized to unity, i.e.\
\bea
\sum\limits_{i=1}^n \alpha_i = 1\,.
\eea 
Thus, the multi--channel mapping $g$ is
\bea 
g(\vec x) = \sum\limits_{i=1}^n \alpha_i g_i(\vec x)\,.
\eea
Since in Eq.~(\ref{variance}) $E(f) = \langle w(\vec x)\rangle$ is
independent of $g(\vec x)$, we see immediately that we need to
minimize  $\left\langle w(\vec x)^2\right\rangle_g$ with respect to
the $\alpha_i$ in order to minimize our error estimate. It is well
known that the overall variance is minimized by distributing it
evenly over the channels used. In practice, the essential part of the
variance per channel can be easily estimated during the sampling
procedure via 
\bea
W_i(\alpha_i) = \int d\vec x\, g_i(\vec x)w^2(\vec x) =
\left\langle \frac{g_i(\vec x)}{g(\vec x)}\,w^2(\vec x)\right\rangle\,.
\eea
Obviously, $\alpha_i$ should increase, i.e.\ the channel $i$ should
contribute more, if its variance is large, i.e.\ if $W_i$ is large. 
In other words, the more wildly fluctuating a channel is, the more
Monte--Carlo points using this channel should be produced to smooth
out the fluctuations of this particular channel and hence of the overall
result. This leads to an iterative improvement of the $\alpha_i$ with 
help of the $W_i$.

In {\tt AMEGIC++}, we therefore iterate an optimization procedure of the
$\alpha_i$ with a limited, user--defined number of steps. In each
optimization step, i.e.\ after a predetermined (and again user--defined) 
number of Monte--Carlo points, the a priori weights are reshuffled 
according to 
\bea
\alpha_i^{\rm new} \sim \alpha_i^{\rm old} 
	W_i\left(\alpha_i^{\rm old}\right)^\beta\,.
\eea
Again, they are normalized in order to cope with the constraint that 
they add up to unity. Note that $\beta$ should obey $0<\beta<1$. In 
\AME its default value is $1/2$. The optimal set of weights yielding the 
smallest overall variance during this procedure is stored and used 
afterwards, i.e.\ in the full evaluation of the cross section. Note,
however, that the results accumulated during the optimization
procedure are added to the final estimate. 

In the following sections we are going to introduce the channels
available within  {\tt AMEGIC++}. They can be roughly divided into two
sets, one set of default channels, where the only process dependence
manifests itself in the number of momenta produced, the other set
consisting of the channels which are constructed by exploring the
phase space structure of specific diagrams. 

\subsubsection{Default channels : {\tt RAMBO} and {\tt SARGE}}
\label{RamSar}
We will start with a brief description of the default channels within 
{\tt AMEGIC++}, namely {\tt RAMBO} and {\tt SARGE}. For a more
detailed discussion of these two integrators we refer to the concise
and clear original publications \cite{RAMBO,Draggiotis:2000gm}.
\begin{enumerate}
\item{{\tt RAMBO}:\\
Essentially, {\tt RAMBO} produces any number $n$ of uniformly distributed 
massless or massive momenta in their center--of--mass frame, i.e.\ a flat 
$n$--particle phase space. Massless momenta are generated in two major 
steps : 
\begin{enumerate}
\item{Energies and solid angles for $n$ ``unrestricted'' momenta are produced
      independently, with a flat distribution of the angles in $\cos\theta$,
      the polar angle, and $\phi$, the azimuthal angle. The energies $E_i$ 
      are chosen according to $E_i\exp(-E_i)$ with no constraints whatsoever.}
\item{The $n$ ``unrestricted'' momenta are rescaled in order to obey the
      conservation of total four--momentum. This changes both the energies
      and angles of the unrestricted momenta when they are mapped to the
      ``restricted'' momenta.} 
\end{enumerate}
For massive momenta another step is now added
\begin{enumerate}
\setcounter{enumi}{2}
\item{The massless momenta are rescaled once more, in order to bring
      them onto their mass--shell. In addition, this induces an 
      extra non--trivial weight (i.e.\ depending on the actual set of
      vectors), compensating for the rescaling of the momenta.}
\end{enumerate}
}
\item{{\tt SARGE}:\\
In contrast, {\tt SARGE} generates $n$ massless momenta distributed
according to the antenna pole structure 
\bea
\frac{1}{(p_1p_2)(p_2p_3)\dots(p_{n-1}p_n)(p_np_1)}\,.
\eea
In principle, the momenta are produced again in two major steps :
\begin{enumerate}
\item{Two massless momenta $q_1$ and $q_n$ are produced in the c.m. 
      frame of the process. Starting from these two momenta the others are
      produced recursively via ``basic antennae'' $dA_{i,j}^k$ according to
      $dA_{1,n}^2dA_{2,n}^3\dots dA_{n-2,n}^{n-1}$. 

   In the expression for the individual $dA_{i,j}^k$, the lower indices $i$ 
   and $j$ denote the two initial momenta and $k$ denotes the generated 
   momentum, respectively. The antennae density is given as 
   \bea
   dA_{i,j}^k = d^4k\delta(k^2)\Theta(k_0)\frac{1}{\pi}
                \frac{p_ip_j}{(p_ik)(p_jk)}
                g\left(\frac{p_ik}{p_ip_j}\right)
                g\left(\frac{p_jk}{p_ip_j}\right)
   \eea
   with
   \bea
   g(\xi) = \frac{1}{2\log\xi_{\rm min}}
             \Theta(\xi-\xi^{-1}_{\rm min})\Theta(\xi-\xi_{\rm min})\,.
   \eea
   It should be noted that $\xi_{\rm min}$ can be related to some
   physical cuts in relative angles of the momenta and transverse momenta.

   The algorithm then is to generate $k_0$ and $\cos\theta$ (the angle 
   between $k$ and one of the two initial momenta) in the c.m. frame of the 
   initial momenta according to $dA$, distribute the polar angle 
   uniformly and boost the three--momentum system back to the frame
   where the initial momenta were given. Obviously, this
   procedure does not satisfy overall four--momentum conservation,
   thus the next step is mandatory:}
\item{The $n$ ``unrestricted'' momenta $q_i$ are boosted and rescaled in
      order to obey the conservation of total four--momentum. Again, this
      changes both the energies and angles of the unrestricted momenta when
      they are mapped on the ``restricted'' momenta, but, of course, it leaves
      the scalar products related to the single antennae invariant.} 
\end{enumerate}
Let us note in passing that the algorithm outlined above constitutes only 
the basic version of {\tt SARGE}, improvements outlined by the authors
of this algorithm have been implemented into {\tt AMEGIC++}.
}
\end{enumerate}

\subsubsection{Building blocks for additional channels}

In addition to the two default channels presented in Sec.~\ref{RamSar},
\AME analyses the phase space structure related to the process and
constructs automatically one channel corresponding to every individual
Feynman diagram. In Tab.~\ref{PSBB} we give some insight into the
building blocks, propagators and decays, of these channels. 

\begin{table}[h]
\begin{enumerate}{
\item{Propagators :\\[2mm]
\begin{tabular}{|l||c|c|}
\hline
Propagator & Distribution & Weight $\vphantom{\frac{|}{|}}$\\\hline
massless   & \begin{minipage}[ht]{7cm}{
             simple pole $1/s^\eta$ with exponent $\eta<1$        
                      }\end{minipage}
                  & 
             \begin{minipage}[ht]{4.3cm}{
              $\displaystyle{\vphantom{\frac{|^|}{|^|}}}
             \displaystyle{\frac{s_{\rm max}^\eta - 
                s_{\rm min}^\eta}{\eta s^\eta}}$
             }\end{minipage}\\[5mm]\hline
massive    & \begin{minipage}[ht]{7cm}{
              Breit--Wigner with mass $M$, width $\Gamma$
                      }\end{minipage}
                  & 
             \begin{minipage}[ht]{4.3cm}{
             $\displaystyle{\vphantom{\frac{|^|}{|^|}}}
             \displaystyle{\frac{1}{\pi}
              \frac{M\Gamma}{(s-M^2)^2+M^2\Gamma^2}}$                      
             }\end{minipage}\\[5mm]\hline
\end{tabular}
}
\item{Decays :\\[2mm]
\begin{tabular}{|l||l|l|}
\hline
Decay               & Distribution $P\to p_i$ 
                    & Weight $\vphantom{\frac{|}{|}}$\\\hline
isotropic 2--body   & flat, $p_i$ have masses
                    & $\displaystyle{\vphantom{\frac{|^|}{|^|}}}
                 \displaystyle
                 {\frac{2P^2}{\pi\sqrt{\lambda(P^2,p_1^2,p_2^2)}}}$\\
\hline
t-channel  & \begin{minipage}[ht]{5cm}{
                $q_1+q_2 \to p_1+p_2$,  $p_{1,2}$ massive,\\
                $\cos\theta$ between $q_1$ and $p_1$
                is distributed like $1/(a-\cos\theta)^\eta$ 
                in the rest frame
                ($a$ depends on mass of the propagator).
                      }\end{minipage} &\\[5mm]
\hline
anisotropic 2--body & \begin{minipage}[ht]{5cm}{
                      $\cos\theta$ between $P$ and $p_1$
                      in the rest frame of $P$
                      is distributed like $1/\cos^\eta\theta$,
                      $p_{1,2}$ are massive }
                      \end{minipage}
                    & \begin{minipage}[ht]{5cm}{
                 $\displaystyle{\vphantom{\frac{|^|}{|^|}}}
                 \displaystyle{\frac{4P^2}{\pi\sqrt{\lambda(P^2,p_1^2,p_2^2)}}
                               \cdot\frac{1}{\cos^\eta\theta}}$\\
                 \hspace*{2mm}$\cdot\displaystyle{\frac{1-\eta}
                     {(\cos^{1-\eta}\theta_{\rm max}-\cos^{1-\eta}\theta_{\rm min})}}$}
                      \end{minipage}\\[5mm]\hline
isotropic 3--body & flat, only $p_2$ is massive 
                  & $\displaystyle{\vphantom{\frac{|^|}{|^|}}}
                 \displaystyle{\frac{4}{\pi^2
                \left(\frac{P^4-p_2^4}{2P^2}+p_2^2 \log\frac{p_2^2}{P^2}\right)}}$\\\hline
\end{tabular}\\[2mm]
}
}\end{enumerate}
\begin{center}
\caption{\label{PSBB}Various decays and propagators used within \AME for the
creation of additional channels. Here, $\lambda(s,s_1,s_2)=
(s-s1-s2)^2-4s_1s_2\,.$}
\end{center}
\end{table}

\subsection{\label{HO}Higher--order effects}

Some higher--order effects have been included in \AME in order to
increase the precision of some of the processes to be calculated. The
first example is the consideration of Coulomb effects for two charged
bosons which have been produced near their mass-shell,
Sec.~\ref{CoCo}. On the other hand higher--order effects can be
included via the running widths of particles, see Sec.~\ref{RMW}.    

\subsubsection{\label{CoCo}Coulomb corrections}

In principle, Coulomb corrections contribute to all processes with 
electrically charged particles. They are particularly important, if two
particles are created close to each other and move with a low relative speed. 
Then the long--range interactions alter the corresponding cross section by
a possibly sizeable factor $\sim 1/v$ with $v$ the relative velocity.

\begin{figure}[h]
\begin{center}
\epsfxsize=12cm\epsffile{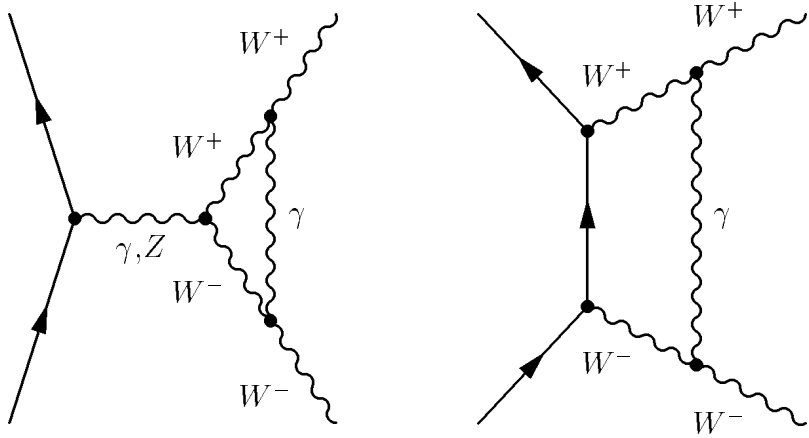}
\caption{\label{Coulamp} Diagrams related to
the Coulomb correction as implemented within {\tt AMEGIC++}.}
\end{center}
\end{figure}

Within {\tt AMEGIC++}, such corrections are included for the production of
$W$--bosons as in \cite{Bardin:1993mc} with a factor
\bea
\label{CoulEq}
C_{\rm coul}^{WW} = \frac{\alpha\pi}{2\beta}\cdot \left[1-
\frac{2}{\pi}\arctan\frac{\left|\left(\beta_M+\Delta\right)\right|^2
        -\beta^2}{2\beta\,\mbox{\rm Im}(\beta_M)}\right]
\eea
with 
\bea
M_\pm &=& p_{W^\pm}^2 - M_W^2\,,\nnb\\
\beta_M &=& \sqrt{\frac{s-4 M_W^2+ 4i\Gamma_W M_W}{s}}\,,\nnb\\
\Delta &=& \frac{\left|M_+^2-M_-^2\right|}{s}\,,\nnb\\
\beta &=& \frac{1}{s}\cdot 
          \sqrt{\left[s-(M_+-M_-)^2\right]\left[s-(M_++M_-)^2\right]}\,.
\eea
This factor just multiplies the corresponding amplitudes with a 
$W^+W^-$--pair for the higher--order correction, i.e.\ effectively a
one plus this factor is multiplied. The amplitudes including the
photon are depicted in Fig.~\ref{Coulamp}, where we have neglected the
decay of the $W$--pair. 

\subsubsection{\label{RMW}Running masses and widths}

Possible schemes for the running of quark masses and boson masses
and widths, respectively, have been implemented in {\tt AMEGIC++}. First
of all, the LO running quark masses are related to the pole mass $m_{\rm
pol}=m(m)$ via the well-known formula:
\bea\label{runningqmass}
m(\mu) &=& m(m)\,\left[\frac{\alpha_S(\mu)}{\alpha_S(m)}
                 \right]^{\gamma_m^{(0)}/2\beta^{(0)}}\,,\nnb\\
\gamma_m^{(0)} &=& 6\,C_F.
\eea
Here, the running of the strong coupling $\alpha_S$ is used. Note that
a leading order running for both, the strong and the electroweak coupling 
$\alpha_{\rm QED}$ is implemented as well.  
The second part belongs to the running of the boson widths.  
Accordingly, a first ansatz uses fixed masses and widths,
whereas the other two try some sophisticated schemes for the
running and appropriate change of these parameters, see
Tab.~\ref{masswidths}. For further details and a comparison between
the advantages and disadvantage of these schemes we refer to the
literature \cite{MassWidths}.  
\begin{table}[h]
\bc
\begin{tabular}{|c|c|c|}
\hline
Scheme & $m_{\rm fl}(s)$ & $\Gamma_{\rm fl}(s)$\\[1mm]\hline
&&\\
0 & $\displaystyle  m_{\rm pole}$ & $\Gamma_{\rm pole}$\\&&\\
1 & $\displaystyle
     \frac{m_{\rm pole}^2}{\sqrt{m_{\rm pole}^2-\Gamma_{\rm pole}^2}}$
  & $\displaystyle\frac{m_{\rm pole}\Gamma_{\rm pole}}
    {\sqrt{m_{\rm pole}^2-\Gamma_{\rm pole}^2}}$\\&&\\
2 & $\displaystyle
     m_{\rm pole}$ & $\Gamma_{\rm pole}\cdot
     \displaystyle{\frac{\sqrt{s}}{m_{\rm pole}}}$
  \\
&&\\
\hline
\end{tabular}
\caption{\label{masswidths} 
Running widths and accordingly modulated masses for the electroweak
gauge bosons within the schemes provided by {\tt AMEGIC++}.}
\ec
\end{table}

\clearpage\newpage
\section{\label{Algorithms}Algorithms}

In this section we describe in some detail the main algorithms
implemented in {\tt AMEGIC++}. We focus on the algorithms employed to
generate Feynman diagrams, Sec.~\ref{GenAmp}, to translate them into
helicity amplitudes, Sec.~\ref{HelAmpGen}, and, finally, on algorithms
employed for the construction of suitable channels for the phase space
integration, Sec.~\ref{GenChan}.

\subsection{Generation of diagrams\label{GenAmp}}

Let us start with the generation of Feynman diagrams for a given process.
Basically, within {\tt AMEGIC++}, this proceeds in four steps, namely
\begin{enumerate}
\item{the creation of empty topologies,}
\item{setting the endpoints (incoming and outgoing particles),}
\item{the determination of suitable intermediate propagators, and, finally}
\item{the elimination of possibly occurring identical diagrams.}
\end{enumerate}
In the following, we will discuss these steps in more detail.

\subsubsection{Creation of empty topologies}

\vspace{0.3cm}

\begin{tabular}{ll}
\hspace*{-2mm}\begin{minipage}[ht]{7cm}{
The topologies which are created and then filled by \AME \ during the
generation of Feynman amplitudes, consist of binary trees of linked
knots, called {\tt Point}s. Each of these knots splits into two
branches and -- correspondingly -- two new knots. Therefore, within
each point there exist links to a {\tt left} and a {\tt right}
point. Consequently, knots which are not splitting any more, point to 
empty right and left knots. Additional information within a point
consists of the flavour of the incoming line and some direction
indicated by $b=~\!\!\!\pm 1$, depending on whether the line is taken as
incoming or outgoing.}\end{minipage} &
\begin{minipage}[ht]{7cm}{
\includegraphics[height=6.8cm]{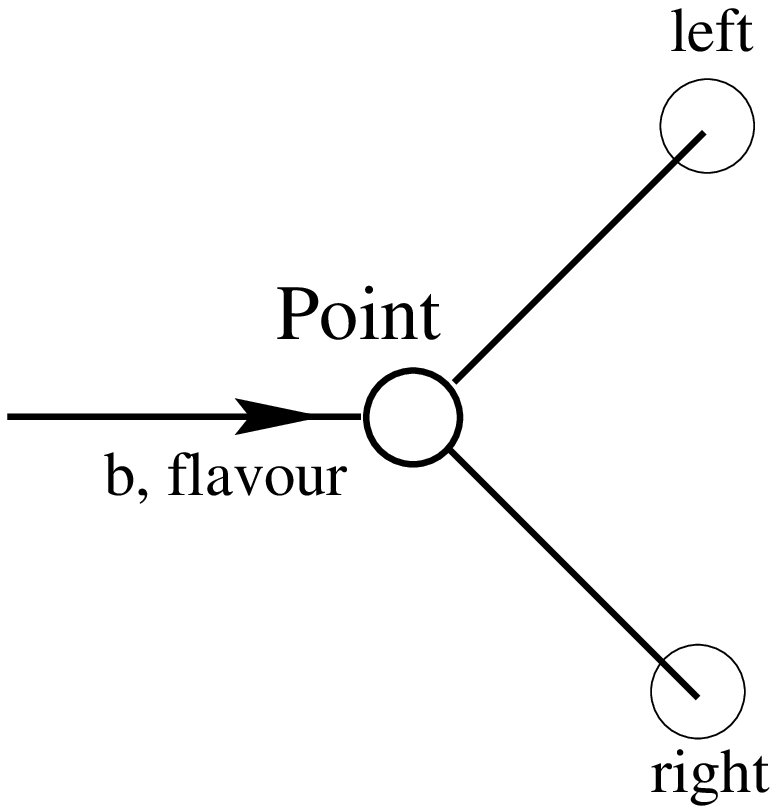}
}\end{minipage} 
\end{tabular}

\vspace{0.3cm}

Let us note that within the Standard Model and its most
popular extensions, interactions between particles are described by
means of vertices with three and four external legs. In all
interesting cases we can understand any four--vertex as two iterated
three--vertices with a shrunk internal line. Therefore it is
sufficient to use only the binary points for the construction
of our topologies.

\begin{figure}[h]
\includegraphics[height=9cm]{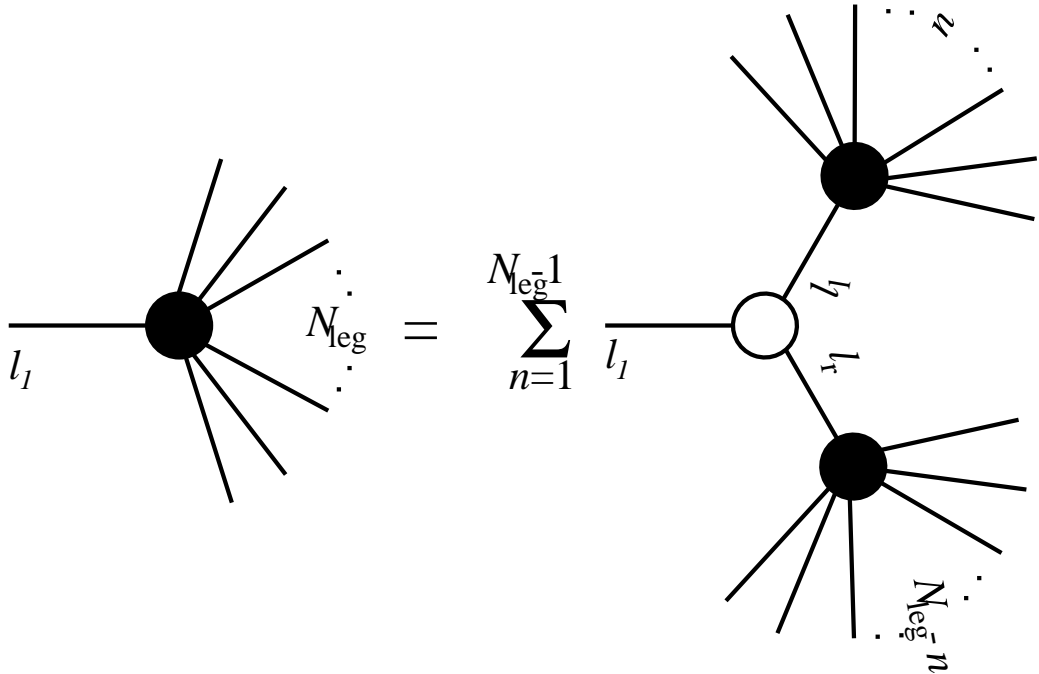}
\caption{\label{Figtopos}Recursive structure for the construction of
topologies.}
\end{figure}

Consider now processes with $N_{\rm in}$ incoming and $N_{\rm out}$
outgoing particles, respectively. Clearly, starting from a single
initial line $l_1$, we need to construct topologies with 
$N_{\rm leg} = N_{\rm in}+N_{\rm out}-1$
additional legs. This can be done recursively, employing all
topologies with $n < N_{\rm leg}$. The idea here is to 
split the initial line $l_i$ into a left and a right branch. The left
line $l_l$ then is replaced with all topologies with 
$l_l\to n$ legs, and the right line $l_r$ with all
topologies with $l_r\to N_{\rm leg}-n$ legs. Graphically,
this can be understood as depicted in Fig.~\ref{Figtopos}.
Within \AME there is a clear distinction of left and
right legs due to the internal structure of physical vertices which
will be mapped onto the blank topologies in later steps. This is why
the sum in Fig.~\ref{Figtopos} extends to $N_{\rm leg}-1$ instead of
only half of it. 

\subsubsection{Setting the endpoints}

\vspace{0.3cm}

\begin{tabular}{ll}
\hspace*{-2mm}\begin{minipage}[ht]{7cm}{
The next step is to set the endpoints, i.e.\ to distribute the 
$n=N_{\rm in}+N_{\rm out}$ incoming and outgoing particles over the
external legs of the corresponding topologies. 
A crucial role plays the way how the vertices are set up internally,
and how the diagrams will be mapped on the helicity amplitudes. 
In fact, this allows us to define a number of 
criteria the endpoints have to meet.
First of all, the first incoming particle is set on position
$1$, i.e.\ it constitutes the ``trunk'' of the topology. All other
particles are then permuted over positions $2$ to $n$.
}\end{minipage} &
\begin{minipage}[ht]{7cm}{
\includegraphics[height=6.8cm]{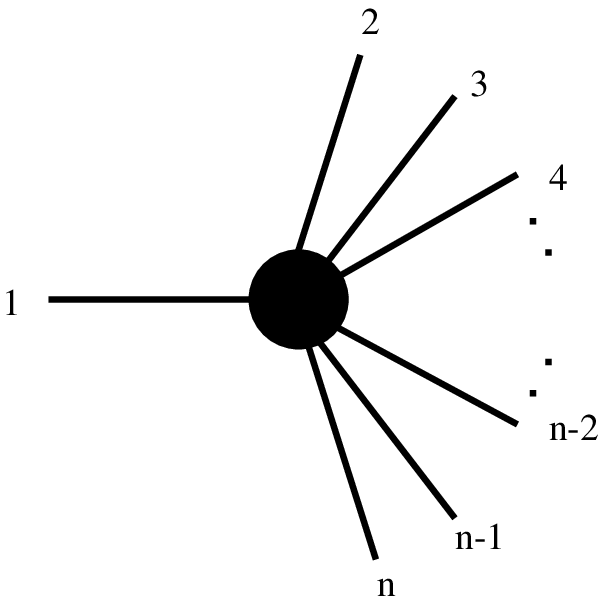}
}\end{minipage} 
\end{tabular}

\vspace{0.3cm}

However, the set of permutations leading to accessible Feynman diagrams
is limited by the intrinsic ordering of incoming and outgoing lines in
{\tt AMEGIC++}, see Tab.~\ref{Vertices}. These
limitations
\footnote{Note that at the moment, Standard Model processes
          only are 
          available within {\tt AMEGIC++}. However, the inclusion of some 
          popular extensions of the Standard Model into \AME \ is under
          way. For these, conserved baryon and lepton 
          generation--number will also be assumed at the beginning.}
are reflected by the
following constraints to be applied to the particles on positions $2$
and $n$
:
\begin{enumerate}
\item{If the particle on position $1$ is an incoming quark 
      $q_{\rm in}$ (lepton of the $k$th generation $l^k_{\rm in}$) the 
      particle on position $n$ has to be either an incoming anti-quark 
      $\bar q_{\rm in}$ ($\bar l^k_{\rm in}$) or an outgoing quark 
      $q_{\rm out}$ ($l^k_{\rm out}$) 
      .}
\item{If the particle on position $1$ is an incoming anti-quark 
      $\bar q_{\rm in}$ (anti-lepton of the $k$th generation 
      $\bar l^k_{\rm in}$) the particle on position $2$ has to be either an
      incoming quark $q_{\rm out}$ ($l^k_{\rm out}$) or an outgoing
      anti-quark $\bar q_{\rm in}$ ($\bar l^k_{\rm in}$) 
      .}
\end{enumerate}
These two conditions reflect the fact that any incoming fermion on
position 1 leads to another fermion (incoming or outgoing) and one
boson. Looking up the ordering of particle types in the vertices (as
specified in Tab.~\ref{Vertices}), it is obvious that spinor lines
stretch from the incoming particle {\tt in[0]} to the outgoing
particle positioned on the right leg {\tt in[2]}, whereas anti-spinor
lines stretch from {\tt in[0]} to {\tt in[1]}. Thus, in addition with
the following algorithms to find intermediate lines (as applied to the
boson line attached to the fermion on position 1), the two conditions
above guarantee that the existence of a first
fermion--fermion--boson vertex is not ruled out {\em a priori}.

For further refinement, there is some bookkeeping of incoming and outgoing
fermions, represented by a variable $q_{\rm sum}$ for the quarks and
similarly for each lepton generation. After each position of the
external legs is equipped with a particle, $q_{\rm sum}$ is calculated 
for each permutation according to
\bea
q_{\rm sum}\to\left\{
\begin{array}{ll} 
q_{\rm sum}+1\,, & \mbox{if particle =} \;\;\bar q_{\rm in}\;\;
                         \mbox{or} \;\; q_{\rm out}\\                
q_{\rm sum}-1\,, & \mbox{if particle =} \;\;\bar q_{\rm out}\;\;
                         \mbox{or} \;\; q_{\rm in}\;.                
\end{array}\right.\nnb
\eea
Permutations under consideration are thrown away, if
\begin{enumerate}
\item{$q_{\rm sum}$ ever exceeds $0$ when a quark sits on position $1$, and
      position $n$ is occupied by either an incoming anti-quark or an
      outgoing quark, or if}
\item{$q_{\rm sum}$ ever drops below $0$, if an anti-quark sits on
      position $1$ and position $2$ is occupied by either an incoming
      quark or an outgoing anti-quark.} 
\end{enumerate}
The same conditions apply generationwise for the leptons. Whenever a 
permutation of the external particles meets the requirements outlined
above, the endpoints are set accordingly, i.e.\ they are supplied with
the position numbers of the specific permutation under consideration. 

Intermediate points receive increasing numbers larger than $99$ to 
distinguish them from endpoints later on. This procedure of giving numbers 
to the individual points is realized recursively, using the
following steps:
\begin{enumerate}
\item{Start from point $1$. Set as its number the number of the first
      particle.}
\item{Go to the left point. If it is an endpoint (i.e.\ with empty left
      and right links), set the next position of the permutation
      sequence as its number and proceed to 3. If it is not an
      endpoint, supply it with an internal number and increment the
      corresponding counter. Repeat this step.}
\item{Go back one step and then to the right point. Again, if it is an 
      endpoint, set the corresponding number of the permutation, and
      repeat step 3. In the complementary case, supply it with an
      internal number and go to step 2. If there is no right point
      left, the topology is equipped with all endpoints set.}
\end{enumerate}
Note that in this procedure, the position numbers of the permutation
are synonymous for the corresponding particle which consists at this
stage of a flavour and the flag $b=\pm 1$, where the two signs
depend on whether the particle is outgoing or incoming. 

To keep track of additional minus signs due to the exchange of two
identical fermions, the original numbers of the particles are compared
with their permuted numbers. Whenever the ordering of the original and
the permuted numbers of two identical fermions does not coincide, the
overall sign of the specific combination of endpoints is multiplied
by $-1$.

\subsubsection{Finding intermediate lines}

Having at hand such a topology with endpoints set and internal numbers 
for the propagators, the aim is now to find suitable intermediate
lines connecting the external points of this specific topology. If any 
set of intermediate lines is found, a Feynman diagram has been
successfully constructed, and if not, the topology can be neglected.  
Again, the determination of the internal lines proceeds recursively
along the following steps:
\begin{enumerate}
\item{Start from point $P$ with a given flavour. Check
      whether either the left or the right or both links are not yet
      equipped with a flavour. If this is the case, test all available
      vertices for a fit to the known flavours of $P$ and its links. 
      For every fit copy the topology with this new
      vertex set. If both flavours were already set, go to the next
      point and start again.}
\item{If the left link was not yet equipped with a flavour, repeat step
      1 with $P\to P_{\rm left}$.}
\item{If the right link was not yet equipped with a flavour, repeat step
      1 with $P\to P_{\rm right}$.}
\end{enumerate}
We will now discuss in some detail, how vertices are tested for a fit
into the topology. 
For this it is indispensable to explain briefly, how vertices are
defined within {\tt AMEGIC++}. For simplicity we will display only
vertices for two fermions and one boson, $f_1\to f_2 b$ plus their
``barred'' combinations. Note, however that the same transformations
apply also for bosons. 

Basically, within \AME \ there are six different ways to group these
three flavours as incoming or outgoing lines, possible double countings
due to identical flavours
will be eliminated during the initialization of the vertices. In
general, during the matching of the vertices onto the topologies,
positions {\tt in[0]}, {\tt in[1]}, and {\tt in[2]} will be mapped
onto a point, its left, and its right link, respectively. As can be
seen in Tab.~\ref{Vertices}, the fermions are ordered in a specific
way, namely the outgoing fermions on position {\tt in[2]} and outgoing
anti-fermions on position {\tt in[1]}.

\begin{table}[h]
\bc
\begin{tabular}{ll}
\hspace*{-2mm}\begin{minipage}[ht]{6cm}{
\begin{tabular}{|lcll|}
\hline
in[0]      & $\to$ & in[1]      & in[2]     \\\hline
$f_1$      & $\to$ & $b$        & $f_2$     \\[1mm] 
$\bar f_1$ & $\to$ & $\bar f_2$ & $\bar b$  \\[1mm] 
$f_2$      & $\to$ & $\bar b$   & $f_1$     \\[1mm] 
$\bar f_2$ & $\to$ & $\bar f_1$ & $\bar b$  \\[1mm] 
$b$        & $\to$ & $\bar f_2$ & $f_1$     \\[1mm] 
$\bar b$   & $\to$ & $f_1$      & $\bar f_2$\\\hline
\end{tabular}
}\end{minipage} &
\begin{minipage}[ht]{7cm}{
\includegraphics[height=6.8cm]{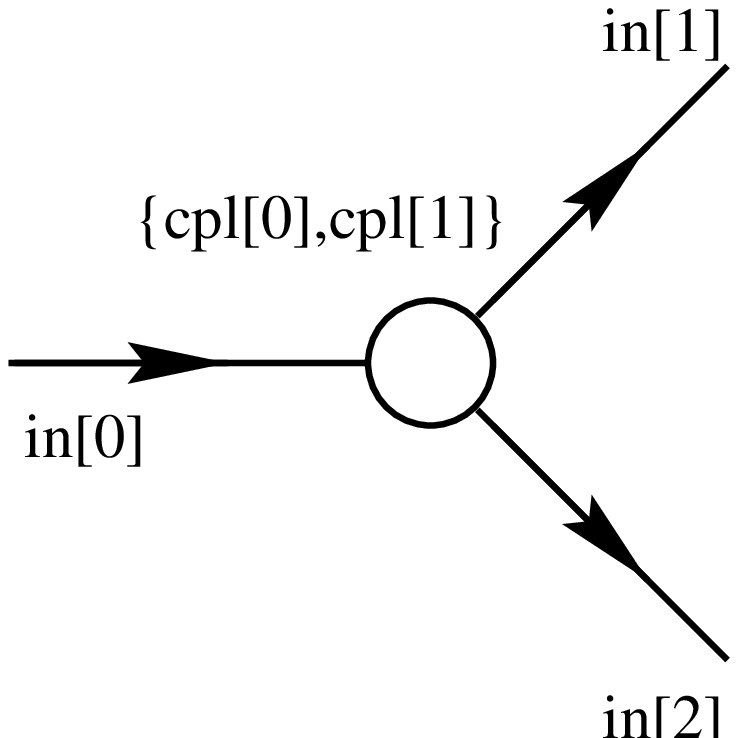}
}\end{minipage} 
\end{tabular}
\caption{\label{Vertices} Vertices within {\tt AMEGIC++}. Note (1) the
similarity with the point structure encountered before and (2) the
specific ordering of fermions and anti-fermions.}
\ec
\end{table}

\begin{figure}[h]
\begin{center}
\includegraphics[width=6.8cm]{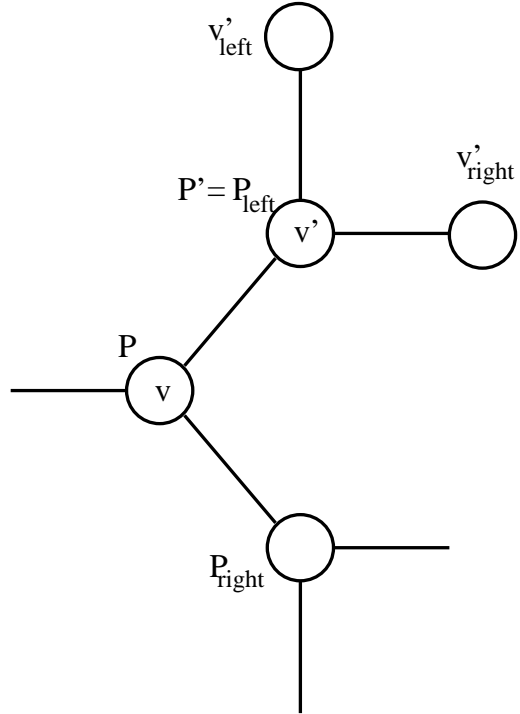}
\caption{\label{SimpleFit} Example topology for the mapping procedure.}
\end{center}
\end{figure}

The test of the vertices for a fit into a given topology now proceeds
on three levels:
\begin{enumerate}
\item{The flavours and $b$-flags of the point under consideration are
      determined. For simplicity, let us introduce {\tt fl[0]}, {\tt
      fl[1]}, and {\tt fl[2]} and similarly {\tt b[0]}, {\tt b[1]},
      and {\tt b[2]} for the flavours and $b$--flags ($b=\pm 1$ for
      incoming and outgoing particles) related to the points $P$,
      $P_{\rm left}$, and $P_{\rm right}$, respectively. Then the
      following adjustments are made (to allow a matching on the
      vertices): 
      \begin{itemize}
      \item A priori, if points $P_{\rm left}$ or $P_{\rm right}$ are
            not yet equipped with a flavour, i.e.\ if they are
            propagators, their respective $b$'s will be set to 0. 
      \item {\tt fl[0]} is a boson : If ${\tt b[1]} = -1$ 
             (${\tt b[2]} = -1$), then 
             ${\tt fl[1]}\to\overline{\tt fl[1]}$ 
             (${\tt fl[2]}\to\overline{\tt fl[2]}$). 
             If {\tt fl[1]} or {\tt fl[2]} are not known yet,  
             {\tt b[1]} = -1 or {\tt b[2]} = -1. 
       \item {\tt fl[0]} is an anti-fermion : If 
              ${\tt b[0]}\cdot{\tt b[1]}=1$, then 
              ${\tt fl[1]}\to\overline{\tt fl[1]}$. 
              If {\tt fl[1]} is not known yet, then 
              ${\tt b[1]} = {\tt b[0]}$.
       \item {\tt fl[0]} is a fermion : If 
              ${\tt b[0]}\cdot{\tt b[2]}=1$, then 
              ${\tt fl[2]}\to\overline{\tt fl[2]}$. 
              If {\tt fl[2]} is not known yet, then 
              ${\tt b[2]} = {\tt b[0]}$.
       \end{itemize}}
\item{Now every vertex available is tested against the -- modified --
      flavours of the specific point, i.e.\ $P$, $P_{\rm left}$, and 
      $P_{\rm right}$. Any vertex $v$ passing this test eventually
      defines the flavours of internal lines which then will have to fit
      themselves. The corresponding test is described in the next step.}
\item{Let us describe this test for flavour {\tt fl[1]} related to the
      point $P'=P_{\rm left}$, c.f. Fig.~\ref{SimpleFit}. If both its
      left and its right link (denoted by $P'_{\rm left}$ and 
      $P'_{\rm right}$) are endpoints, all available vertices are
      tested against them after the following transformations: 
      \begin{itemize}
      \item If $P'$ denotes a boson, the flavours of $P'_{\rm left}$ 
            and $P'_{\rm right}$ will be treated as barred and if the
            corresponding particles are incoming (${\tt b}=-1$).
      \item If $P'$ is an anti-fermion and if 
            ${\tt b'[0]}\cdot{\tt b'[1]}=1$, then 
            the flavour of $P'_{\rm left}$ will be regarded as barred.
      \item If $P'$ is a fermion and if 
            ${\tt b'[0]}_{v'}\cdot{\tt b'[2]}=1$, then 
            the flavour of $P'_{\rm right}$ is barred.
      \end{itemize}
      If now the modified flavours of any vertex $v'$ fit to the flavours
      of the points $P'$, $P'_{\rm left}$, and $P'_{\rm right}$ then
      this vertex makes an acceptable choice. If any such vertex is
      found, the flavour of $P'$ has passed the test.

      Of course, if either of the two points $P'_{\rm left}$ and 
      $P'_{\rm right}$ is not an endpoint, the test described above is
      unnecessary. }
\end{enumerate}

\subsubsection{Eliminating identical diagrams}

Unfortunately, the algorithms described in the preceeding sections
might produce identical diagrams, mainly because there is no really
sophisticated treatment of identical particles and topologies. Within
\AME \ therefore, the cure to potential double counting by identical
diagrams is explicit comparison of each pair of diagrams.
This again proceeds recursively by following the point lists of both
diagrams under test with the following algorithm 
\begin{enumerate}
\item If both points {\tt p$_1$} and {\tt p$_2$} to be compared do not
      have the same number of offsprings or have different flavour, the 
      recursion terminates. 
\item The test of step 1 is repeated with {\tt p$_1\!\to$left} and
      {\tt p$_2\!\to$left}, and possibly after passing with
      {\tt p$_1\!\to$right} and {\tt p$_2\!\to$right}.
\item If the test of step 2 yields a negative result, then there is
      still the possibility of double counting due to identical
      particles. To catch this the flavours of 
      {\tt p$_{1,2}\to\{$left, right$\}$} are checked
      crosswise. If this test is passed, i.e.\ if the flavours of  
      {\tt p$_1\!\to$right} and {\tt p$_2\!\to$left} and of
      {\tt p$_1\!\to$left} and {\tt p$_2\!\to$right} mutually agree,
      step 1 is repeated with both combinations.
\item Finally, if the recursion does not terminate before all points
      were tested, the diagrams are identical and the second one will
      be discarded. 
\end{enumerate}

\subsection{Translation into helicity amplitudes\label{HelAmpGen}}

In this section, we will turn our focus on the translation of the
generated Feynman diagrams into corresponding helicity amplitudes and
on some items concerning their evaluation once a set of external
momenta is given. The following issues will be discussed in some
detail:
\begin{enumerate}
\item The treatment of the momentum flow along the spinor lines,
\item the piecewise conversion of the diagrams into corresponding
      $Z$--functions,
\item and the evaluation of the helicity amplitudes.
\end{enumerate}

\subsubsection{Spinor direction and momentum flow}

\begin{figure}[h]
\begin{center}
\begin{tabular}{cc}
\parbox{7cm}{\includegraphics[width=6cm]{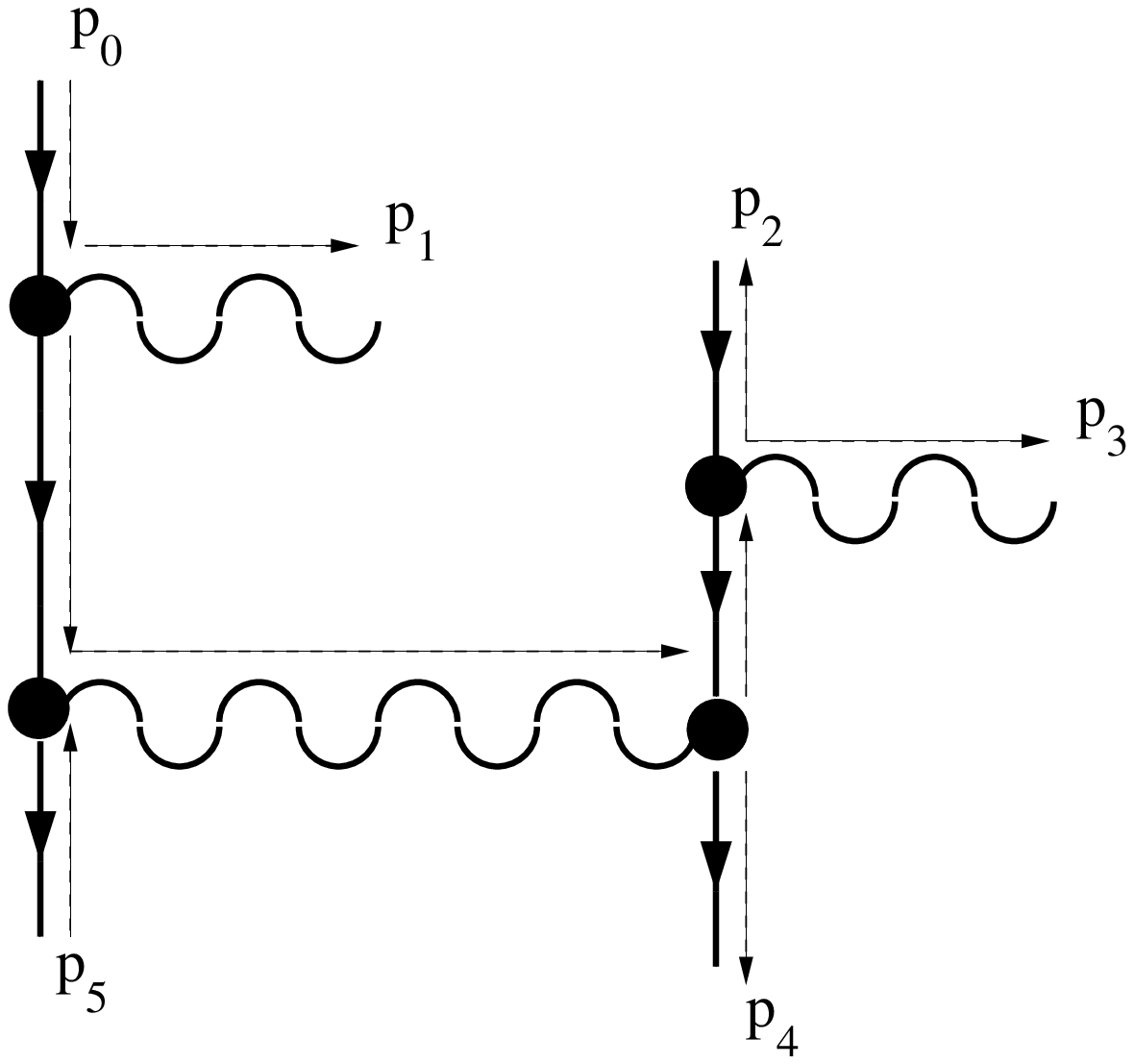}} &
\parbox{7cm}{\includegraphics[width=6cm]{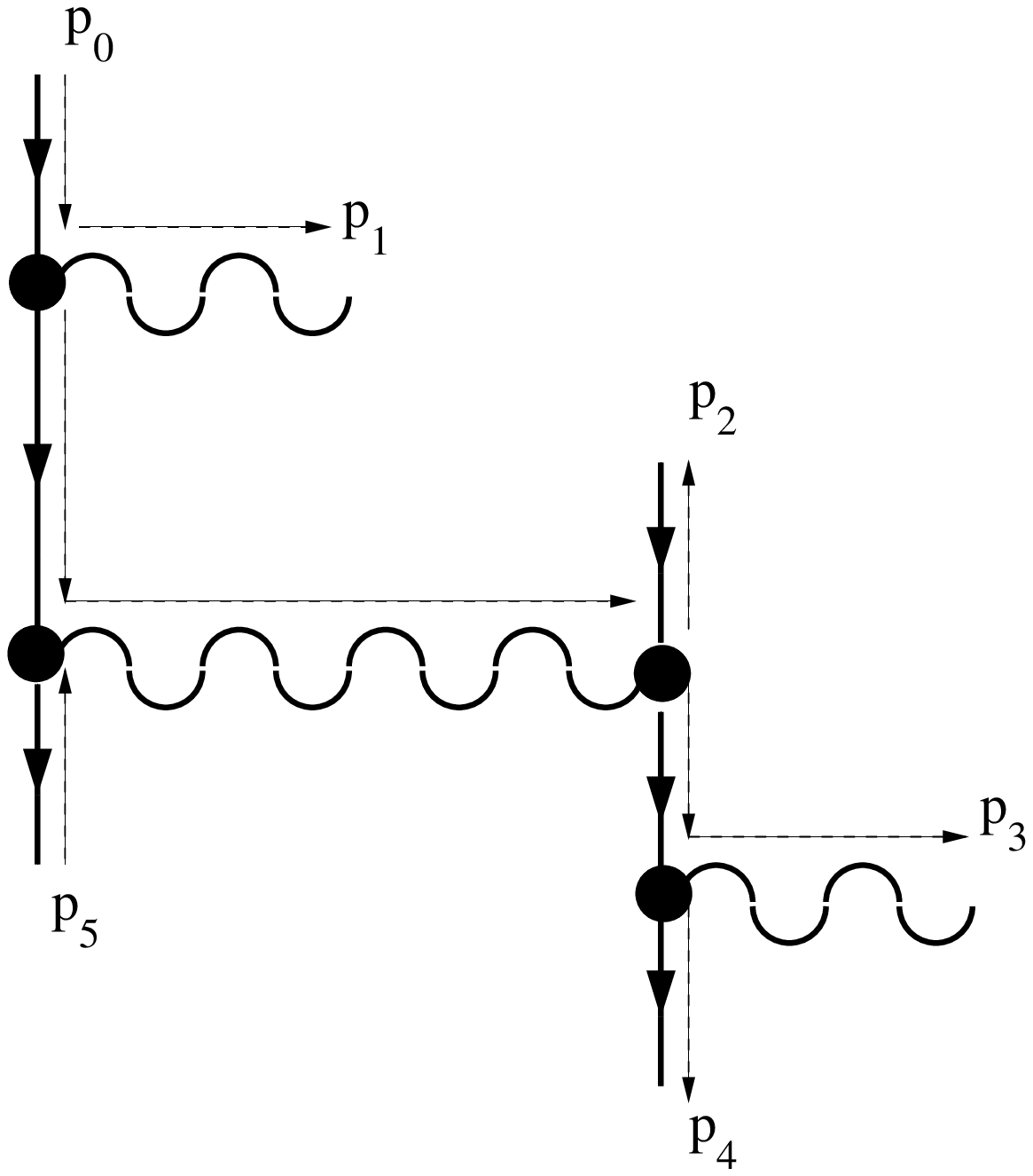}} \\
$\displaystyle\bar v(-p_5)\gamma_\mu\frac{p\slaa_2+p\slaa_3+p\slaa_4-p\slaa_5}
                            {(p_2+p_3+p_4-p_5)^2}
             \gamma_\nu u(p_0)$ &
$\displaystyle\bar v(-p_5)\gamma_\mu\frac{p\slaa_2+p\slaa_3+p\slaa_4-p\slaa_5}
                            {(p_2+p_3+p_4-p_5)^2}
             \gamma_\nu u(p_0)$ \\
$\displaystyle\times
 \bar u(p_4)\gamma^\mu\frac{-(p\slaa_2+p\slaa_3)}
                           {(p_2+p_3)^2}
            \gamma_\rho v(-p_2)$ &
$\displaystyle\times
 \bar u(p_4)\gamma_\rho\frac{(p\slaa_3+p\slaa_4)}
                            {(p_3+p_4)^2}
            \gamma^\mu v(-p_2)$ 
\end{tabular}
\caption{\label{FigSpinlines1} Example graphs for the process
$e^-(p_0)e^+(p_5)\to e^-(p_2)e^+(p_4)\gamma(p_1)\gamma(p_3)$ and the
spinorial part of the amplitudes. Clearly, there is a difference in the
sign of both diagrams due to the misalignment of momentum and
spin--flow in the left diagram. Note that $t$--channel like
topologies emerging from exchanging incoming $e^+$ and outgoing $e^-$
do not alter the overall signs of the propagators. Instead, they merely
lead to a replacement $p_2\leftrightarrow -p_5$.}
\end{center}
\end{figure}

Having generated Feynman diagrams including relative factors of $-1$
for the exchange of two identical fermions, we are now in the position
to construct the corresponding helicity amplitudes. However, there's
a caveat here, since another source of relative phases of $-1$ has
not been discussed yet. These additional factors stem from a mismatch
of momentum and spinor flow along a line. To illustrate this point,
consider the diagrams exhibited in Fig.~\ref{FigSpinlines1}. Obviously,
the sign of the momentum of each fermion line, either external or
intermediate, is defined with respect to the spin direction. This
results for instance in spinors $\bar v(-p_5)$ for the incoming
positron in the graphs depicted in Fig.~\ref{FigSpinlines1}. 

\begin{figure}[h]
\begin{center}
\begin{tabular}{cc}
\parbox{7cm}{\includegraphics[width=6cm]{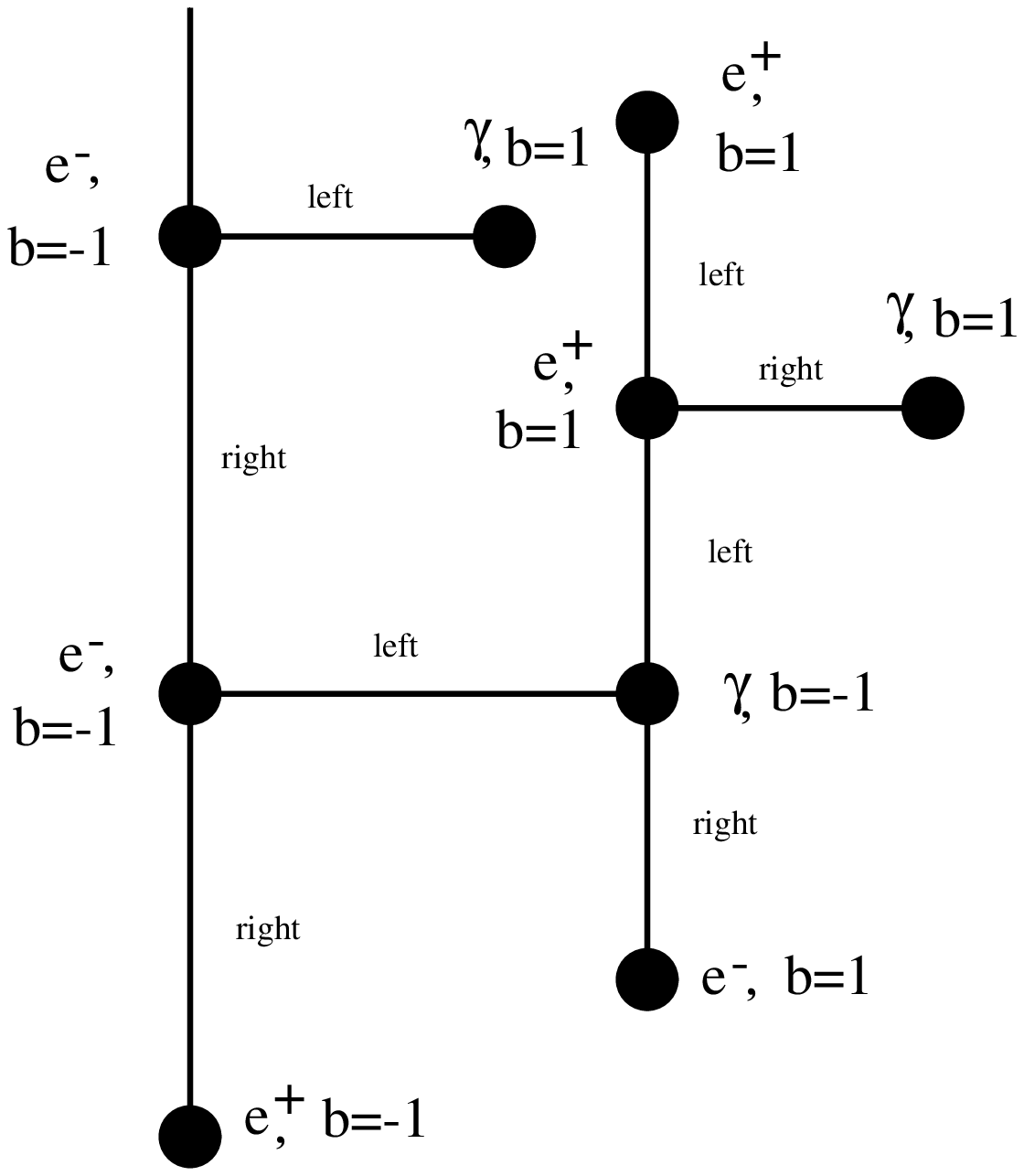}} &
\parbox{7cm}{\includegraphics[width=6cm]{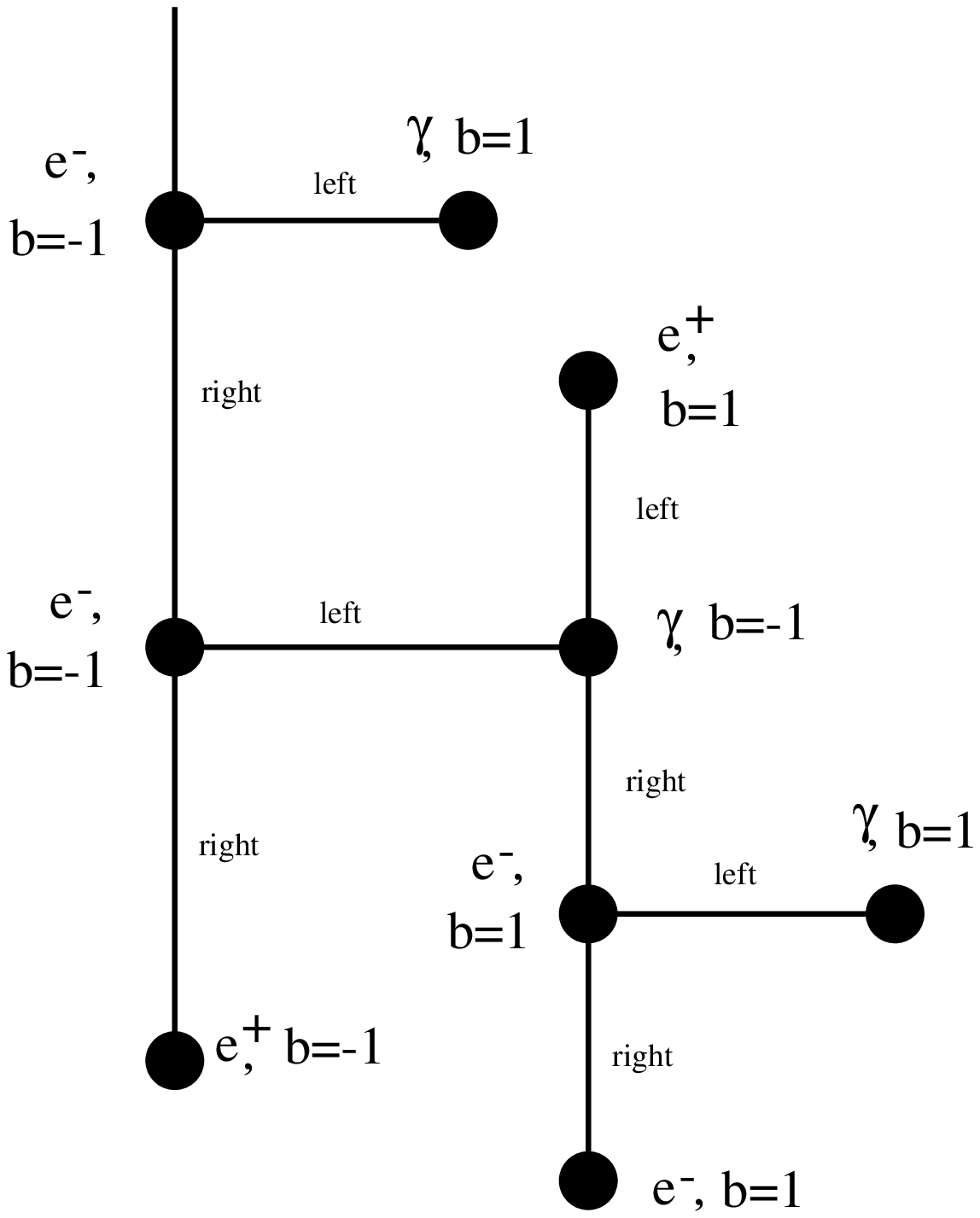}} 
\end{tabular}
\caption{\label{FigSpinlinesP} Assignments for the point lists
corresponding to the graphs in Fig.~\ref{FigSpinlines1}. Note that on the
level of point lists, the $t$--channel topologies differ only in their
endpoints.}
\end{center}
\end{figure}

Coming back to the construction of Feynman amplitudes described above,
the flavours of the point lists are given as in Fig.~\ref{FigSpinlinesP}.
For all but the leftmost fermion lines, a general rule can be
formulated as follows:
\begin{itemize}
\item If a point {\tt p} has an incoming boson as flavour, and if
      {\tt p$\to$left} is an intermediate fermion line (with number 100 or
      larger), then the overall sign of the diagram is multiplied by $-1$. 
\end{itemize}
This rule relies only on the fact that for incoming boson lines, there
is a strict ordering of outgoing fermion lines with respect to
particles/anti-particles. This ordering immediately allows to extract
information about the alignment of spin and momentum flow along the
lines. 

However, the rule above is of somewhat limited use only, since
apparently the leftmost fermion line, i.e.\ the one starting at the
point {\tt p[0]} is not dealt with. According to the
algorithms employed for the construction of the amplitudes, the first
point of this line, {\tt p[0]}, is occupied by an incoming particle.

The leftmost fermion line, i.e.\ the line starting at the first point, 
might contribute a change of sign, if it starts with an incoming particle
represented by an $u$--spinor or with an outgoing anti-particle
represented by a $v$--spinor. Then the corresponding endpoint of the
spinor line has to be found. This endpoint is then represented by
either an $\bar u$ or a $\bar v$--spinor, i.e.\ it is either an outgoing
particle or an incoming anti-particle. 

\subsubsection{Finding the appropriate Z--functions}

Now we are in the position to present the algorithms leading finally
to the helicity amplitudes, or in other words, to the appropriate
products of $Z$--functions. This proceeds in two steps :

\begin{enumerate}
\item Projecting the point list onto the larger building blocks,
      i.e.\ pieces with a number of intermediate vector and scalar
      bosons, and
\item flipping the arguments of the emerging $Z$--functions, such
      that the particles associated with a ``barred'' spinor are always
      followed by particles with ``unbarred'' spinors. This point
      applies for the spinors constituting the polarization vectors of
      spin--1 bosons as well.
\end{enumerate}
\begin{tabular}{ll}
\hspace*{-2mm}\begin{minipage}[ht]{9cm}{
We will now elucidate the two steps above. For a better understanding
of the projection algorithm mapping the point lists onto helicity
amplitudes we recall briefly the structure of the basic building
blocks available within {\tt AMEGIC++}. Schematically, they can be represented
like in the figure on the right. They all consist of a number of boson
propagators -- the wiggly lines -- between fermion propagators or
external spinors, related to fermions or their polarization vectors,
represented by the arrowed straight lines. The blob in the middle then
represents all possible Lorentz--structures connecting the boson
propagators.
}\end{minipage} &
\begin{minipage}[ht]{7cm}{
\includegraphics[height=4cm]{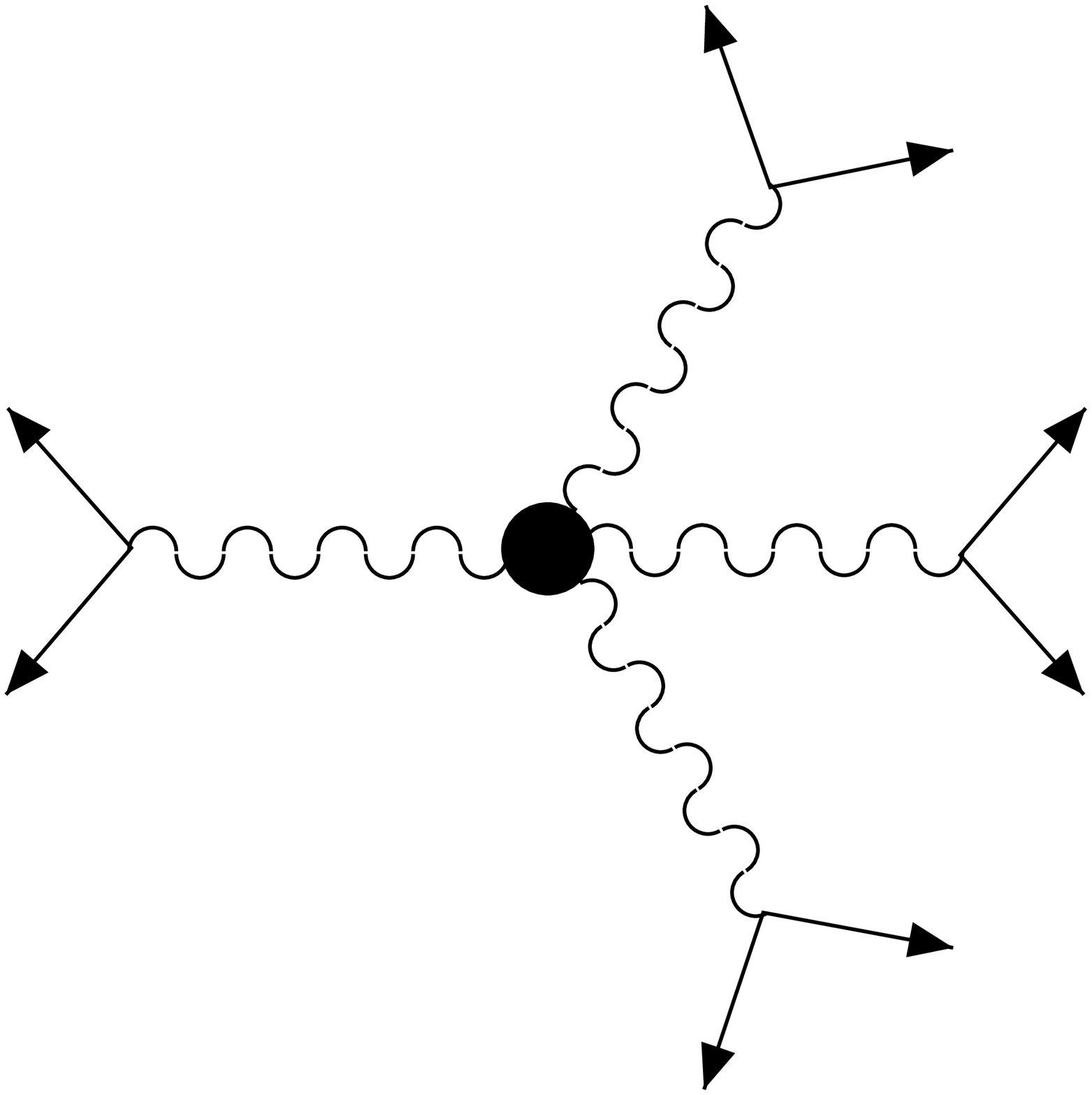}
}\end{minipage} 
\end{tabular}

\vspace{0.3cm}

\begin{figure}
\begin{tabular}{lcl}
\begin{minipage}[ht]{6cm}{
\includegraphics[width=6cm]{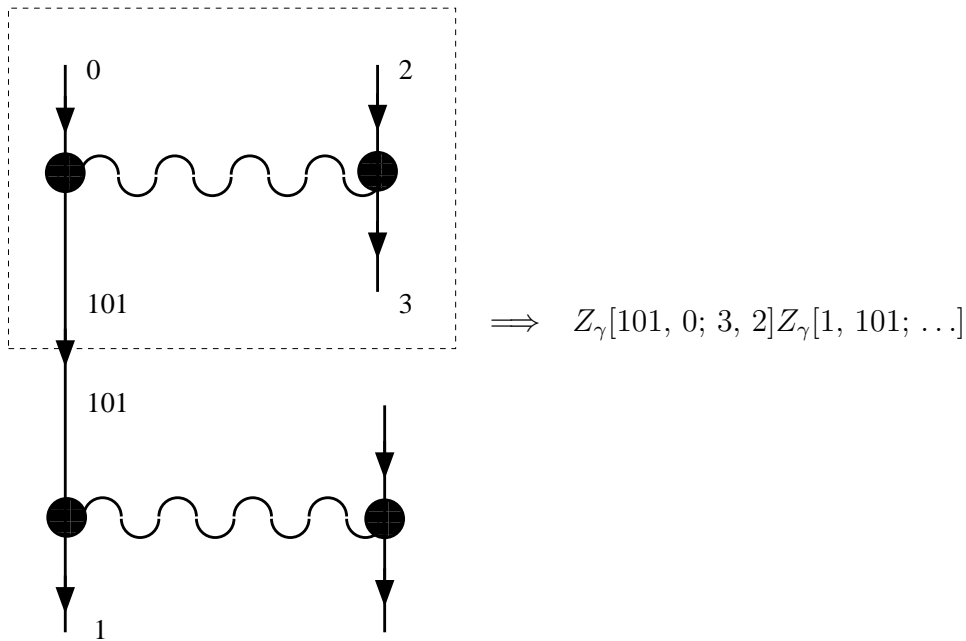}
}\end{minipage} &
$\Longrightarrow$ &
\begin{minipage}[ht]{6cm}{
$Z_\gamma[101,\,0;\,3,\,2] Z_\gamma[1,\,101;\,\dots]$
}\end{minipage} 
\end{tabular}
\caption{\label{ZflipEx} Example diagram for flipping the ordering
of spinors within $Z$--functions.}
\end{figure}

\noindent These building blocks naturally emerge, because intermediate
fermion lines can be cut by use of the completeness relation, cf.\
Eq.~(\ref{completeness}), while vector boson lines cannot be cut in such
a way. These considerations lead straightforward to the algorithm
employed in the mapping procedure sketched below:
\begin{enumerate}
\item Start with a point {\tt p}. Check, if the particle related to
      this point is either an external or an intermediate fermion (with
      {\tt p$\to$left} a boson and {\tt p$\to$right} a fermion), an
      external or intermediate anti-fermion (with {\tt p$\to$left} an
      anti-fermion and {\tt p$\to$right} a boson) or an external boson.
      This condition ensures that 
      \begin{enumerate}
      \item a reasonable translation is possible at all, where the
            ordering of the connected points for incoming fermions or 
            anti-fermions is just another check, and that
      \item intermediate boson lines are not double counted.  
      \end{enumerate}
      If the conditions above are met, the boson is selected and the
      steps 2 and 3 are executed.
\item Starting from the boson, the links of the subsequent points are
      followed recursively along possibly occurring further
      bosons. In every boson line branching into two fermions this
      recursion terminates. By this procedure, the number of bosons,
      their type and their topological connections are explored. These
      characteristics determine which of the building blocks
      exhibited in Figs.~\ref{Figv3}, \ref{Figv4}, and \ref{Figv5} are
      to be used. 
\item Having determined the structure of the building block, the task
      left for the translation of this particular piece is to fill in
      the corresponding arguments and couplings. 
\item The procedure is repeated with the points {\tt p$\to$left} and
      {\tt p$\to$right}.
\end{enumerate}\noindent
Unfortunately, this procedure does not care about any ordering
of spinors coming into play either via external particles or their
polarizations or by using the completeness relation on the
intermediate fermion lines. Let us illustrate this by considering the
amplitude depicted in Fig.~\ref{ZflipEx}

\noindent
We see immediately that the particle indices occur in ordered pairs of
the form $\{\bar u,\, u\}$. However, within {\tt AMEGIC++}, this
ordering of the spinor pieces against the spin flow is mandatory but
is not necessarily realized immediately after the construction of the
helicity amplitudes. Hence, after being constructed, the Z--functions   
are searched for external fermions or anti-fermions not obeying the
ordering $\bar u,\,u$. If found, the indices within the corresponding
pair are switched. If the partner--index denotes a propagator, the
index occurs twice due to the completeness relation. Consequently it
has to appear exactly once on the first and on the second position of
such a pair. This is taken care of by switching the sequence in the
other Z--function in which this particular index appears.  

\subsubsection{Evaluation of the amplitudes}

As the last item related to the helicity amplitudes, we 
discuss briefly, how for a given set of external momenta and
helicities a particular amplitude is evaluated. The key point to
notice here is that we cut open the fermion lines by use of the
completeness relation Eq.~(\ref{completeness}). Thus for each
intermediate fermion line a sum over both helicities and both particle 
and anti-particle type spinors ($u\bar u$ and $v\bar v$, respectively) 
has to be performed. For massive particles, the masses of the $u$--
and the $v$- spinors are connected with different signs, entering
finally the $Z$--functions. Therefore, the internal fermion lines are
counted first, then for each internal fermion line two summation
indices are introduced, labeling the two helicities and the particle
and anti-particle components, respectively. For each combination of
these indices, the corresponding spinor labels are filled into the
Z--functions, and a minus sign internally labels anti-particles and will be 
recognized at the level of the building blocks to alter the sign of
the corresponding mass. Finally, for each combination of internal
helicities and particle anti-particle labels all building blocks are
multiplied separately to be summed.

\subsection{Generation of integration channels\label{GenChan}}

Let us turn now to the generation of additional channels for the
multi--channel phase space integration performed by {\tt AMEGIC++}. As
already explained in Sec.~\ref{TB}, for the definition of efficient
integration channels it is essential to know as much as possible about
the structure of the integrand, i.e.\ possible (in the Monte Carlo
sense) singularities in phase space. For {\tt AMEGIC++}, this proves
to be the case, because the Feynman diagrams are already at hand. So,
basically, the steps performed within \AME for the construction of
channels are: 
\begin{enumerate}
\item For each Feynman diagram the internal lines are identified and 
      it is decided, whether they are in the $s$-- or in the
      $t$--channel, 
\item then the propagators and decays are translated according to their
      properties into the building blocks listed in Tab.~\ref{PSBB},
\item finally a subset of the individual channels (each corresponding
      to one specific diagram) is selected.
\end{enumerate}

\subsubsection{Properties of the internal lines}

\begin{figure}[h]
\bea
\nonumber
\includegraphics[width=6.5cm]{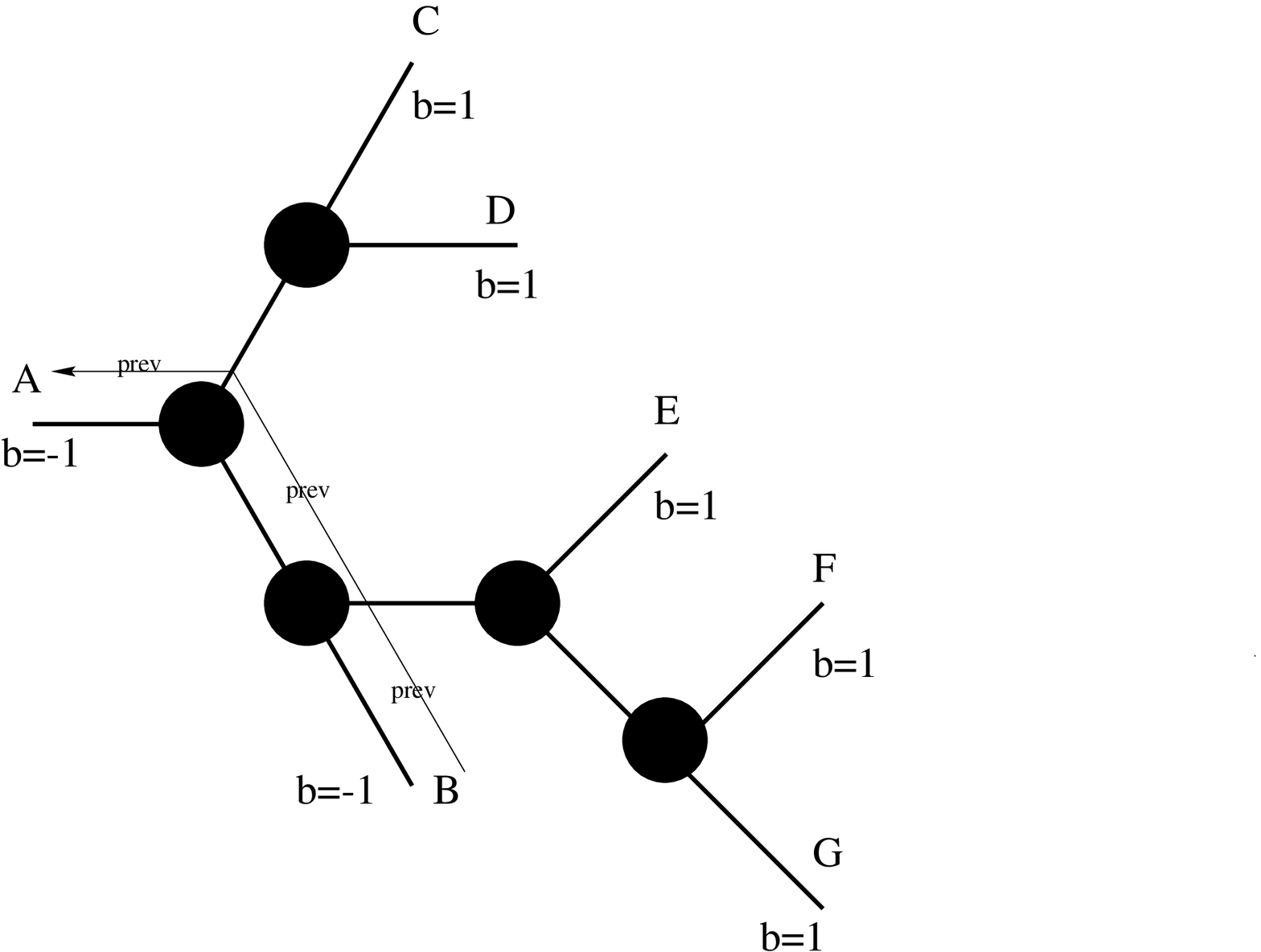}
\;\;\;\mbox{\raisebox{4cm}{$\Longrightarrow$}}\;\;\;
\mbox{\raisebox{2cm}{\includegraphics[width=5cm]{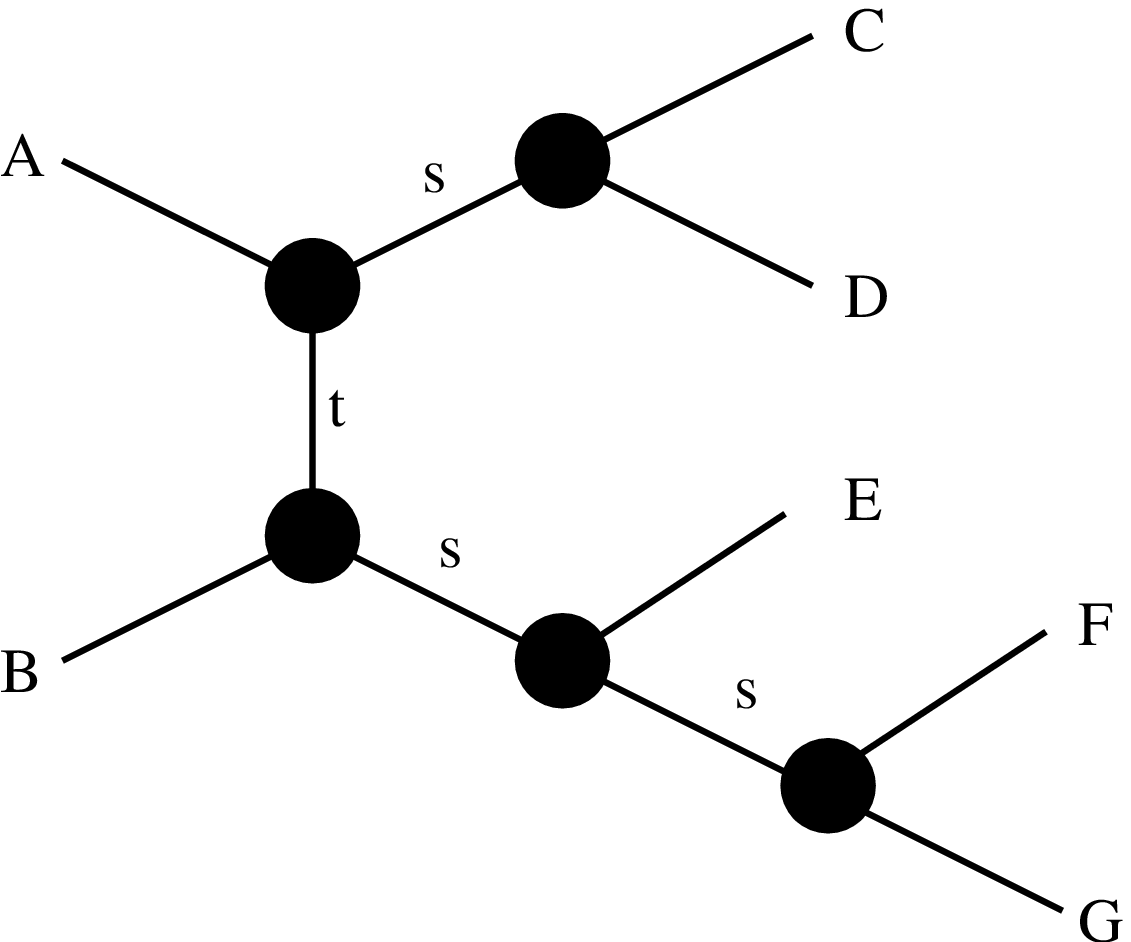}}}
\eea
\caption{\label{st} Identifying the propagator types for a process
$A+B \to C+D+E+F+G$.}
\end{figure}

The first step in the construction of an integration channel for a
particular Feynman diagram is to decide, in which kinematical region
the individual propagators are, i.e.\ whether they are $s$-- or
$t$--channel propagators. This again is done recursively starting
from position $1$ in the list of linked points constituting the
diagram. By iterative steps to the left and right links the other
endpoint with $b=-1$, i.e.\ the other endpoint related to the second
incoming particle is found. In these iterative steps, connections to
the corresponding previous points are set. It is then straightforward to
follow their track back to the starting point, equipping each
propagator on the way with a flag indicating that this is a
$t$--channel propagator. More graphically, this identification
amounts to a ``redrawing the diagram with $t$--channels running
vertically and $s$--channels running horizontally'', see Fig.~\ref{st}.

\subsubsection{Construction of channels}

\begin{figure}[h]
\bea
\includegraphics[width=5cm]{picts/ConstructChannel2.eps}
\;\;\;\mbox{\raisebox{2cm}{$\Longrightarrow$}}\;\;\;
\mbox{\raisebox{0cm}
{\includegraphics[width=5cm]{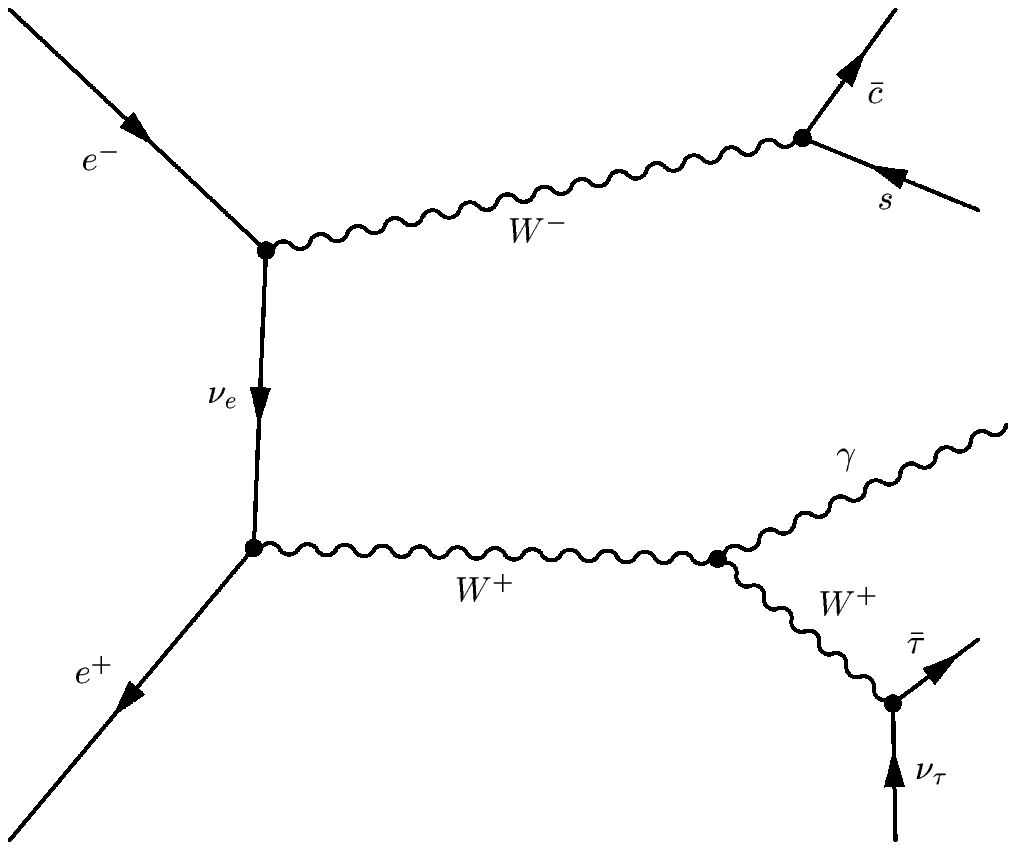}}}\nnb
\eea
\caption{\label{st3} One of the diagrams related to the process 
$e^+e^-\to s\bar c\bar\tau\nu_\tau\gamma$.}
\end{figure}

Now we are in the position to construct appropriate channels. The
steps here are the following:
\begin{enumerate}{
\item{Depending on the number of $t$--channel propagators, the
      construction of channels in principle starts either with an
      isotropic two--body decay, with a $t$--channel propagator or
      with an isotropic three--body decay (for $n_t=0,\,1,\,2$,
      respectively).} 
\item{The first task when implementing any decay then is to determine
      the (virtual) masses of the decay products. This is achieved in
      the following way: 
     \begin{itemize}
     \item{First, decay products which are outgoing particles in the
           process under consideration are put on their mass--shell.}
     \item{Then, decay products which are propagators are equipped
           with a virtual mass squared $s$. It satisfies
           $s_{\rm min}\le s\le s_{\rm max}$ and is distributed
           either according to a single pole (massless propagators,
           i.e.\ propagators with mass $m_p^2<s_{\rm min}$) or
           according to a Breit--Wigner distribution (massive 
           propagators, i.e.\ $m_p^2>s_{\rm min}$).

           The limits of the virtual mass squared $s$ are given by the 
           following considerations: 
           $\sqrt{s}$ obviously should be larger
           than the sum of the masses of all outgoing particles
           produced in the subsequent decays of the propagators. In
           turn, $\sqrt{s}$ running in a propagator should be smaller
           than the energy squared entering the production vertex of
           the propagator minus the masses of the other propagators 
	   which are already chosen, and minus the minimum masses of
           the propagators with yet undefined masses.} 
     \end{itemize}}
\item{After having chosen in the first step the basic form of the
      channel, i.e.\ the form of the first decay, the subsequent decays
      are filled in recursively. By default, for the secondary decays,
      only isotropic and anisotropic two--body decays are
      available. Again, the momentum transfer squared along the
      propagators is chosen according to the step above. This
      procedure of decays and propagators is iterated until only
      outgoing particles remain.}  
}\end{enumerate}
As an example, let us consider the diagram depicted in Fig.~\ref{st3},
contributing to the process $e^+e^-\to s\bar c\bar\tau\nu_\tau\gamma$.
Let us assume for better illustration that both propagators $CD$ and
$EFG$ are  resonating with $M_{CD}\Gamma_{CD} > M_{EFG}\Gamma_{EFG}$
(in fact, in our example process the two products are
identical).

The corresponding channel is constructed in the following way:
\begin{enumerate}
\item{Counting the number of $t$--channel propagators, we find
      $n_t=1$. Therefore the basic form of the channel is determined
      to be of the simple $t$--channel form.}
\item{With $s=(p_A+p_B)^2$ we find the following minimal and maximal
      momentum transfers:
      \bea\begin{array}{lcllcl}
      s^{CD}_{\rm min} &=& (m_C+m_D)^2\,,\;\;\;\;\;\;\;&
      s^{CD}_{\rm max} &=& 
          \left(\sqrt{s}-\sqrt{s_{\rm min}^{EFG}}\right)^2\nnb\\[5mm]
      s^{EFG}_{\rm min} &=& (m_E+m_F+m_G)^2\,,\;\;&
      s^{EFG}_{\rm max} &=& 
          \left(\sqrt{s}-\sqrt{s_{\rm min}^{CD}}\right)^2\,.\nnb
      \end{array}\\
      \eea
      Then $s^{EFG}$ is chosen first within its limits as
      given in the equations above, and only then  $s^{CD}$ is
      determined, with $s_{\rm min}^{EFG}$ replaced by
      $s^{EFG}$. These virtual masses for the propagators enter the
      $t$--channel decay of step 1.} 
\item{We continue along the $CD$--propagator. Since both
      particles $C$ and $D$ are fermions, the decay of $CD\to C+D$ is
      translated as an isotropic two--body decay.} 
\item{Let us follow now the $EFG$--line, and consider the radiation of
      particle $E$, a photon, off this line. As it is well
      known, massless vector bosons tend to have an enhancement in the 
      collinear and soft region of emission. Therefore, this decay
      will be described via an anisotropic two--body decay within the
      channel.} 
\item{Finally, since particles $F$ and $G$ are fermions,
      as depicted in our example, their decay is described by an
      isotropic two--body decay.}  
\end{enumerate}
The corresponding code generated by \AME for this specific channel can
be found in Appendix \ref{SampleCode}.

\subsubsection{Selection of channels}
Having generated one channel for each diagram, two things become
pretty obvious during integration:
\begin{enumerate}
\item{Not all diagrams are equally singular, i.e.\ some of the channels 
      are more effective, whereas others almost do not contribute.}
\item{Having too many channels, the multi--channel method is not
      awfully efficient. The various -- and often irrelevant --
      channels are ``stealing'' a priori weights from each other and
      from the good, relevant channels.}
\end{enumerate}
To alleviate this situation, in {\tt AMEGIC++}, some selection of good
channels is performed before the integration starts. Depending on the
c.m. energy available for this process, the propagators which yield the
singular behavior, are examined for each channel. The maximal number
of potentially resonant propagators is determined for each channel,
and the most ``successful'' channels in this category are used in the
phase space integration.

\clearpage\newpage
\section{Implementation\label{Classes}}

Basically, the determination of cross sections which is the main
task of {\tt AMEGIC++}, can be divided into three major stages:
\begin{enumerate}
\item{The input has to be classified, i.e.\ the process(es) to 
      be calculated as well as the framework of the model must 
      be determined. At this stage incoming and outgoing particles
      are specified, particle spectra and couplings are determined,
      the Feynman rules are established and eventually, using these
      Feynman rules, particle widths are calculated. Actually, this 
      last step already invokes the other two stages, i.e.\ 
      generation of Feynman amplitudes and their integration.}
\item{Now, having established the model framework, the Feynman 
      diagrams related to each single process have to be constructed
      and translated into helicity amplitudes. Hence, the scattering
      amplitude is nearly ready for integration. But since some of the 
      helicity combinations yield exactly zero, the calculation 
      can be much alleviated, when these parts are eliminated 
      beforehand. Additionally, the amplitude can be compactified
      analytically by common factors. These manipulations are performed
      by translating the expression for the amplitude into a character 
      string and simplify this character string accordingly. Now,
      evaluating an amplitude for given external momenta means to 
      interpret a character string. Saving this very string into a 
      {\tt C++} library file and linking this file together with the
      main program speeds up the calculation considerably.
      This is the last refining of the amplitudes, and they are ready 
      for integration.}
\item{Finally, the scattering amplitude has to be integrated by Monte
      Carlo methods. Hence, the final step consists of the random 
      generation of phase space points, i.e.\ of sets of four-momenta 
      for the outgoing particles and their summation in order to
      obtain the cross section. This task could be easily achieved, if
      the amplitude was uniform in the whole phase space and would not
      possess any divergencies in the Monte Carlo sense, i.e.\ sharp
      peaks. However, as usual life is much more difficult than one
      might wish and scattering amplitudes tend to have an abundance
      of peaks in regions, where particles become either soft,
      collinear or resonant. The mapping of the uniform and flat
      distribution of phase space points onto a structure which suits
      the one of the amplitude, is crucial for the phase space
      integration.}
\end{enumerate}
Following these three major steps, this chapter is structured as
follows: In analogy to the first stage, we first describe in
Sec.~\ref{Organization}, how the overall organization of the whole
program works out, and how the processes are set up. Then, in
Sec.~\ref{Model} we discuss the way, the model under consideration
enters. In Sec.~\ref{Amplitude}, we then turn our attention to the
generation of Feynman diagrams and their translation into helicity
amplitudes. The next section, Sec.~\ref{Strings}, is dedicated to the
subsequent treatment of these amplitudes in terms of character
strings. Here some of the secrets of the ancient {\tt kabbalistic
system} will be disclosed. The final integration over the whole phase
space will be discussed in the next section, Sec.~\ref{PhaseSpace}.
Last but not least a lot of small helper routines must be provided for
accomplishing all tasks. Their explanation as well as the description
of the system of parameters and switches will complete this chapter
with its last two sections.

\subsection{Organization\label{Organization}}

\begin{figure}
\includegraphics[height=13cm]{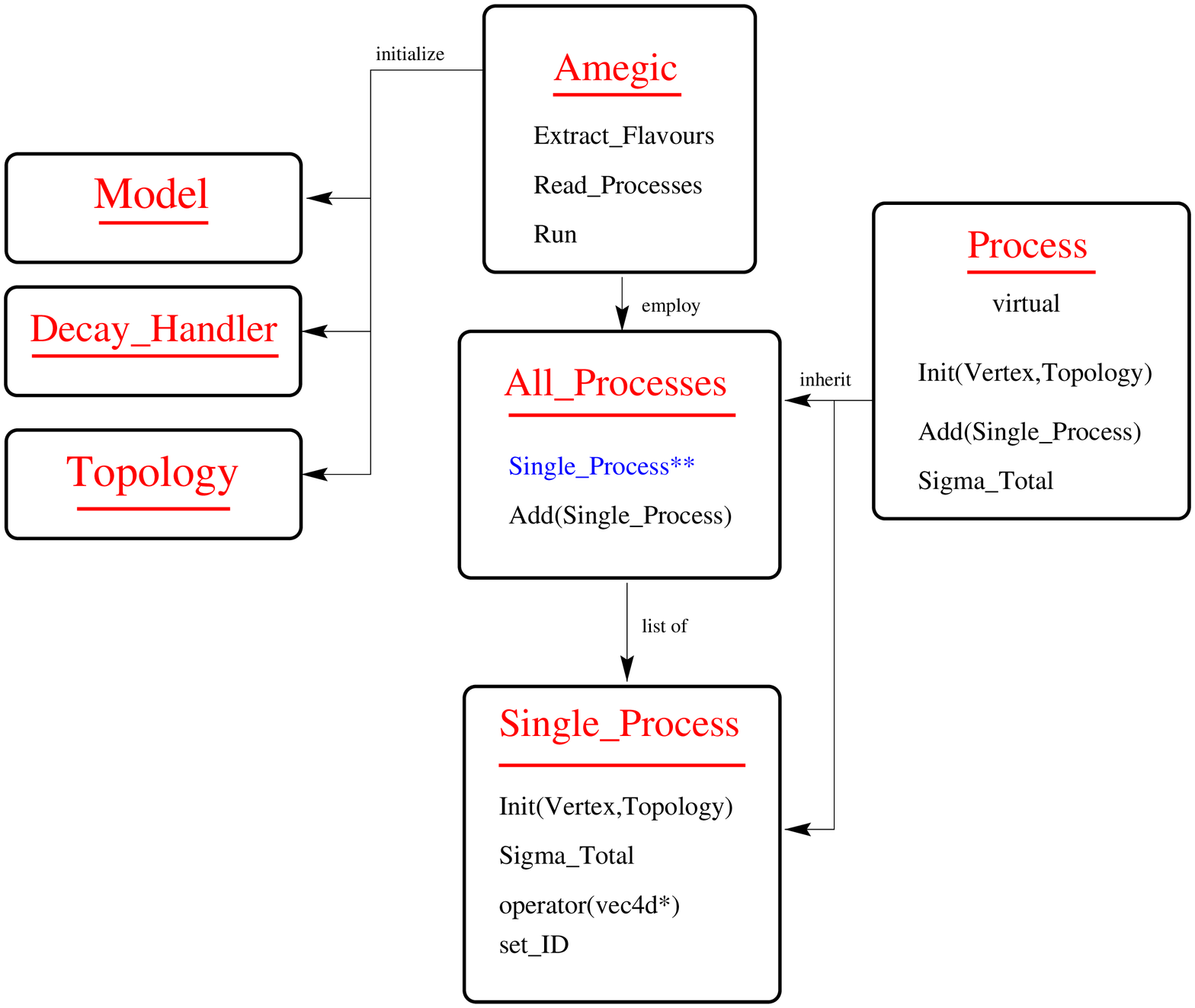}
\caption{\label{MainStruc}Main structure of the program.}
\end{figure}

The central class steering all processes, i.e.\ the class which is used
within the {\tt main} routine, is called unsurprisingly {\tt Amegic}. 
During the initialization part it specifies the model, i.e.\ the
Feynman rules, starts the calculation of the decay widths and 
branching ratios of unstable particles, determines all possible
topologies up to a certain maximum number of outgoing legs and
initializes the individual processes by reading in the incoming and
outgoing particles. Therefore it employs all the other major classes
of the program, see Fig.~\ref{MainStruc}. 

Specific details of the initialization of the chosen model can be
found in Sec.~\ref{Model}, whereas the other classes will be
described in this section. A brief outline of the classes discussed 
in this section can be found in Tab.~\ref{MainClass}.
\begin{table}[h] 
\bc 
\begin{tabular}{l|l} 
Class/Struct & Purpose\\ 
\hline
&\\ 
Amegic           & Main class, reads in the process parameters.\\               
Decay\_Handler   & Handles all decays of the unstable particles,\\
                 & calculates widths and branching ratios.\\ 
DecayTable       & The table of decay chains.\\
Process          & Mother class of all processes.\\
All\_Processes   & Maintains a number of processes.\\
Single\_Process  & Maintains the initialization and calculation of\\
                 & one single process.\\
Amplitude\_Base  & Container for all integrable amplitudes.\\
Helicity         & Determines the helicity combinations yielding\\
	         & a non--zero result. Explores and uses symmetries,\\
	         & i.e.\ identical results for different helicity\\
                 & combinations.\\		 
struct signlist  & List of helicity signs, contains {\tt on}--switches.\\
Polarisation     & Provides the polarisation vectors of external
                   bosons.\\     
Point            & A point structure for building binary trees.\\  
Topology         & Calculates all possible topologies.\\ 
Single\_Topology & Contains a single topology.\\
&\\
\end{tabular}
\ec
\caption{\label{MainClass}A short description of the major classes handling the
         organization of {\tt AMEGIC++}.}
\end{table}                                                                   
\subsubsection{Organizing the Organization - {\tt Amegic}}

The class {\tt Amegic} contains the following methods :
\begin{enumerate}
\item The {\tt Constructor} of this class runs the whole
      initialization procedure of the program along the following
      steps: 
      \begin{itemize}
      \item The model file which has been chosen from the input
            parameters, is specified and initialized. During this
            stage the spectrum of the model is determined. 
      \item The vertices which are in fact the Feynman rules deduced
            by the model are initialized.
      \item The processes to be calculated are read in from the {\tt
            Process.dat} data file with the help of {\tt
            Read\_Processes()}. The incoming and outgoing
            particles for the different processes are fixed.
      \item Having at hand the maximum number of external legs the raw
            topologies can be determined by constructing a {\tt
            Topology} object.  
      \item For every particle which is marked as unstable in the 
            {\tt Particle.dat} data file, the total width and the
            branching ratios have to be calculated using the {\tt
            Decay\_Handler}. With {\tt
            Decay\_Handler::Find\_Unstable()} all unstable particles
            are detected and listed in a {\tt DecayTable}. Now, {\tt
            Decay\_Handler::Calculate()} evaluates the width and the
            branching fractions of the particles.      
      \item Last but not least the individual processes will be
            initialized via\\ {\tt Process::Init()}. Consequently the
            raw topologies are combined with the vertices yielding the
            Feynman diagrams which are then translated into helicity
            amplitudes. 
      \end{itemize} 
\item {\tt Read\_Processes()} reads in the list of processes from
      {\tt Process.dat}, extracts the flavours with the help of
      {\tt Extract\_Flavours()}, and groups the processes into
      {\tt All\_Processes}. If only one process is to be
      considered, a {\tt Single\_Process} will be created
      instead. 
\item In {\tt Extract\_Flavours()} the raw text given in the
      {\tt Process.dat} data file is translated into the numbers
      and flavours of the incoming and outgoing particles. 
\item {\tt Run()} calculates the total cross section of all
      involved processes via\\ {\tt Process::Sigma\_Total()}.
\end{enumerate}

\subsubsection{Decays - {\tt Decay\_Handler}}

The primary aim of the {\tt Decay\_Handler} is to generate a {\tt
DecayTable} containing all information concerning the decays of
unstable particles, i.e.\ total widths, branching ratios, the flavour
of the decay products etc.. The {\tt DecayTable} is constructed as a
two--fold list in the following way.

\begin{tabular}{ll}
\hspace*{-5mm}\begin{minipage}[h]{10cm}{
Every entry into the {\tt DecayTable} itself is a {\tt DecayTable},
having pointers into two directions (see Figure to the right):
\begin{enumerate}
\item{{\tt Next}, pointing to the next decaying particle or the next
decay channel, and}
\item{{\tt More}, pointing to a list of decay channels for the same
      decaying particle.}
\end{enumerate}
Furthermore, each {\tt DecayTable} has a varying (limited by the user)
number of flavours, the flavour of the decaying particle plus the 
flavours of its decay products. The user defines the maximal number of
particles created in a decay. If, in the process of setting up the 
{\tt DecayTable}, a particle decays into a final state consisting of
one or more unstable particles, these can be then replaced by their
own decay products. 
}\end{minipage} &
\begin{minipage}[h]{14cm}{
\includegraphics[height=10cm]{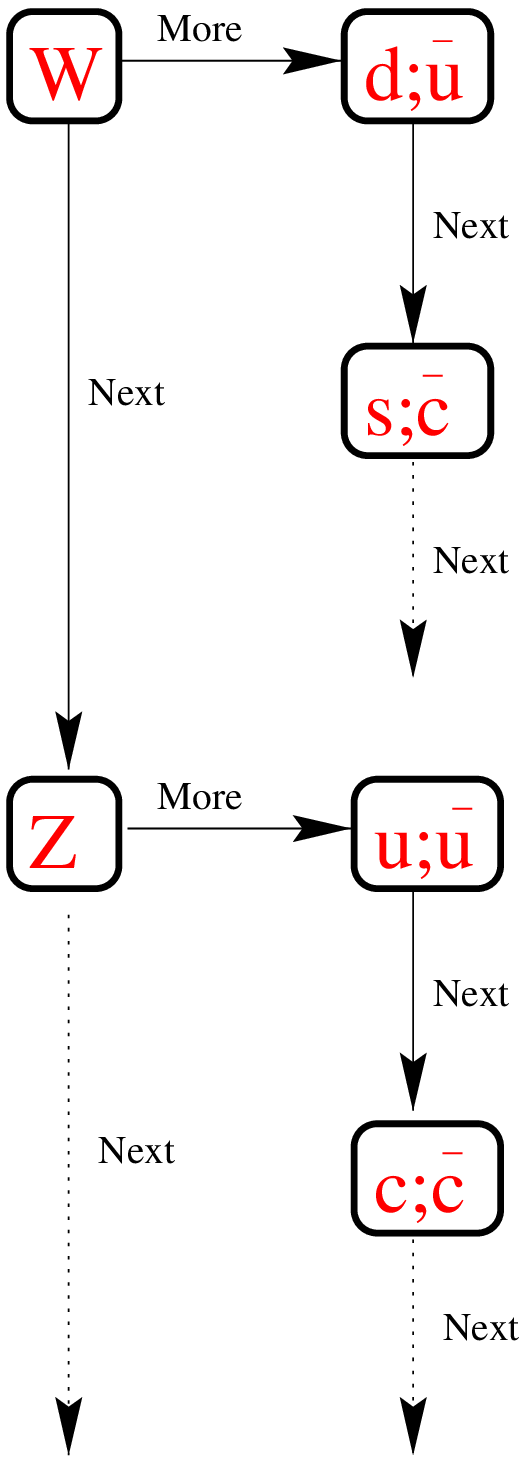}
}\end{minipage} 
\end{tabular}
Now we are in the position to understand, how the {\tt Decay\_Handler}
fills this table via the following methods : 
\begin{enumerate}
\item{{\tt Find\_Unstable()} produces the list of all unstable
      particles, where the decay width has to be calculated. A simple
      loop over all particles probes their on-switch which is read in from
      the {\tt Particle.dat} data file, i.e.\ whether the user wants
      to include them or not. Discarded particles are neglected
      completely, so they do not show up, neither as internal nor as
      external lines. Having at hand all decaying particles the
      products will be determined with the routines {\tt
      Find\_Decay\_Products\_2()} for the $1\to2$ decays and 
      {\tt Find\_Decay\_Products\_3()} for any further decays.}
\item{{\tt Find\_Decay\_Products\_2()} finds every $1\to2$ decay for
      the unstable particles. Naively, one could take every possible
      pair of  particles as decay products. But applying a number of
      constraints, only decays with non--zero width are taken into
      account. Thus every pair of decay products has
      to obey the following restrictions: 
      \begin{itemize}
      \item The incoming flavour has to interact at all, so it has to 
            show up in the list of vertices as an incoming or
            outgoing particle, respectively. This test is performed
            with {\tt Check\_In\_Vertex()}.  
      \item Due to energy--momentum conservation none of the decay
            products can be the same as the incoming particle.
      \item The masses of the decay products must be separately
            smaller than the mass of the incoming particle. The sum
            of both masses can be larger, if one or both of the decay
            products is unstable and can be understood as an internal 
            line. Later on, the decays into such virtual particles is
            taken into account via three or more body decays.
      \item All quantum numbers, i.e.\ charge and spin, have to be
            conserved. Hence, a decay of a fermion into bosons only is
            disallowed.  
      \end{itemize}
      If it fulfills all these constraints the decay channel is added
      to the {\tt DecayTable} with the help of {\tt Add}, i.e.\ another
      item is added to the {\tt More} list of the corresponding
      decaying particle.}
\item{In {\tt Add()} the last check for a particle and its decay
      products is performed. With the help of {\tt Check\_Vertex()} 
      the three flavours are compared with the flavours of
      all existing vertices in the 
      actual model. Only if this check is passed, a new {\tt
      DecayTable} is added to the list.} 
\item{{\tt Find\_Decay\_Products\_3()} finds any further decay beyond
      two-body decays. Therefore a loop over all already determined
      decays is performed. In the case a virtual particle is found,
      i.e.\ the sum of the masses of all decay products is larger than
      the mass of the incoming particle, a further decay of unstable
      decay products is realized with {\tt Next\_Decay()}.} 
\item{{\tt Next\_Decay()} is used for the determination of further decays. 
      Now, every decay product is analysed once more. If it is an unstable 
      particle itself, one can of course find it in the list of decaying 
      particles. A further loop over all possible decays of this particle 
      and its replacement by these decay products completes the creation 
      of a new decay mode which then will be attached to the 
      {\tt DecayTable}.}     
\item{{\tt Check\_In\_Vertex()} checks, if a certain flavour is included in 
      the list of vertices.}
\item{{\tt Check\_Vertex()} checks, if a vertex with the given flavour
      combination exists.}
\item{{\tt Calculate()} calculates the width of the different decay
      modes and sets the branching ratios. It uses the list of decays
      which has already been constructed in {\tt Find\_Unstable()}, to
      determine the width of the different decay modes using {\tt
      Rec\_Calc()}. Having at hand all decay widths it is
      straightforward to calculate the different branching ratios with
      {\tt Branching\_Ratios()}.} 
\item{{\tt Rec\_Calc()} evaluates the widths of the different decay
      modes recursively in order to take into account the further
      decay of virtual particles. For the calculational part, the
      routines of {\tt Process} are utilized.}  
\item{{\tt Branching\_Ratios()} performes a loop over all decaying
      particles and determines the branching ratios with the help of
      {\tt Rec\_BR()}.}  
\item{{\tt Rec\_BR()} calculates the branching ratios recursively,
      since all further decays have to be taken into account.}
\item{{\tt Print()} prints the table of decays including the decay
      widths and branching ratios.}
\end{enumerate}

\subsubsection{Helicities and Polarisations}

The structure {\tt signlist} contains one helicity combination, an {\tt
On}--switch and a multiplicity. The latter one can be used to reduce
the number of helicity combinations by dropping equivalent ones,
i.e.\ those yielding the same numerical result. The organization is
done by the class {\tt Helicity} via the following methods: 
\begin{enumerate}
\item{The {\tt Constructor} determines all helicity combinations using
      a loop over loops technique as described in Appendix~\ref{Loop}. 
      Having at hand all possible helicity
      combinations with two helicities per outgoing particle, the second 
      helicity for scalar particles (which have no helicity at all) and 
      massive vector bosons (where the sum over the polarizations is 
      carried out differently to the massless vector bosons) is
      switched off.}  
\item{{\tt Max\_Hel()} returns the maximum number of helicities.}
\item{{\tt operator[]()} returns a certain helicity combination.}
\item{{\tt switch\_off()} switches a certain helicity combination off.}
\item{{\tt On()} returns the status of a certain helicity combination.}
\item{{\tt Inc\_Mult()} increases the multiplicity of a combination by one.}
\item{{\tt Multiplicity()} returns the multiplicity of a certain
      helicity combination.} 
\end{enumerate}
The class {\tt Polarisation} maintains all methods needed in
connection with polarisations for massless and massive vector bosons:
\begin{enumerate}
\item{{\tt Spin\_Average()} The normalization constant for averaging  
      the spin, the color and the polarisations of all incoming
      particles is determined at this stage. Accordingly, every
      fermion yields a factor of $1/2$, every quark an extra factor of 
      $1/3$, every gluon a factor of $1/8$ and the massless and
      massive vector bosons gain a factor of $1/2$ and $1/3$,
      respectively.} 
\item{{\tt Massless\_Vectors()} determines whether there is any
      massless vector boson in the in- or outstate. If this is the
      case, an extra four-momentum vector for the construction of the
      polarisation vector is attached to the list of vectors.}
\item{{\tt Massive\_Vectors()} counts the number of massive vector
      bosons in the in- and outstate. For every hit two new
      four--momenta are attached, see Eq.~(\ref{PolSpinMass}). The
      change of the normalization is performed according to 
      Eq.~(\ref{PolSpinMassNorm}).} 
\item{In {\tt Attach()} a connection between the massive vector boson
      and its two polarisation vectors is drawn, ensuring that the
      latter ones add up to the vector boson momentum.}
\item{{\tt Reset\_Gauge\_Vectors()} changes the extra gauge vector,
      when massless vector bosons are present. Usually, this routine
      is used during the gauge test.}  
\item{{\tt Set\_Gauge\_Vectors()} sets the gauge vector of the
      massless vector boson and calculates the two polarization
      vectors for each massive vector boson. The latter ones are 
      determined according to Eq.~(\ref{PolSpinMass}) along the following 
      algorithm :\\ 
      The four-momentum of the massive vector boson is boosted into
      its rest frame and two massless four--momenta are generated as 
      uniformly distributed massless pseudo--decay products of the
      vector boson. Afterwards they are boosted back into the lab frame.}
\item{{\tt Massless\_Norm()} calculates the normalization due to
      massless vector bosons, see Eq.~(\ref{PolSpin}).}
\item{{\tt Massive\_Norm()} yields the normalization due to massive
      vector bosons, see Eq.~(\ref{PolSpinMassNorm}).}
\item{{\tt Replace\_Numbers()}: During the translation of the Feynman
      diagrams into helicity amplitudes all polarisation vectors obtain
      dummy numbers. The correct connection to the appropriate polarisation
      vector is performed in this routine utilizing 
      {\tt Single\_Amplitude::MPolconvert()}. As described above,
      massive vector bosons are treated as propagators decaying into
      their -- massless -- polarisation vectors, the respective
      changes are carried out with the help of  
      {\tt Single\_Amplitude::Prop\_Replace()}.}  
\end{enumerate}

\subsubsection{Handling of processes}

\begin{figure}
\includegraphics[height=12cm]{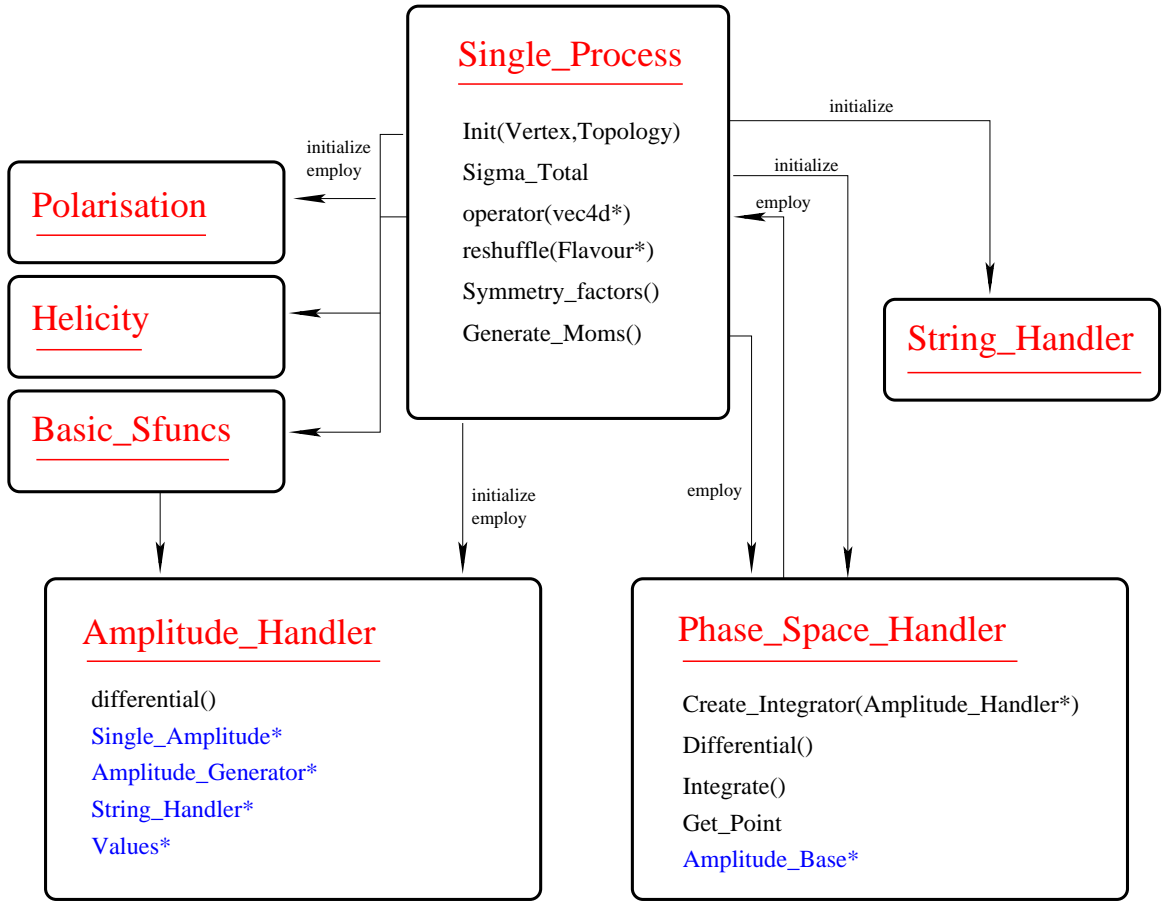}
\caption{One single Process containing an amplitude and a phase space
         evaluation.}
\end{figure}
The two main classes concerning the overall handling of processes are 
{\tt All\_Processes} and {\tt Single\_Process}. Both classes provide
the amplitudes squared, either for a list of processes or for a single
process, respectively. In order to integrate both with the {\tt
Phase\_Space\_Handler}  they have to be derived from the class {\tt
Amplitude\_Base} which is the base of an (phase space) integrable
class. Therefore, its purely virtual operator {\tt operator()()} must
be overwritten by the appropriate inherited class. On the other hand,
the two process classes have to provide a similar interface which
simplifies the usage from outside. Accordingly, both classes are
derived from a mother class called {\tt Process}, a virtual class
which contains a minimal set of routines, i.e.\ {\tt Init()}, for the
initialization, {\tt Sigma\_Total()} for the calculation of the total
cross section and {\tt Add()} for adding a new process to {\tt
All\_Processes}. Since all these methods have to be overwritten by the
derived class, a detailed description can be found there.

The container for a list of processes is {\tt All\_Processes}:
\begin{enumerate}
\item{Within the {\tt Constructor()} the maximum number of processes 
      is set and a blank list of {\tt Single\_Process}es is generated.} 
\item{{\tt Init()} performs a loop over all processes and initializes
      them via\\ {\tt Single\_Process::Init()}.}  
\item{{\tt Sigma\_Total()} evaluates the individual cross sections for
      all processes and adds them up. At this stage a parallelization
      of the calculation for the different processes would enhance the
      performance. This is part of a future project.}  
\item{The {\tt operator()()} is inherited from {\tt Amplitude\_Base}
      and is used for integrating the sum over all cross
      sections. Accordingly a loop and a sum over all 
      {\tt Single\_Process::operator()()}s is done.}  
\item{{\tt Add()} appends another process to the list.}   
\end{enumerate}
The organization, i.e.\ the initialization as well as the calculation
of the amplitude squared for single process lies in the hands of the
class {\tt Single\_Process}:
\begin{enumerate}
\item{The {\tt Constructor()} of the class covers a large part of
      the initialization procedure which is accomplished along :
      \begin{itemize}
      \item{The sequence of the flavours for the incoming and outgoing
            particles is reshuffled using the routine {\tt reshuffle()}.}
      \item{A string which identifies the given process unambiguously,
            is generated via {\tt set\_ID()}.} 
      \item{A directory is created named with this ID. It will contain 
            all the output from this process.}
      \item{The overall normalization is calculated using 
            {\tt Symmetry\_factors()} and 
            {\tt Polarization::Spin\_Average()} taking into account
            the symmetry factor due to permutations of outgoing
            particles and the average over the incoming quantum
            numbers, respectively.}   
      \item{The number of polarization vectors is determined,
            exploring the massless and massive vector bosons with 
            {\tt Polarization::Massless\_Vectors()} and 
            {\tt Polarization::Massive\_Vectors()}, respectively. Now, 
            the number of four-momenta is simply the sum of the number
            of incoming and outgoing particles plus the number of
            polarization vectors needed.}
      \item{With {\tt Polarization::Attach()} the polarization vectors 
            of the massive vector bosons are connected with the
            appropriate vector boson.} 
      \item{The normalization due to the special treatment of massive
            vector bosons is calculated via 
            {\tt Polarization::Massive\_Norm()}.}
      \item{Now, all helicity combinations are determined by
            constructing the {\tt Helicity} object.}
      \item{The basic functions for the determination of the building
            blocks, i.e.\ the S-functions (see Eq.~(\ref{Sfunctions})),
            will be initialized by constructing a {\tt Basic\_Sfuncs}
            object.}
      \item{Last but not least the {\tt String\_Handler} for the
            generation and translation of the helicity amplitudes into
            character strings is constructed.}
      \end{itemize}}
\item{Now, the method {\tt Init()} has only a small task left, since
      the bulk of initializations has been already carried out. This
      task is the construction of the {\tt Amplitude\_Handler} object
      and the initialization of the S-functions and the 
      {\tt String\_Handler} via the appropriate routines 
      {\tt Basic\_Sfuncs::Initialize()} and {\tt String\_Handler::Init()},
      respectively.}
\item{In {\tt reshuffle()} the user--given sequence of the input
      flavours is changed to an intrinsic standard. The internal order
      in \AME is fermions before bosons. Then, 
      particles are first ordered according to the absolute value of
      their kf--code and particles are always listed before
      anti-particles. Massless vector bosons are listed at the end of
      the line which is closely related to the treatment of
      polarization vectors within the program.}
\item{{\tt set\_ID()} constructs the uniquely defined process
      name. The simple rule is that at first the number of incoming
      and outgoing particles is translated to the character string. At next
      the flavours of these particles follow. An example can be found
      in the Test Run Output, see Appendix \ref{TRO}.} 
\item{The {\tt Symmetry\_factors()} for outgoing particles are
      constructed from the multiplicity $n$ of a flavour, where every
      multiple occurrence gains a factor of $1/n!$.} 
\item{{\tt Tests()} contains two different tests of the derived
      amplitude. First of all, a gauge test can be performed, if
      massless vector bosons are present. In this case, the amplitude
      is calculated with two different choices of the corresponding
      gauge vectors and the results will be compared. Of course, if
      the program is doing everything correct, this test should be
      passed, i.e.\ both results should coincide for every single
      combination of external, physical momenta. Taken by itself, this 
      test provides a powerful tool for checking amplitudes,
      vertices etc.. Therefore it might be useful sometimes to add
      a photon or a gluon to a process 
      to have this kind of test at hand. The next test
      concerns the translation of the helicity amplitudes into character
      strings. During this translation numerically small parts will be
      neglected and the emerging string will be reorganized and
      simplified. Therefore, a test for the correctness of the string
      is urgently necessary. Again, the amplitudes calculated with and 
      without the string method are to be compared.}       
\item{{\tt Sigma\_Total()} calculates the total cross section for the
      given process utilizing the class 
      {\tt Phase\_Space\_Handler}. After constructing this object the
      method  {\tt Phase\_Space\_Handler::Create\_Integrator()} builds
      all the possible integration channels. With 
      {\tt Phase\_Space\_Handler::Integrate()} the calculation of the
      total cross section is achieved.}  
\item{The {\tt operator()()} yields the amplitude squared for a given 
      configuration of four-momenta for the incoming and outgoing
      particles. The first step in this calculation consists of
      setting the gauge vector. Then, a loop over all possible
      helicity combinations is performed and the different values for
      the individual amplitudes are summed up. Multiplying the result
      with the normalization for the gauge vector derived from 
      {\tt Polarization::Massless\_Norm()} and the overall
      normalization gives the amplitude squared.}    
\end{enumerate}

\subsubsection{Construction of topologies}

The class {\tt Point} contains all information about a specific
vertex, i.e.\ pointers to the {\tt left} and to the {\tt right} 
next {\tt Point},
the flavour, the particle or propagator number etc.. In this way a
tree structure can be generated by means of linked {\tt Point}s. 
Accordingly, all topologies with the same number of legs are contained
in a {\tt Single\_Topology}, other variables are the number of
legs, the depth as well as the list of {\tt Point} lists. The class
{\tt Topology} creates and handles a list of {\tt Single\_Topology}s
(which differ by the number of legs) via: 
\begin{enumerate}
\item{The {\tt Constructor()} constructs all topologies with the
      method {\tt Build\_All()}.}  
\item{{\tt Build\_All()} generates all topologies up to a maximum
      number of legs. It starts by initializing the first simple
      topology with one external leg only. Since every topology
      contains all other topologies with a smaller number of external
      legs, it is created recursively out of those by calling
      {\tt Build\_Single()}.}   
\item{{\tt Build\_Single()} builds all topologies for a certain number
      of external legs, for details concerning the algorithm see
      Sec. \ref{GenAmp}.}  
\item{{\tt Get()} returns a list of topologies with a certain number
      of legs.} 
\item{{\tt Copy()} copies a given point list into another.}
\item{{\tt Print()} prints a topology.}
\end{enumerate}
    
\subsection{The Model\label{Model}}

Cross sections are calculated within the framework of a specific
model, i.e.\ they depend on a set of assumptions and parameters like
the particle spectrum and the properties (quantum numbers) of the
particles as well as their interactions defined via vertices and the
corresponding coupling constants. Strictly speaking, this framework is 
defined by the underlying Lagrangian, from which the parameters can be 
either read off or calculated. However, for the sake of implementing
such a model in terms of computer algorithms, it is sensible to divide 
the information defining the model into several pieces, reflected by
the class structure. In general, in \AME particle properties are
handled by the class {\tt Flavour}, interactions are governed by the
classes {\tt Model} and {\tt Vertex}, and the appropriate
(possibly running) coupling constants and the particle  masses are
generated by {\tt Coupling} and {\tt Spectrum} classes, respectively.  

A brief outline of all the classes mentioned above, can be found in
Tab.~\ref{ModelTab}. The different methods of the classes are
described in detail in the subsequent sections, covering the
complexes Flavour, Vertices, Model, Spectrum and Couplings.  

\begin{table}
\bc 
\begin{tabular}{l|l} 
Class/Struct & Purpose\\ 
\hline
&\\ 
kf             & Connects the kf-code of every particle with a name.\\ 
kf\_to\_int    & Translates the kf--code into an integer value.\\  
part\_info     & Contains all the information about a particle\\
               & which is read in from the {\tt Particle.dat} data file.\\ 
Flavour        & All particle properties are handled from here.\\ 
fl\_iter       & An iterator over all possible flavours is often useful.\\ 
Single\_Vertex & Contains all information about a Feynman rule.\\ 
Vertex         & Maintains the initialization of the vertices,\\ 
               & i.e.\ the Feynman rules of the chosen model.\\ 
Model          & Is the virtual mother class of all models.\\ 
Model\_QCD     & Contains the QCD Feynman rules.\\
Model\_EE\_QCD & Contains the QCD Feynman rules plus a coupling of the\\
               & QCD particles to electron and positron.\\
Model\_EW      & Contains all electroweak Feynman rules.\\ 
Model\_SM      & The complete Standard Model can be found here.\\ 
Spectrum\_EW   & The electroweak spectrum, i.e.\ all masses of the leptons\\
               & and gauge bosons are maintained here.\\  
Spectrum\_QCD  & The QCD spectrum, i.e.\ the masses of the quarks is\\
               & read in within this class.\\  
Couplings\_EW  & The electroweak coupling constants as well as their\\
               & running is performed within this class. \\ 
Couplings\_QCD & The running of the strong coupling constant is calculated.\\ 
               & \\
\end{tabular}
\caption{\label{ModelTab} Principal classes to set up and organize the
physical framework.}
\ec
\end{table}                                                                   

\subsubsection{The Flavour}
{\tt kf} contains the kf--code of every particle according to the PDG
\cite{Garren:2000st}, whereas the small class {\tt kf\_to\_int} translates this
kf--code into an integer value and back. All necessary particle
information is stored in {\tt part\_info}, namely the kf--code, the
mass and width, the charge and the isoweak charge, a flag for strong 
interacting particles, the spin, a flag for the majorana character of a 
particle, an on--switch, a flag for stable particles and the name of the 
particle. The class {\tt Flavour} handles all properties a particle
could have. Therefore, it mainly provides routines for obtaining
information about a particle type:
\begin{enumerate}                  
\item {\tt texname()} gives the name of a particle in a LaTeX
      format. This is used for instance by {\tt Vertex::Tex\_Output()}.
\item {\tt kfcode()} gives the kf--code of a particle.
\item The {\tt operator int()} returns the integer value of a
      kf--code. A minus sign covers the possible anti-particle nature.  
\item {\tt bar()} returns the anti-particle of a particle.
\item {\tt charge()} returns the electromagnetic charge in units of
      the proton charge. 
\item {\tt icharge()} returns as an integer three times the
      electromagnetic charge.  
\item {\tt isoweak()} returns the value of $T_3$, the third component
      of the weak isospin. 
\item {\tt strong()} yields $1$, if the particle interacts via the
      strong force.
\item {\tt spin()} returns the spin of a particle.
\item {\tt ispin()} returns as integer twice the spin of a particle.
\item {\tt ison()} returns $1$, if a particle should take part in the
      generation of the Feynman rules.
\item {\tt isstable()} is the switch which returns $1$ in case the
      particle width should be calculated by the program, otherwise the
      width stored within the {\tt Particle.dat} data file is used.
\item {\tt set\_on()} switches a particle on.
\item {\tt mass()} returns the mass of a particle.
\item {\tt set\_mass()} sets the mass of a particle.
\item {\tt width()} returns the width of a particle.
\item {\tt set\_width()} sets the width of a particle.
\item {\tt name()} returns the character name of a particle.
\item {\tt isfermion()} returns $1$, if the particle is a fermion.
\item {\tt isboson()} returns $1$, if the particle is a boson.
\item {\tt isscalar()} returns $1$, if the particle is a scalar, i.e.\ if it
      has spin zero.
\item {\tt isvector()} returns $1$, if the particle is a vector, i.e.\ if it
      has spin one.
\item {\tt isquark()} returns $1$, if the particle is a (anti-) quark.
\item {\tt isgluon()} returns $1$, if the particle is a gluon.
\item {\tt islepton()} returns $1$, if the particle is a (anti-) lepton.
\item {\tt isuptype()} returns $1$, if the particle is up--type. Note
      that this extends to leptons and neutrinos as well.
\item {\tt isdowntype()} returns $1$, if the particle is
      down--type. Note that this extends to leptons and neutrinos as
      well. 
\item {\tt isanti()} returns $1$, if it is an anti-particle.
\item {\tt hepevt()} returns the HEPEVT kf--code.
\item {\tt from\_hepevt()} sets the flavour according to the HEPEVT kf--code.
\end{enumerate}
The class {\tt fl\_iter} provides an iterator over all particles in
the particle data table. Two methods, i.e.\ {\tt First} and {\tt Next}
return the first and the appropriate next particle, respectively. 

\subsubsection{The Vertices}

A {\tt Single\_Vertex} contains all information about an $1\to 2$
Feynman rule, i.e.\ the three flavours, the left-- and right--handed
coupling constants and a possible representation by a character string.    
The class {\tt Vertex} handles the generation of Feynman rules
utilizing some of the methods from the class {\tt Model}. 

\begin{figure}
\includegraphics[height=5cm]{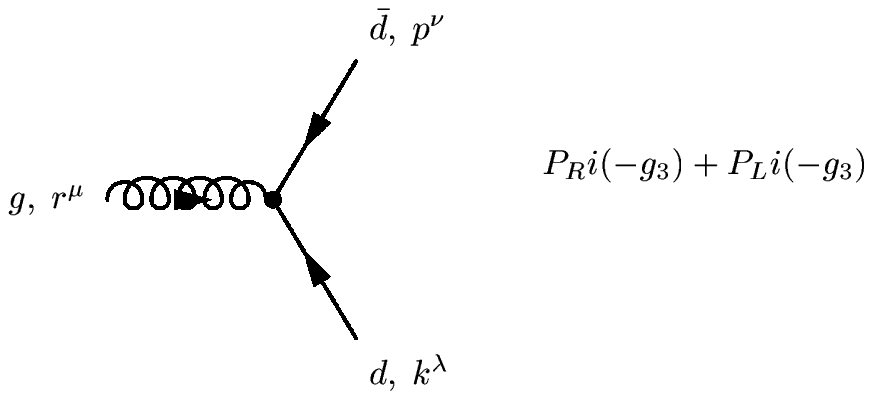}
\caption{\label{TexVertex}A sample vertex in LaTeX output form. Note
that $P_R$ and $P_L$ are the right and left handed projectors
including a $\gamma_\mu$.}
\end{figure}

\begin{enumerate}
\item The {\tt Constructor()} builds the list of vertices 
      ({\tt Single\_Vertex}) and produces a LaTeX output file. First
      of all, the list is filled via the Model routines {\tt c\_FFV},
      {\tt c\_FFS}, {\tt c\_VVV},{\tt c\_SSV}, {\tt c\_VVS} and 
      {\tt c\_SSS}, see Tab.~\ref{CouplStrucTab}. 
      Having at hand all the different vertices, the LaTeX output is
      generated with {\tt Tex\_Output()}. Since during the matching of
      the vertices onto the raw topologies not only the standard form
      of a vertex is used, but also its rotations (obtained as a
      combination of permutating the flavours and taking their
      anti-flavours), every vertex is transformed into the six
      possible combinations with an appropriate change of the
      couplings, see Tab.~\ref{Vertices}. A test with 
      {\tt Check\_Equal()} prevents any double counting of already
      generated vertices.    
\item {\tt Check\_Equal()} compares any combination of three flavours
      generated in the process of initialization with all already
      generated vertices and prevents any double counting.  
\item {\tt Print()} prints the list of vertices.
\item {\tt operator[]()} returns a {\tt Single\_Vertex} from its number.
\item {\tt Max\_Number()} returns the maximal number of vertices.
\item {\tt Tex\_Output()} generates a LaTeX output of all Feynman
      rules. The character strings generated for each vertex in the
      model classes will be combined with a FeynMF 
      \cite{Ohl:1995kr} picture of a vertex, for
      example see Fig.~\ref{TexVertex}.  
\item {\tt AddVertex()} adds a {\tt Single\_Vertex} to the list of vertices.
\item {\tt FindVertex()} searches for a given {\tt Single\_Vertex} in
      the list of vertices and returns its number.
\end{enumerate}

\bc
\ec

\subsubsection{The Model}

The {\tt Model} class represents the basis for all other model
classes which are simply derivatives. Consequently, most 
methods are purely virtual. However, they play a similar role in every 
model and therefore, they will be described only once. The first part
is dedicated to the initialization of the model and the vertices:  
\begin{enumerate}
\item {\tt Init\_Vertex()} generates the list of vertices by
      constructing the {\tt Vertex} object.
\item {\tt Get\_Vertex()} returns the list of vertices.
\item Usually {\tt Init()} generates the masses and couplings of the
      given model with the help of an appropriate {\tt Spectrum} and
      {\tt Coupling} class, respectively. 
\end{enumerate}
All Feynman rules can be classified according to their Lorentz structure
which depends on the incoming and outgoing flavours. This
classification scheme of the vertices in \AME is depicted in 
Tab.~\ref{CouplStrucTab}, the methods listed there are then filled in 
a model dependent way.
\begin{table}[t] 
\bc
\begin{tabular}{l|l|l|l} 
Method & Incoming & Left Outgoing & Right Outgoing\\ 
\hline
&&&\\ 
c\_FFV()& Fermion      & Vector boson & Fermion\\
c\_FFS()& Fermion      & Scalar boson & Fermion\\
c\_VVV()& Vector boson & Vector boson & Vector boson\\
c\_SSV()& Scalar boson & Vector boson & Scalar boson\\
c\_VVS()& Vector boson & Scalar boson & Vector boson \\
c\_SSS()&Scalar boson  & Scalar boson & Scalar boson \\
&&&\\
\end{tabular}
\caption{\label{CouplStrucTab} Connection of methods for the 
initialization of vertices with the external particles.}
\ec
\end{table}                                                                   
Since the Model class contains not only all the Feynman rules, but
also the different coupling constants some helper methods permit to
gather this information:
\begin{enumerate}
\item {\tt SinTW()} returns the $\sin{\theta_W}$.
\item {\tt CosTW()} returns the $\cos{\theta_W}$. 
\item {\tt Aqed()} returns $\alpha_{\rm QED}$ at the CM energy scale.
\item {\tt Aqed(double)} returns $\alpha_{\rm QED}$ at the given scale.
\item {\tt Aqcd()} returns $\alpha_{\rm QCD}$ at the CM energy scale.
\item {\tt Aqcd(double)} returns $\alpha_{\rm QCD}$ at the given scale.
\end{enumerate}
The different model classes which are derived from the main class
{\tt Model}, can be characterized via the Feynman rules (i.e.\ vertices) they 
initialize:
\begin{enumerate}
\item {\tt Model\_QCD} contains all pure QCD Feynman rules, i.e.\ the
      interaction of quarks via gluons and the gluon self--interaction
      in the axial gauge.
\item {\tt Model\_EE\_QCD} includes all vertices from {\tt Model\_QCD}
      plus the interaction of quarks with electrons and positrons via
      photons and $Z$ bosons.  
\item {\tt Model\_EW} includes all electroweak vertices, i.e.\ the
      interaction of all Standard Model particles via photons, $Z$,
      $W$ and Higgs bosons.  
\item {\tt Model\_SM} combines {\tt Model\_QCD} and {\tt Model\_EW} to
      yield all interactions of the Standard Model.
\end{enumerate}
\subsubsection{The Spectrum}
The two different spectra of the Standard Model, i.e.\ the strong and
the electroweak sector, are represented by the classes 
{\tt Spectrum\_EW} and {\tt Spectrum\_QCD}. Both have only the method
{\tt Fill\_Masses()} in order to set or calculate the masses of the
different particles, when called. However, within the Standard Model,
all fermion masses are fundamental parameters, therefore, they are
only read in from the {\tt Const.dat} data file and set with the
method {\tt Flavour::set\_mass()}, shadowing the values given in {\tt
Particle.dat}.  

In contrast, the masses of the $Z$ and $W$ bosons are 
generated via spontaneous symmetry breaking and can be calculated to
leading order, i.e.\ without quantum corrections, with the help of the
coupling constants $\alpha_{\rm QED}$ and $\sin{\theta_W}$ and the 
Higgs vacuum expectation value (vev). In {\tt AMEGIC++}, 
an option is provided to calculate these masses and set them, thus
overwriting their values given in {\tt Particle.dat}. 

Note that in the framework of the Standard Model, these classes seem
somewhat over--engineered, but for later extensions to models beyond
the Standard Model spectra should be calculable in dependence to
some set of parameters which might or might not be different from the 
sheer listing of particle masses. As an example take the Minimal
Supersymmetric Standard Model, where -- neglecting quantum corrections 
-- the masses of the partner particles in general depend on the mass
parameters of the Standard Model particles plus some symmetry breaking
terms.  
\subsubsection{The Couplings}
{\tt Couplings\_EW} reads in the electroweak coupling constants and
maintains the calculation for the running of $\alpha_{\rm QED}$:
\begin{enumerate}
\item {\tt Init()} reads in the coupling constants of the electroweak
      model, i.e.\ $\alpha_{\rm QED}$ and $\sin{\theta_W}$ at the $Z$ pole,
      the Higgs vacuum expectation value and the four parameters of
      Wolfenstein's parameterization of the CKM matrix. Accordingly, these
      values are set and the CKM matrix is generated. In order to calculate
      the running of $\alpha_{\rm QED}$ all thresholds are
      determined with {\tt Thresholds()}. Then, the coupling constant as
      well as the $\beta$--function of the leading order running
      which depends on the number of active charged fermions is
      calculated via {\tt Init\_AQED()}.  
\item {\tt Thresholds()} determines all particle thresholds which
      contribute to the running of $\alpha_{\rm QED}$,
      i.e.\ they have to have an electric charge.
\item {\tt Init\_AQED()} determines $\alpha_{\rm QED}$ at the scales
      introduced by the different particle thresholds. In order to obtain an
      efficient evaluation of the running coupling constant, the
      $\beta$--function is precalculated as well. 
\item {\tt aqed()} returns $\alpha_{\rm QED}$ at a given scale.
\item {\tt VEV()} returns the Higgs vacuum expectation value.
\item {\tt SinThetaW()} returns the $\sin{\theta_W}$.
\item {\tt CosThetaW()} returns the $\cos{\theta_W}$.
\item {\tt CKM()} returns the specified entry of the CKM matrix.  
\end{enumerate}
Note that the running of electroweak parameters other than
$\alpha_{\rm QED}$ has not been implemented so far. 
{\tt Couplings\_QCD} reads in the strong coupling $\alpha_{\rm QCD}$
and calculates its running:
\begin{enumerate}
\item {\tt Init()} reads in the only coupling constant of the strong
      interaction, $\alpha_{\rm QCD}$ at the $Z$ pole. Similarly to
      the procedure in {\tt Couplings\_EW} the thresholds as well as
      the coupling constants at these thresholds are determined using
      the methods {\tt Thresholds()} and {\tt Init\_As()},
      respectively.     
\item {\tt Thresholds()} specifies the thresholds for all strong
      interacting particles. 
\item {\tt Init\_As()} calculates the value of the strong coupling at
      the different threshold scales as well as the $\beta$--function
      in order to perform a one loop running of the coupling.
\item {\tt as2()} returns $\alpha_{\rm QCD}$ at a given scale.
\end{enumerate}

\subsection{The Amplitude\label{Amplitude}}

\begin{figure}[h]
\includegraphics[height=14cm,angle=90]{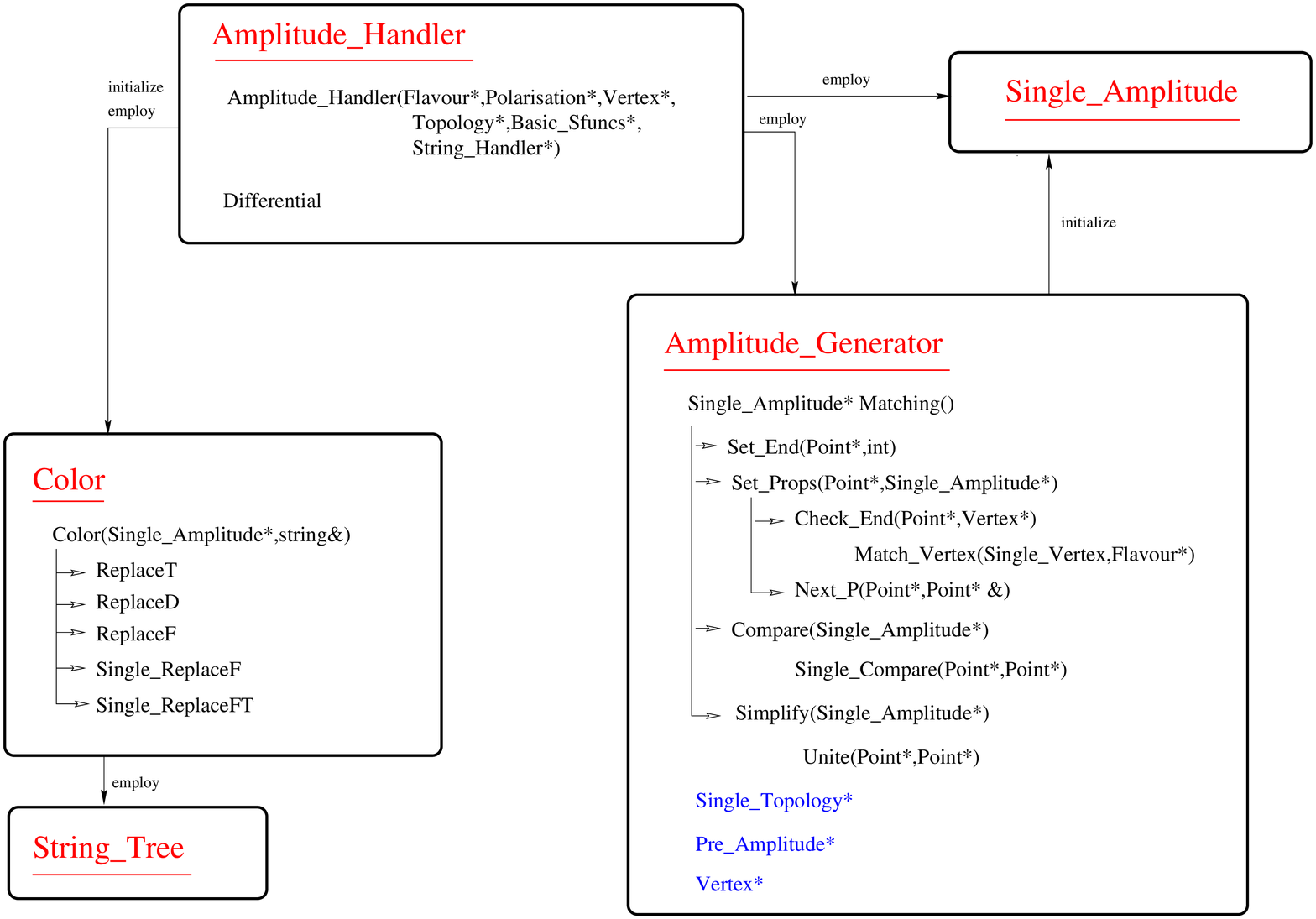}
\caption{\label{FigAmpl}The interplay between the organizing class
{\tt Amplitude\_Handler} and the classes for generating, storing and
calculating a single amplitude (Feynman diagram).}
\end{figure}

In this section we describe the generation of Feynman diagrams, 
their translation into the helicity amplitudes and their calculation.
Accordingly, we start with the basic handling of amplitudes in
Sec.~\ref{AmplHand}, then dwell on the generation of the Feynman
diagrams in Sec.~\ref{AmplGen} and on their representation within a
{\tt Single\_Amplitude} in Sec.~\ref{AmplSing}. In this section the
translation into helicity amplitudes is performed as well. We end with
the description of the tools for the calculation of the amplitude in
Sec.~\ref{AmplCalc} as well as the Color structure and Coulomb factors
in Sec.~\ref{AmplCoCo}. A short outline of the involved classes and
structures can be found in Tab.~\ref{AmplTab}, whereas a pictorial
overview is given in Fig.~\ref{FigAmpl}.
\begin{table}[h]
\bc 
\begin{tabular}{l|l} 
Class/Struct & Purpose\\ 
\hline
&\\ 
Amplitude\_Handler     & Handles all amplitudes, i.e.\ the sum of all
                         Feynman\\
	               & diagrams.\\    
Amplitude\_Generator   & Generates all amplitudes, i.e.\ matches the
                         vertices \\
                       & onto the topologies.\\  
struct Pre\_Amplitude  & Is a {\tt Point} list.\\ 
\hline
Single\_Amplitude      & Handles one Amplitude, i.e.\ Feynman diagram.\\
Zfunc\_Generator       & Translates the Feynman diagrams into\\ 
	               & helicity amplitudes.\\ 
Zfunc                  & Is a structure for storing one $Z$--function.\\ 
Prop\_Generator        & Finds and labels all propagators.\\
Pfunc                  & Is a structure for storing one propagator.\\
Color\_Generator       & Determines the color structure.\\      
Cfunc                  & Is a structure for storing one color matrix.\\
\hline
Building\_Blocks       & The different building blocks of a helicity
                         amplitude.\\       
Mathematica\_Interface & Interfaces to Mathematica, calculating some\\
                       & building blocks.\\ 
Composite\_Zfuncs      & Composed helicity amplitudes.\\     
Elementary\_Zfuncs     & All elementary helicity amplitudes,\\
	               & i.e.\ $Z$, $X$- and $Y$ functions.\\       
Basic\_Sfuncs          & The $S$ functions.\\         
Momfunc                & Is a structure which stores a four-momentum
                         for\\ 
                       & incoming and outgoing particles as well as\\ 
	               & for propagators.\\ 
\hline
Color                  & Calulates the sum of color matrices.\\                
Coulomb                & Calculates Coulomb corrections for $W$-pair
                         creation.\\               
&\\ 
\end{tabular}
\caption{\label{AmplTab} All classes and structures which are used for
the determination and calculation of a Feynman amplitude.}
\ec
\end{table}                                                                   
\subsubsection{\label{AmplHand}Handling of Amplitudes}

The organization of the generation as well as the calculation of a
Feynman amplitude is governed by the class {\tt Amplitude\_Handler}:
\begin{enumerate}
\item The {\tt Constructor} governs the whole initialization
      procedure, i.e.\ the generation of Feynman diagrams and their
      translation into helicity amplitudes in the following way:
      \begin{itemize}
      \item All Feynman diagrams are generated with the construction
            of an\\ {\tt Amplitude\_Generator} object and the call to
            its method {\tt Matching()}. This method
            constructs a list of {\tt Single\_Amplitude}s, each of them
            representing an individual Feynman diagram through a list
            of linked -- and filled -- {\tt Point}s.   
      \item A loop over all amplitudes transforms them into the helicity
            amplitudes with 
            {\tt Single\_Amplitude::Zprojecting()}. The couplings are
            filled into the {\tt String\_Handler} by 
            {\tt Single\_Amplitude::Fill\_Coupling()}.  
      \item Now, the dummy numbers for the polarization vectors can be
            replaced by the proper number within the $Z$--functions by
            calling\\ {\tt Polarization::Replace\_Number()}. 
      \item All the $S$--, $\eta$-- and $\mu$--functions depending on
            the four-momenta, including the propagators, are
            calculated within {\tt Basic\_Sfuncs}. Having at hand all
            $Z$--functions and the list of propagators {\tt Pfunc},
            they can be initialized with 
            {\tt Basic\_Sfuncs::Build\_Momlist()}.   
      \item Last but not least the matrix composed of colour factors
            for each pair of amplitudes is calculated by constructing
            the {\tt Color} object. 
      \end{itemize}  
\item {\tt Get\_Graph()} returns the {\tt Point} list of a Feynman
      diagram for a given number.
\item {\tt Fill\_Coupling()} performs a loop over all 
      {\tt Single\_Amplitude}s and fills the couplings into the 
      {\tt String\_Handler} via {\tt Single\_Amplitude::Fill\_Coupling()}.   
\item {\tt Get\_Graph\_Number()} returns the number of graphs.
\item {\tt Get\_Pointlist()} returns the {\tt Point} list of a Feynman
      diagram for a given number. Note that this method considers the full
      list of graphs, not the shortened version of the 
      {\tt Single\_Amplitude} list, where some graphs are united which
      differ only by a photon, $Z$ or Higgs boson propagator.  
\item {\tt Get\_PointlistNumber()} returns the number of the full 
      {\tt Point} list. 
\item {\tt differential(int*,int)} calculates the differential cross
      section by using the standard calculation methods, i.e.\ no character
      strings. The arguments of the method are a list of signs for
      the different helicities of incoming and outgoing particles,
      whereas the second argument corresponds to the counting number
      of the current helicity combination. A loop over all amplitudes
      and a call to {\tt Single\_Amplitude::Zvalue()} fills the
      complex value of one  
      diagram for a specific helicity combination into an array. At
      the end, taking into account the elements of the color matrix 
      ({\tt Color::Mij()}), all diagrams are summed up. Sometimes,
      single diagrams yield zero depending on the structure of
      the couplings. This is taken care of by storing the value of
      each diagram and dropping those which yield no result at all
      after a user defined number of calls, typically one thousand.    
\item The second {\tt differential(int)} method differs from the other
      one by two points. First, the amplitude is calculated using the
      already generated character strings for the evaluation of each
      diagram via {\tt String\_Handler::Zvalue()}. Accordingly, only
      one argument is needed for this method, i.e.\ the current
      counting number of the helicity combination. Second, since these
      character strings are already shortened, i.e.\ zero parts are dropped,
      no special care has to be taken for zero diagrams. 
\end{enumerate}
\clearpage

\subsubsection{\label{AmplGen}Generating the Feynman diagrams - 
{\tt Amplitude\_Generator}}

The structure {\tt Pre\_Amplitude} stores a list of points. Therefore
it can be regarded as a predecessor of a proper 
{\tt Single\_Amplitude}. In fact, until this {\tt Pre\_Amplitude} has
not passed certain tests it cannot become a full amplitude. The
generation of these {\tt Pre\_Amplitude}s as well as the handling of
the full amplitudes is governed by the class 
{\tt Amplitude\_Generator} by means of the following methods:
\begin{enumerate}
\item The {\tt Constructor()} obtains the topology which fits the
      actual number of legs via {\tt Topology::Get()} and generates an
      empty list of {\tt Pre\_Amplitude}s.  
\item {\tt Matching()} is the main method within this class. It constructs
      the list of {\tt Single\_Amplitude}s and simplifies them. For the
      first step all permutations of the incoming and outgoing particles
      are constructed. This is achieved by a loop over loops
      technique (cf.\ Appendix \ref{Loop}), 
      where the permutations are generated and then have to obey
      certain constraints, listed in Sec.~\ref{GenAmp}.
      Having passed these constraints a permutation of incoming and
      outgoing particles can be matched on all possible
      topologies. First of all, these particles are set to the
      endpoints of a certain topology with {\tt Set\_End()}. After this, all
      possible propagators in between are tested with {\tt Set\_Props()} and
      eventually diagrams are generated. At the very end of the loop
      over loops all diagrams are stored into a list of 
      {\tt Single\_Amplitude}s. However, similar diagrams could still
      occur and have to be dropped. This is achieved with the method 
      {\tt Compare()}. Having at hand the full list of distinguishable
      diagrams, they are stored ({\tt Save\_Pointlist()}) for the later use
      in the construction of {\tt Channel}s employed during phase
      space integration. Now the list of diagrams can be shortened by
      unifying those which differ only in a photon, $Z$ or Higgs
      boson propagator with the method {\tt Simplify()}. 
\item {\tt Set\_End()} recursively sets all endpoints of a topology
      onto a particle permutation, i.e.\ the particle number, its
      flavour and the {\tt b} flag which marks incoming and outgoing
      particles. Propagators are labeled with numbers beginning with
      100. 
\item {\tt Set\_Props()} performs the matching procedure for one
      topology and one combination of endpoints. Basically, it tries to find
      every possible flavour combination for the inner propagators in
      the following way: \\
      At first, the raw topology with the endpoints already set is
      copied into a {\tt Pre\_Amplitude}. Then a loop over all 
      {\tt Pre\_Amplitude}s is performed. Within the loop, every time,
      an internal line is equipped with a flavour, i.e.\ each time a
      new propagator is tested, a new {\tt Pre\_Amplitude} is created
      and appended. Consequently, the loop over {\tt Pre\_Amplitude}s
      ends, if no more propagators can be tested. The test is
      performed along the following steps: 
      \begin{itemize}
      \item The next free propagator in the actual {\tt Pre\_Amplitude},
            i.e.\ a {\tt Point} where the flavour is not set already,
            is selected via {\tt Next\_P()}. If no free {\tt Point}
            can be found, the loop advances to the next 
            {\tt Pre\_Amplitude}.     
      \item A loop over all possible vertices is performed. Each
            vertex is compared with a flavour combination which is
            manipulated in the way described in Sec.~\ref{GenAmp}. This
            is done with {\tt Match\_Vertex()}. 
      \item If the comparison was successful, the flavours of the
            outgoing particles are set onto this vertex. If they are 
            connected with endpoints, the left and the right 
            {\tt Point}s are checked by ({\tt Check\_End()}). 
      \item Having passed all these tests, the actual 
            {\tt Pre\_Amplitude} is copied. It is
            attached at the end of the
            list of {\tt Pre\_Amplitude}s and the appropriate vertex
            is changed in the copy. This method ensures that all
            other vertices can be tested as well for this particular
            {\tt Point}.      
      \item If the loop over the vertices has finished, the loop over
            the {\tt Pre\_Amplitude}s proceeds to the next amplitude
            and the actual one is switched off. Therefore, every new
            {\tt Point} which has matched onto a vertex, yields a new 
            amplitude. In this way, only amplitudes which are fully
            filled, reach the end of the loop. If no new amplitude
            exists, the loop terminates.  
      \end{itemize}
      In this fashion, a number of pre--amplitudes has been generated
      which will be subjected to some simple tests in order to avoid
      double counting of Feynman diagrams: 
      \begin{itemize}
      \item If the chosen model is QCD (all QCD Feynman rules
            plus a coupling to electrons and positrons), only diagrams
            can survive which have just one electroweak (i.e.\ photon,
            $Z$ boson) propagator. This is used for the generation of
            QCD final states in electron--positron annihilations. 
      \item In multiple boson vertices a certain ordering of the
            particles has to be demanded in order to avoid a double
            counting of, for instance, three gluon vertices. 
      \end{itemize}
      Passing these tests the located {\tt Pre\_Amplitude} can be
      translated into a proper Feynman diagram, i.e.\ a 
      {\tt Single\_Amplitude}. An additional minus sign for every
      exchange of two fermions is included and stored as well.
\item {\tt Next\_P()} searches recursively through the list of 
      {\tt Point}s for the next undetermined propagator, i.e.\ 
      for a {\tt Point} with the {\tt Flavour} not yet set.
\item {\tt Print\_P()} prints a {\tt Point} list recursively.
\item In {\tt Match\_Vertex()} a given combination of {\tt Flavour}s
      is compared with a given {\tt Single\_Vertex}, i.e.\ all {\tt Flavour}s
      which are set will be compared and if they match, the unset 
      {\tt Flavour}s as well as the coupling constants can be filled.
\item {\tt Check\_End()} checks for a given {\tt Point} with already
      set flavour whether the linked left and right 
      {\tt Point}s are already endpoints and, if this is the case, whether 
      they match. For the matching part, a loop over all possible
      vertices is performed and every vertex is compared with a
      slightly modified flavour combination of the incoming and the two
      outgoing particles. The flavours are changed according to the
      rules presented in Sec.~\ref{GenAmp}. Having at hand the proper
      flavour combination it is checked against the actual vertex via
      {\tt Match\_Vertex()}. If the checks are passed, the couplings
      of the {\tt Point} are set.  
\item {\tt Compare()} performs a comparison of all amplitudes. Since
      the derived list of Feynman diagrams could still have some double
      countings, this step is necessary. Within a double loop over all 
      {\tt Single\_Amplitude}s the appropriate {\tt Point} lists are
      compared with {\tt Single\_Compare()}, where one of them is marked for
      deletion in case they are equal. This is achieved by using the 
      {\tt On} switch of a {\tt Single\_Amplitude}. Finally, all marked
      amplitudes are deleted with the method {\tt Kill\_Off()}.   
\item {\tt Single\_Compare()} compares two {\tt Point} lists
      recursively which means that in one recursion step always two 
      {\tt Point}s are inspected. They are checked for equal {\tt Flavour}
      and number. After this, the left and right {\tt Point}s are
      compared with a recursive call to {\tt Single\_Compare()}. However, 
      if the regarded {\tt Point} belongs to a triple gluon vertex,
      the left and right branch could be exchanged and compared
      again. Thus, all possible combinations of exchanging the
      different branches are considered.       
\item {\tt Kill\_Off()} deletes all marked {\tt Single\_Amplitude}s
      which are switched off. A loop over all possible amplitudes and an
      appropriate reordering of the list of {\tt Single\_Amplitude}s
      is sufficient for this purpose. 
\item {\tt Save\_Pointlist()} saves all {\tt Single\_Amplitude}s into
      a list of {\tt Point} lists. This is necessary for the later use
      within the {\tt Channel\_Generator} for the phase space integration.
\item {\tt Simplify()} compares all different amplitudes. If two of
      them differ only by a photon, $Z$ or Higgs boson propagator they can
      be united and regarded as the same amplitude. Later on, during the
      calculation of the different pieces of helicity amplitudes, the
      contributions of the different propagators will be summed
      up. Therefore, always two {\tt Point} lists are compared. In case they
      differ only be the regarded boson propagators, they are unified
      with the method {\tt Unite()}.   
\item {\tt Unite()} unites two amplitudes which differ only by a
      photon, $Z$ or Higgs boson propagator. Therefore, the arrays of
      couplings for the start and end {\tt Point}s of the propagator
      have to be enhanced to include the appropriate coupling of the
      other propagator types. The {\tt Flavour} of the propagator is
      exchanged with a list of {\tt Flavour}s for all occurring flavour
      types.    
\end{enumerate}

\subsubsection{\label{AmplSing}A single amplitude}

\begin{figure}[t]
\includegraphics[height=12cm]{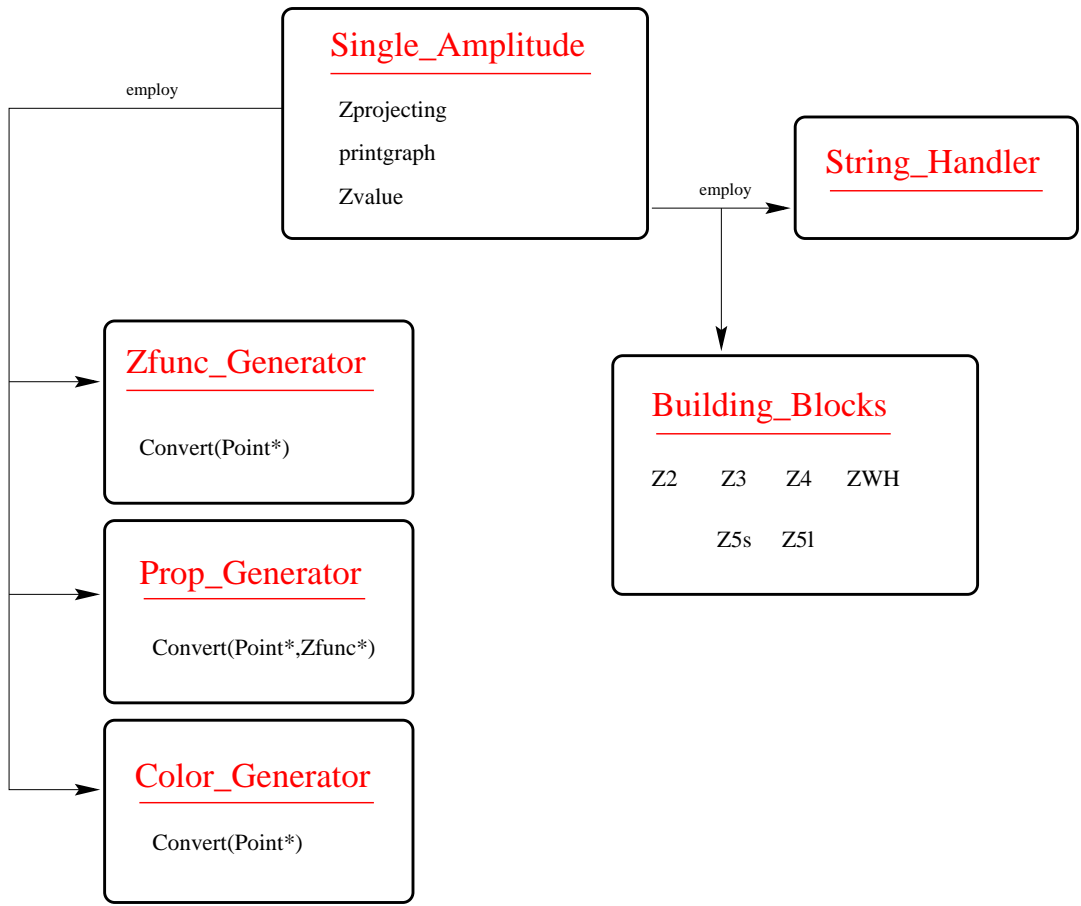}
\caption{\label{FigSingAmp}One Feynman diagram.}
\end{figure}

The {\tt Single\_Amplitude} is an implementation of a particular
Feynman diagram. Therefore, it includes the {\tt Point} list for the
graph, a list of {\tt Zfuncs} for the translation of the graph into
helicity amplitudes, a list of propagators and a list of color
matrices. The class governs the generation of all these lists and
calculates the amplitude with the help of {\tt Building\_Blocks}, for
a pictorial overview of the connection to other classes see
Fig.~\ref{FigSingAmp}.
Furthermore, {\tt Single\_Amplitude} provides a number of methods for
manipulating the generated lists.     
\begin{enumerate} 
\item The {\tt Constructor()} initializes the list of 
      {\tt Points}. However, during the generation of this list in the
      {\tt Amplitude\_Generator} not much care has been taken to the
      correct direction of the spin flow for the first fermion
      line. Therefore, a number of propagators might still have the
      wrong spinor direction, i.e.\ they are considered to be particles
      instead of anti-particles or vice versa. In order to cure this
      problem, two methods ({\tt Initial\_Fermionline()} and 
      {\tt Reverse\_Fermionline()}) are used in the following way: 
      \begin{itemize}
      \item If the first particle is an incoming fermion and not an
            anti-particle, the initial spinor line is marked with
            {\tt Initial\_Fermionline()}. Furthermore, the endpoint of
            this line is determined as well. 
      \item If the endpoint of this fermion line is an outgoing
            particle, the spin flow of this first spinor line is
            reversed beginning with the last particle in {\tt
            Reverse\_Fermionline()}.    
      \end{itemize}       
      During these two steps a special flag in the {\tt Point} list ($t$) is
      set. Later on, during the generation of the list of
      propagators, this flag determines the particle or anti-particle
      character of the appropriate propagator. 
\item {\tt Initial\_Fermionline()} determines recursively the initial
      fermion (i.e.\ spinor) line and its endpoint. The pointer to
      the previous {\tt Point} is set as well. 
\item {\tt Reverse\_Fermionline()} starts from the endpoint of the
      first fermion line. The $t$ flag is set equal to one in case the
      propagator is an anti-particle.  
\item {\tt Zprojecting()} translates the Feynman diagram into the
      helicity amplitude language in the following steps:
      \begin{itemize}
      \item The colour structure of the Feynman diagram is determined
            using the {\tt Color\_Generator}. With 
            {\tt Color\_Generator::Convert()} every vertex gains a
            colour matrix and {\tt Color\_Generator::C2string()}
            translates the whole term into a character string which will
            be used for later evaluation.  
      \item The various pieces of a helicity amplitude are
            generated within the \\ {\tt Zfunc\_Generator}. Therefore, 
            {\tt Zfunc\_Generator::Convert()} interprets the Feynman
            diagram, i.e.\ the list of {\tt Points} in terms of
            helicity amplitudes and {\tt Zfunc\_Generator::Flip()}
            changes the sequence of pairs of barred and unbarred
            spinors, so that in each pair, barred spinors come before
            unbarred ones in accordance with the definition of the 
            $X$--, $Y$-- and $Z$--functions.  
      \item The {\tt Prop\_Generator} constructs a list of propagators,
            i.e.\ determines the momenta of the different propagators
            in terms of incoming and outgoing particles. With 
            {\tt Prop\_Generator::Convert()} the list is generated,
            {\tt Prop\_Generator::Fill()} fills in the ingredients
            (dependencies on external particles) of every propagator
            and kills the ones which are not used 
            ({\tt Prop\_Generator::Kill()}). In this method, the list
            of {\tt Zfuncs} is explored and every propagator which
            could not be found, is deleted.     
      \end{itemize}  
      Now, lists of {\tt Zfuncs}, {\tt Cfuncs} and {\tt Pfuncs} for the
      helicity amplitudes, the color matrices and the list of
      propagators, respectively, are at hand.  
\item {\tt Fill\_Coupling()} fills all couplings into the 
      {\tt String\_Handler}. For this purpose, a loop over all 
      {\tt Zfunc}s is performed and every particular coupling is set 
      via {\tt String\_Handler::Get\_Cnumber()}.   
\item {\tt MPolconvert()} converts polarizations, i.e.\ replaces every
      number in the list of {\tt Zfunc}s with a given new number for the
      polarization.    
\item {\tt Prop\_Replace()} adds a new propagator to the list of 
      {\tt Pfunc}s. This method is commonly used for massive
      polarization vectors, where an external massive vector boson is
      treated like a propagator decaying into its polarizations.  
\item {\tt Zvalue()} calculates the helicity amplitude for one
      Feynman diagram
      \footnote{Note, that by construction, the boson propagators
                are already part of the $Z$--function.}:
      \begin{itemize}
      \item The maximum number of fermionic propagators (i.e. \ those
            which are switched on) is determined and their numbers are stored
            into an array. 
      \item If this number of propagators equals zero, only one {\tt Zfunc}
            has to be calculated. The arguments of the $Z$ function as
            well as their helicities are filled into a list of arguments via 
            {\tt Fill\_Args()}. Then, the appropriate couplings are set via 
            {\tt Building\_Blocks::Set\_Arg\_Couplings()}. Now the $Z$
            function can be calculated with {\tt Single\_Zvalue()}. 
      \item If any fermionic propagator exists, a loop over all possible
            states for every fermion propagator has to be performed. Since the
            number of propagators is not known in advance, a loop over
            loops technique has to be applied (cf.\ Appendix~\ref{Loop}). 
 	    In the external loop
            the sign of the propagators which correspond to a
            particle or anti-particle state, and the two helicity
            states have to be altered. Having at hand one such sample,
            the product of all $Z$ functions for this amplitude has to
            be calculated using a loop over all {\tt Zfunc}s. 
            The evaluation is performed in the same way as described
            above for a single {\tt Zfunc}. Now, the product of all
            $Z$'s is multiplied with an extra mass term 
            ({\tt Mass\_Terms()}) which corresponds to the real and
            virtual propagator mass, see Eq.~(\ref{FermionPropsEq}). The
            contributions for every set of propagator states is summed
            up and multiplied at the very end with all emerging
            fermion propagators ({\tt Building\_Blocks::Set\_P()}) and
            with a factor of one half taking into account the sum over
            particles and anti-particles.  
      \end{itemize}
      Finally, the calculated value multiplied by the overall sign of the
      amplitude is returned.      
\item {\tt Fill\_Args()} constructs the list of arguments for one $Z$
      function. For this purpose it uses the list of helicities for the
      incoming and outgoing particles as well as the list of propagators
      with their appropriate helicity states. A loop over the arguments of
      the $Z$ function is performed and translated in the following way:
      \begin{itemize}
      \item If the argument is a propagator, i.e.\ the number is larger
            than $99$, a signed propagator number and the appropriate
            helicity is added to the list of arguments. The sign
            corresponds to a particle or an anti-particle state
            propagating.   
      \item If the argument is an incoming or outgoing particle, its number
            and helicity are taken over unchanged.
      \item If the argument corresponds to a polarisation of a
            massless vector boson, its number is larger then the
            maximum number of particles and smaller than $20$. In this
            case, the helicity of the polarisation is the same as the
            helicity of the appropriate massless vector boson and the
            number of the momentum is translated via subtracting $10$.     
      \item The number of the polarisation for a massive vector boson is
            always between $20$ and $99$. These polarisations get a 
            helicity of minus one by construction and the momentum
            number can be gained by subtracting $20$.  
      \end{itemize}   
\item {\tt Single\_Zvalue()} calculates one {\tt Zfunc} using the
      various possibilities of the {\tt Building\_Blocks}.  
\item {\tt Mass\_Terms()} calculates the mass term of
      Eq.~(\ref{FermionPropsEq}) for the decomposition of fermionic
      propagators using {\tt Building\_Blocks::Set\_Mass()}.     
\item {\tt printgraph()} prints all lists for one Feynman diagram,
      i.e.\ the list of {\tt Zfunc}s, {\tt Pfunc}s and {\tt Cfunc}s.   
\item {\tt Get\_Pointlist()} returns the list of {\tt Points}. 
\item {\tt Get\_Clist()} returns the list of colour matrices,
      i.e.\ {\tt Cfunc}s. 
\item {\tt Get\_Plist()} returns the list of propagators, i.e.\ {\tt Pfunc}s.   
\end{enumerate}
The structure {\tt Zfunc} includes all informations which are
necessary to calculate an arbitrary $Z$ function, namely the type
($-1..9$), the list of arguments and the list of couplings, see
Tab.~\ref{ZfuncType}. 
\begin{table}[t] 
\bc
\begin{tabular}{l|c|c|c|l}
Type & \#arg & \#coupl& \#prop &Description\\
\hline 
&&&&\\
-1& 2& 2& 0& A scalar boson which is connected\\
  &  &  &  & to two fermions.\\
0 & 4&12& 0& An outgoing or incoming vector boson \\ 
  &  &  &  & attached at a fermion line\\
  &  &  & 1& or two fermion lines connected via\\ 
  &  &  &  & a vector boson.\\
1 & 6&10& 3& A three vector boson vertex.\\
2 & 6& 7& 3& A two vector and one scalar boson vertex.\\
3 & 6&10& 3& A one vector and two scalar boson vertex.\\
4 & 0& 1& 0& A three scalar boson vertex.\\
5 & 8&10& 5& The four vector boson vertex.\\
9 &10&13& 7& The five vector boson vertex.\\
&&&&\\
\end{tabular}
\caption{\label{ZfuncType}All possible types of a {\tt Zfunc} and their
description as well as the number of arguments, couplings and
propagators.}
\ec
\end{table}                                                                   
Last but not least every bosonic propagator which is
involved in this $Z$ function is stored as well in a special list. 
This will become necessary later on for the calculation of massive
vector boson propagators.  

The class {\tt Zfunc\_Generator} constructs the list of {\tt Zfuncs},
i.e.\ it translates a Feynman diagram into the helicity language by
decomposing it into a number of $Z$ functions. This is one of the
central parts of the program. Note that during the determination of
the list of arguments for every $Z$ function, a certain internal
numbering scheme is applied, for details see Tab.~\ref{PolNumb}. 

\begin{table}[t]
\bc
\begin{tabular}{l|l} 
Particle type & Number\\ 
\hline
&\\ 
Incoming or outgoing & 0-9\\
Fermion propagator & 100-120\\
Boson propagator & 200-\\
Polarization of a massless vector boson & Number of the boson plus 10.\\
Polarization of a massive vector boson & Number of the boson plus 20.\\
Dummy numbers for the calculation&120-140\\
of the Color matrices & \\
&\\ 
\end{tabular}
\caption{\label{PolNumb}The particle numbering scheme in the list of
arguments for every $Z$ function.}
\ec
\end{table}

\begin{enumerate}
\item{\tt Convert()} translates a given graph, i.e.\ a list  of {\tt
Point}s into the list of {\tt Zfunc}s recursively. A new {\tt Zfunc}
will be produced, if the current particle is a fermion, a scalar boson
or an incoming vector boson. Depending on the {\tt Flavour} type of
this particle, two pointers will be set. In case it is a fermion, one
pointer is set to the outgoing boson and one to the outgoing
fermion. If the particle is a boson, only the boson pointer is set,
whereas the fermion pointer is set to zero. With the current {\tt Point} and
these two pointers as arguments the type of the {\tt Zfunc} can be
determined ({\tt Determine\_Zfunc()}) and the arguments can be filled
in ({\tt Fill\_Zfunc()}). At the end two recursive calls to 
{\tt Convert()}, with the left and the right hand {\tt Point} as
arguments, ensure the recursive examination of the whole graph.   
\item{\tt Determine\_Zfunc()} determines the type of the {\tt Zfunc}
according to Tab.~\ref{ZfuncType}, where the method {\tt IsGaugeV()}
is used to count the number of vector and scalar bosons at one vertex.
\item{\tt IsGaugeV()} determines the number of vector and scalar
bosons within one vertex.
\item{\tt Fill\_Zfunc()} specifies all arguments, couplings and
propagators of one {\tt Zfunc}. First of all, the sizes of all three lists
are determined (see Tab.~\ref{ZfuncType}). Now, the different lists
can be initialized and filled by calling the appropriate method. Note
that each type of {\tt Zfunc} has its own method ({\tt Zm1()}, 
{\tt Z0()}, etc.). 
\item{\tt Set\_In()} sets the arguments and couplings for the incoming
part of a {\tt Zfunc}. If the fermionic pointer is set (see 
{\tt Convert()}), the arguments are set on the current and the
fermionic pointer, where the sequence is always anti-particles before
particles. The two couplings will be filled with the couplings from
the current pointer. Note that the sign of the graph gains an extra minus for
a fermionic anti-particle propagator. The second case is addressed to
an incoming boson. If it is scalar, both arguments are set on the
number of this boson and the couplings are set to zero (this is only the
dummy part of a {\tt Zfunc}). For a vector boson the first argument is
set on the number of the boson and the second is set to the number
plus $20$ or $11$ for massive or massless vector bosons, respectively.
Here, the couplings are set to unity.
\item{\tt Set\_Out()} sets the arguments, couplings and propagators
for one outgoing end of a {\tt Zfunc}. At the beginning, the
propagator will be set onto the current boson. If the boson is not an
outgoing one the arguments of the {\tt Zfunc} are set onto the numbers
of the left and right hand particle. The sign of the graph will be
changed for a fermionic propagator on the left hand side (i.e.\ an
anti-particle propagator). Outgoing bosons will be managed in the same
way like incoming bosons (see {\tt Set\_In()}), but with an reversed
sequence of arguments. 
\item{\tt Set\_Scalar()} sets the left-- and right--handed couplings
to a scalar boson within a {\tt Zfunc}. This means that both arguments
are set to the number of the scalar boson and the couplings are set to
zero. The list of propagators is enhanced by the scalar boson propagator.  
\item{\tt Get\_Flav\_Pos3()} determines the position of the three
electroweak vector bosons in the method {\tt Z1()}. 
\item{\tt Zm1()} sets the arguments with {\tt Set\_In()} or {\tt
Set\_Out()} for an incoming or outgoing scalar boson, respectively.
\item{\tt Z0()} specifies the arguments and couplings for a vector
boson exchange between two fermion lines or an incoming or outgoing
boson:
\begin{itemize}
\item For an incoming or outgoing scalar boson the incoming part will
be set on position $0$ ({\tt Set\_In()}) and the outgoing part at
position $1$ ({\tt Set\_Out()}).
\item The same holds true for an incoming vector boson.
\item In the case of outgoing vector bosons or the exchange of a
vector boson between two fermion lines the outgoing part is set on
position $0$ and the incomings are set on position $1$ with 
{\tt Set\_Out()} and {\tt Set\_In()}, respectively.     
\end{itemize}
Note that the sequence of the arguments plays an important role
during the calculation of the $Z$ function. 
However, multiple propagators ($\gamma$, $Z$, Higgs--boson) will be 
considered by enhancing the list of couplings of the {\tt Zfunc}.
\item The methods {\tt Z\{1,2,3,4\}()} correspond to a three boson vertex,
where the positions within the list of arguments depend on the
appropriate vertex, for details see Tab.~\ref{ThreeVertex}. Note
that all incoming bosons are placed with {\tt Set\_In()}, outgoing
bosons are set with {\tt Set\_Out()} or {\tt Set\_Scalar()} for vector
or scalar bosons, respectively. However, multiple propagators can occur
in the case of vector bosons, too. Then, the list of couplings will be
enhanced by the extra couplings.
\begin{table}[h] 
\bc
\begin{tabular}{l|c|c|c|l}
Method & Incoming & Left & Right & Condition\\
\hline 
&&&&\\
{\tt Z1()} & 2 & 0 & 1 & For a three gluon vertex.\\
	   & 2 & 0 & 1 & For $\gamma/Z,W,W$ vertex.\\
	   & 0 & 2 & 1 & For $W,\gamma/Z,W$ vertex.\\
	   & 0 & 1 & 2 & For $W,W,\gamma/Z$ vertex.\\
{\tt Z2()} & 2 & 0 & 1 & For $H,Z/W,Z/W$ vertex.\\
	   & 0 & 2 & 1 & For $Z/W,H,Z/W$ vertex.\\
           & 0 & 1 & 2 & For $Z/W,Z/W,H$ vertex.\\
{\tt Z3()} & 2 & 0 & 1 & For $V,S,S$ vertex.\\
	   & 0 & 2 & 1 & For $S,V,S$ vertex.\\
	   & 1 & 0 & 2 & For $S,S,V$ vertex.\\	
{\tt Z4()} &   &   &   & For $S,S,S$ vertex (is only a coupling).\\
&&&&\\
\end{tabular}
\caption{\label{ThreeVertex}The positions in the list of arguments for
the different three boson vertex types and constellations.}
\ec
\end{table}                                                                     
\item{\tt Z5()} sets the arguments for the four vector boson vertex,
see Tab.~\ref{FourVertex}. Apart from that, all other options are the
same like in the three boson vertex case. 
\begin{table}[h] 
\bc
\begin{tabular}{l|c|c|c|c|c|c|c|l}
Method     & I& L&LL&LR& R&RL&RR& Condition\\
\hline 
&&&&&&&&\\
{\tt Z5()} & 0&  & 3& 2& 1&  &  & For $\gamma/Z;W;\gamma/Z;W$ vertex.\\
           & 1&  & 3& 0& 2&  &  & For $W;W;\gamma/Z;\gamma/Z$ vertex.\\
           & 0& 1&  &  &  & 2& 3& For $\gamma/Z;W;\gamma/Z;W$ vertex.\\
           & 1& 0&  &  &  & 2& 3& For $W;\gamma/Z;\gamma/Z;W$ vertex.\\
           & 0& 1&  &  &  & 2& 3& For $G;G;G;G$ vertex.\\
&&&&&&&&\\
\end{tabular}
\caption{\label{FourVertex}The positions in the list of arguments for
the four vector boson vertex with different constellations. Note that
$I$ stands for incoming, $L$ and $R$ for left and right, respectively.}
\ec
\end{table}
\item{\tt Z9()} considers the calculation of the five vector boson
vertex. Note that in the current version only the
five gluon vertex is used, i.e.\ no five electroweak gauge boson
vertices are taken implemented yet. This is part of future work.
\item{\tt Flip()} takes care for the correct order of particle and
anti-particles in the list of arguments in the {\tt Zfunc}s, i.e.\ the
sequence of spinors. Here, anti-spinors always have to stand before
spinors in the appropriate two-particle blocks of the {\tt Zfunc}s. 
Therefore a loop over all {\tt Zfuncs} is performed and the
arguments of every function are examined. Now, a change will occur in
case an incoming fermion or an outgoing anti-fermion is placed
before its partner. The same change will be enforced, if the second
particle is an incoming anti-fermion or an outgoing fermion. Both
checks are necessary, since one of the arguments could be a
propagator, where the spinor or anti--spinor property is not fixed 
before. If a change should be performed and the partner of the
examined particle is a propagator, not only the current {\tt Zfunc}
but also a second one will be altered. This is due to the fact that
a fermionic propagator always emerges in two {\tt Zfunc}s. Now, if the
propagator stands on the first position after the change, the second
occurrence of the propagator in another {\tt Zfunc} has to be as a
second argument. Therefore, the other {\tt Zfunc} will be searched and
changed accordingly. However, this second occurrence of the propagator
can have a propagator as a partner as well. Consequently, the change
of the position for this partner results in another change for its
second appearance and so on. In this way, the whole fermion line will
be flipped.      
\item{\tt Get()} returns the list of {\tt Zfunc}s.
\end{enumerate}
The structure {\tt Pfunc} contains all informations to characterize a
propagator. The necessary arguments are the number of the propagator
and all those incoming and outgoing momenta which sum up to its 
momentum. A  {\tt Flavour} gives the type of the propagator and a
complex number stores the calculated value. Last but not least an on
switch allows the dropping of individual propagators.  

The class {\tt Prop\_Generator} generates the list of propagators,
i.e.\ the {\tt Pfunc}s, and calculates their complex value for one 
individual graph:
\begin{enumerate}
\item {\tt Convert()} translates a topology, i.e.\ a list of 
{\tt Point}s into a list of propagators ({\tt Pfunc}s). A propagator
is characterized through a number which is larger than $99$ (this
corresponds directly to our numbering scheme, see
Tab.~\ref{PolNumb}). Now, a recursive
examination of the whole tree yields all possible propagators. Note
that some of the propagators are combined for different
boson types (i.e.\ photon, $Z$, and Higgs boson). Since different boson
types yield different propagator values, a new propagator will be
generated for every boson. For every 
propagator a {\tt Pfunc} is created and attached to the list. The
first argument of the {\tt Pfunc} is the number of the appropriate
propagator and the two next arguments correspond to the numbers of the
left and right handed particles in the graph. At the end of the method
two recursive calls to {\tt Convert()} will be performed.
\item {\tt Fill()} completes the list of arguments for the
propagators. In {\tt Convert()} every propagator is filled into a 
{\tt Pfunc}. The arguments of this {\tt Pfunc} are the flavour of the
propagator and the left and right handed particles. However, in order
to calculate the four-momenta of the propagators later on, the
dependence on the incoming and outgoing particles is needed (since
only the four--momenta of these particles are available). The task of
this method is to replace the dependence of the propagator on the
left and right handed particle by the dependence on all (incoming and
outgoing) particles which are attached at the same branch. Therefore, a loop
over all {\tt Pfunc}s is enclosed by a loop which ends, when all
dependencies are filled in. Now, every {\tt Pfunc} is examined.  
If a propagator is found in the arguments, the
number of this propagator is replaced by the list of dependencies for
this propagator itself. These replacements are performed until every
propagator has a list of dependencies which includes only the numbers
of incoming and outgoing particles.     
\item {\tt Kill()} deletes all propagators which do not appear as an
argument in the list of {\tt Zfunc}s. A loop over all propagators and
$Z$ functions searches for the number of the current propagator in the
list of arguments of the current $Z$ function. If a propagator could
not be found, it will be switched off.  
\item {\tt Calculate()} evaluates the complex value of all
propagators. Consequently, a loop over the list of {\tt Pfunc}s is
performed. Within the loop first of all, the four--momentum of the
current propagator has to be determined. This can be achieved by
adding up all the depending four-momenta of incoming and outgoing
particles, i.e.\ all four--momenta which are attached at the same
branch of the tree as the propagator. Note that the four--momenta of
the incoming particles are multiplied with a $-1$. Now, the complex
value of the propagator can be calculated using the mass and the width
of the current particle. According to our Feynman rule conventions
every fermionic and scalar particle is multiplied with an extra complex
unit and the vector bosons are provided with minus a complex unit.       
\item {\tt Get()} returns the list of {\tt Pfunc}s. 
\end{enumerate}
The structure {\tt Cfunc} represents an element of the color matrices,
i.e.\ a generator of the $SU(3)$ color group ($T^a_{bc}$) or a
structure constant of the group ($f_{abc}$). They are indicated by
different types within {\tt Cfunc}, where a $10$ represents a delta
function, a $0$ corresponds to a generator of the fundamental
representation and the number $1$ is connected with a structure
constant. The appropriate labels
($abc$) are saved in a list of arguments. 

The {\tt Color\_Generator} generates a list of these {\tt Cfunc}s
which belong to every graph. A string representation will be created
as well, in order to simplify the calculation of the color factors
later on: 
\begin{enumerate}
\item {\tt Convert()} generates the list of {\tt Cfunc}s. Therefore,
the whole topology of a graph is searched recursively and every vertex
yields a new color matrix {\tt Cfunc} depending on the incoming and
the two outgoing particles of this vertex (see Tab.~\ref{ColorGen}).
\begin{table}[h] 
\bc
\begin{tabular}{l|l|l|l|l|l|l|l} 
Incoming & Left & Right & Type&arg0&arg1&arg2&Condition\\
&outgoing&outgoing&&&&&\\
\hline 
&&&&&&&\\ 
Quark&Boson&Quark&0/10&L &R &I &Incoming anti-quark.\\
     &     &     &    &L &I &R &Incoming quark.\\
Quark&Quark&Boson&0/10&R &L &I &Incoming anti-quark.\\
     &     &     &    &R &I &L &Incoming quark.\\
Boson&Quark&Quark&0/10&I &L &R &\\
Boson&Boson&Boson&1   &I &L &R &\\
&&&&&&&\\ 
\end{tabular}
\caption{\label{ColorGen}In this table the different allocations for
the arguments of a {\tt Cfunc} are shown. The left, right and the
incoming particle are denoted as {\tt L}, {\tt R}, and {\tt I},
respectively. Note that the type depends on the nature of
the boson. In case it is a gluon, the type is $0$ and for all other
bosons the type is $10$.}
\ec
\end{table}                                                                   
At the end two recursive calls to {\tt Convert()} (with the left and
right hand side {\tt Point}) ensure that every
vertex in the topology is translated into a {\tt Cfunc}. 
\item {\tt Kill()} performes a first simplification of the color
structure, i.e.\ all delta functions which involve propagators, are
deleted and the appropriate replacements are done within the list of
{\tt Cfunc}s. Accordingly, a loop over all {\tt Cfunc}s is performed
and every delta function (i.e.\ type equals $10$) is examined. If one
of the two arguments is a propagator, all emerging numbers of
this propagator will be replaced by the second argument in all 
{\tt Cfunc}s. At the end these {\tt Cfunc}s which correspond to a
delta function with propagator indices only, are deleted. Note that delta
functions remain which have incoming or outgoing particles in the
argument.
\item {\tt C2string()} translates the list of {\tt Cfunc}s into a character
string. During the generation of the string, every type of {\tt Cfunc}
will be translated accordingly, i.e.\ type $0$ into a $T[A,b,c]$, type
$1$ into a $F[A,B,C]$ and type $10$ into a $D[a,b]$ function. Note
that every upper case letter lies in the range of $1\dots 8$ and the lower
case letters in the range of $1\dots 3$. Furthermore, some care has to be
taken that propagators which appear in different {\tt Cfunc}s, gain
the same letter. However, at this stage not only the color string will
be created, but also its complex conjugate. Here, every propagator has
to get a totally new letter with respect to the propagators in the
original string, since these letters are only inner summation
indices. Consequently, the color factor can now be calculated by just
merging those two strings and no special care has to be taken for
inner summation indices any more. This simplifies the calculation
significantly later on (see the class {\tt Color}).    
\item {\tt Get()} returns the list of {\tt Cfunc}s.
\end{enumerate}
\subsubsection{\label{AmplCalc}Calculating the Amplitude}
The calculation of the amplitudes is organized on different
abstraction levels. On the highest level stands the class 
{\tt Building\_Blocks} which evaluates the different composed $Z$
functions and will be called from within 
{\tt Single\_Amplitude}. The building blocks themselves rely on the 
{\tt Composite\_Zfuncs} and the {\tt Mathematica\_Interface}s, where
the latter one uses the methods of the {\tt Composite\_Zfuncs} as
well. Strictly speaking, these composed $Z$ functions translate a
simple notation of the different $Z$ functions into a call with all
arguments, where the methods of the class {\tt Elementary\_Zfuncs}
will be used. These methods calculate all the $Z$,$X$,$Y$,
etc. functions with the help of a set of basic spinor products and
normalizations which are established in the class 
{\tt Basic\_Sfuncs}. This is the lowest level class which directly
uses the four--momenta of the incoming and outgoing particles.     

The {\tt Building\_Blocks} are the basic elements of a
calculation. They calculate one single $Z$ function and are called
from within {\tt Single\_Amplitude::Zvalue()}:
\begin{enumerate}
\item With the {\tt Constructor()} of this class a new object of
{\tt Composite\_Zfuncs} is initialized. Together with the 
{\tt Mathematica\_Interface}s they form the basic calculational
tools of this class.
\item {\tt Reset()} resets the {\tt Composite\_Zfuncs} object (via the
appropriate {\tt Reset()} method).
\item {\tt Set\_Arg\_Coupl()} is used to set the list of arguments and
couplings within this class and the {\tt Composite\_Zfuncs} object. 
\item {\tt Set\_P()} calls {\tt Composite\_Zfuncs::Set\_P()} and
calculates the {\tt Kabbala} value of the given propagator ({\tt Pfunc}).
\item {\tt Set\_Mass()} calculates the mass term which occurs during
the decomposition of the fermionic propagator into products of spinors
(via\\ {\tt Composite\_Zfuncs::Set\_Mass()}).
\item {\tt Z2()} calculates the {\tt Kabbala} value of the original
$Z$ function. Here, the different contributions of multiple
propagators (photon, $Z$ and Higgs boson) are summed up as well. The
$Z$ function is calculated via {\tt Composite\_Zfuncs::Z\_Z()} and the
propagator via {\tt Composite\_Zfuncs::Z\_P()}. Note that the
contribution of a scalar boson exchange is composed of the product of
two $Y$ functions ({\tt Composite\_Zfuncs::Z\_Y()}). The last missing
piece is the mass term which appears in the numerator of the
propagator for massive vector boson exchanges ($p^\mu p^\nu/M^2$). This
term can be composed out of two $X$ functions 
({\tt Composite\_Zfuncs::Z\_X()}) and a constant which is directly handled
using the method  {\tt String\_Generator()::Get\_Enumber()}. Note
that this method translates a complex number into a {\tt Kabbala}
object and saves the value in the list of {\tt Zfunc}s (see
Sec.~\ref{Strings}). 
\item All other methods calculate an appropriate $n$ boson vertex,
usually with the help of a {\tt Mathematica\_Interface()}. A new
object will be created the first time this method is called. 
The result is a product of the list of the inner propagators 
(gained via {\tt Composite\_Zfuncs::Z\_P()}),
the value of the vertex ({\tt Mathematica\_Interface()::Calculate()}) 
and the coupling constant\\ ({\tt String\_Handler::Get\_Enumber()}). 
A further complication can arise when multiple propagators occur 
(i.e.\ a photon and a $Z$ boson). Then the contributions will be simply
summed up. If more than three vector bosons appear in one vertex, a
sum of the multiple iterated three vector boson vertices and a
quadruple vector boson vertex has to be performed. This is due to our
handling of quadruple vector boson vertices. However, a detailed list
of every method can be found in Tab.~\ref{BBVert}.   
\end{enumerate}
\begin{table}[t] 
\bc
\begin{tabular}{l|l} 
Method & Purpose\\
\hline 
&\\ 
Z3V()   & Three vector boson vertex,\\ 
        & sums up multiple propagators with {\tt Z31}.\\
Z31()   & Three vector boson vertex.\\
ZVVS()  & Two vector, one scalar boson vertex.\\
ZSSV()  & Two scalar, one vector boson vertex,\\ 
        & sums up multiple propagators with {\tt ZSSV1()}.\\
ZSSV1() & Two scalar, one vector boson vertex.\\
ZSSS()  & Three scalar boson vertex, simply a coupling.\\
Z4V()   & Four vector boson vertex.\\
Z5V()   & Five vector boson vertex.\\
&\\ 
\end{tabular}
\caption{\label{BBVert} A selected list of methods for the calculation
of an $n$ boson vertex.}
\ec
\end{table}                                                                   
The {\tt Mathematica\_Interface} can be used to implement algebraic
expressions which are produced by {\tt Mathematica}. Here, the output
of the appropriate {\tt Mathematica} session will be translated into a
proper {\tt C++} file and linked to the object. This process can be
done in a semi-automatic way. However, usually all necessary
calculations have been already performed and the linked {\tt C++} files
are ready to use:
\begin{enumerate}
\item The {\tt Constructor()} can be used with two different
options. The first one which is the usual option, is used for the
calculation of already translated methods. An identity string
indicates the method to be used. In the second option a 
{\tt Mathematica} file will be read in and translated into a {\tt C++}
file. First of all, the file will be read in and casted into a
string. The method {\tt Shorten()} deletes all occurrences of $p$ and
$q$ characters which are necessary in the {\tt Mathematica} file, but
are now obsolete. With {\tt C\_Output()} the string will now be
translated into a {\tt C++} file. All other parts, i.e.\ the linkage of
this new method and the calling sequence have to be included by hand,
a sample could be found in Appendix~\ref{SampleMath}.   
\item {\tt C\_Output()} transforms an algebraic expression in the form
of a character string into a {\tt C++} method of this class. Therefore a
new file will be created and the output of the string expression is
performed in the same way as the output of the libraries, see 
Sec.~\ref{Strings}, i.e.\ a maximum number of characters per line will be
enforced.   
\item {\tt Shorten()} uses {\tt Kill\_String()} to delete all $p$ and
$q$ characters  
\item {\tt Kill\_String()} kills the occurrence of a given character in
a string.  
\item {\tt Calculate()} is the steering routine of the calculational
part of this class. According to the given string identity, this
routine choses the appropriate method for the evaluation.  
\item The methods {\tt Z(), X(), S(), Y()} and {\tt M()} calculate an
elementary $Z$, $X$, $S$, or $Y$ function or a mass term, respectively. The
appropriate methods {\tt Z\_Z(), Z\_X(), Z\_S(), Z\_Y()} and 
{\tt Z\_M()} of the class {\tt Composite\_Zfuncs} are used accordingly.
\item All other methods calculate a multiple boson vertex directly,
see Tab.~\ref{MathBos}.
\end{enumerate}
\begin{table}[t] 
\bc
\begin{tabular}{l|l} 
Method & Purpose\\
\hline 
&\\ 
vGGG()      & Three vector boson vertex without any mass terms.\\ 
vZWW()      & Three vector boson vertex including all mass terms.\\  
vVVS()      & Two vector one scalar boson vertex.\\
vSSV()      & Two scalar one vector boson vertex.\\
vGGGG\_it() & Two iterated three boson vertices as a part of the\\ 
            & four vector boson vertex, masses are neglected.\\ 
vGGGG\_q()  & Quadruple vector boson vertex, masses are neglected.\\
vZZWW\_it() & Like {\tt vGGGG\_it()}, but including all mass terms.\\
vZZWW\_q()  & Like {\tt vGGGG\_q()}, but including all mass terms.\\
v5G\_it()   & Three iterated three boson vertices as part of the\\ 
            & five vector boson vertex, masses are neglected.\\
v5G\_itq()  & One triple plus one quadruple vector boson vertex,\\
            & neglecting masses.\\
&\\ 
\end{tabular}
\caption{\label{MathBos} A selected list of methods for the direct
calculation of a part of a multiple boson vertex.}
\ec
\end{table}                                                                   
The class {\tt Composite\_Zfuncs} translates a simplified form of a
call to the elementary $Z$ functions ({\tt Elementary\_Zfuncs}) into a
full calling sequence. Therefore this class mainly exists for a better
readability of the {\tt Building\_Blocks} and the\\ 
{\tt Mathematica\_Interface}:  
\begin{enumerate}
\item The {\tt Constructor()} creates a new {\tt Elementary\_Zfuncs}
object. 
\item {\tt Reset()} sets the pointers of the arguments and the
couplings of a $Z$ function to zero.
\item {\tt Set\_Arg\_Coupl()} sets the arguments and couplings of a $Z$
function. 
\item {\tt Set\_P()} calls the method {\tt Elementary\_Zfuncs::P()} in
order to calculate the value of a propagator. 
\item {\tt Set\_Mass()} calls the method 
{\tt Elementary\_Zfuncs::Mass\_Term()} in order to evaluate the extra mass
term of a decomposed fermionic propagator, see Eq.~(\ref{FermionPropsEq}).
\item {\tt Z\_P()} calculates the product of propagators for a $Z$
function. Therefore, a list of propagator numbers is used to extract
all propagators taking part in the current $Z$ function. Their {\tt
Kabbala} values will be multiplied up and returned. 
\item {\tt Z\_X()} calculates an $X$ function and translates a simple
combination of arguments into the full list and calls 
{\tt Elementary\_Zfuncs::X()}. Note that this method takes into
account a sign convention which permits the proper handling of the
four-momentum current.
\item {\tt Z\_S()} works in the same way like {\tt Z\_X()}, but calls\\
 {\tt Elementary\_Zfuncs::S()}.
\item {\tt Z\_Z()} translates the simple notation of an elementary $Z$
function into the full call of {\tt Elementary\_Zfuncs::Z()}.
\item {\tt Z\_Y()} works like {\tt Z\_Z()} with a slight
difference. Here, it can happen that both couplings are zero. In this
case the value $1$ will be returned.    
\item {\tt Z\_M()} calls directly {\tt Elementary\_Zfuncs::M()}
without any translations.
\end{enumerate}
The {\tt Elementary\_Zfuncs} form the last step on the ladder to the
calculation of the different $Z$ functions. In this class they will be
finally calculated and the\\ {\tt String\_Generator} is filled with the
different $Z$ functions. At least two methods exist
for each function, one for building the string part and the second
to calculate the appropriate $Z$ function. Sometimes different
methods exist for different calling sequences within the string
libraries.
\begin{enumerate}
\item The methods {\tt Z()}, {\tt Y()}, {\tt X()} and {\tt S()}
calculate an elementary $Z$, $Y$, $X$ function or a scalar product of
two four-momenta and provide its value for the 
{\tt String\_Generator()}. Therefore all arguments and
couplings will be filled into one list. Now, all
propagators which have numbers larger than $99$ will be mapped onto
real four-momenta (i.e.\ with numbers smaller than $100$) via 
{\tt Map()}. Then, the value of the function can be calculated via the
appropriate {\tt \{Z,Y,X,S\}calc()} and translated into a {\tt Kabbala}
type value with\\ {\tt String\_Generator::Get\_\{Z,Y,X,S\}number()}. Note
that in case a string is to be built, the function with all arguments
is stored within the {\tt String\_Generator}.
\item The methods {\tt Zcalc()}, {\tt Ycalc()}, {\tt Xcalc()} and 
{\tt Scalc()} calculate an elementary $Y$, $X$, $Z$ function or a
scalar product according to Tabs.~\ref{Yfuncs}, \ref{Xfuncs} and
\ref{Zfuncs}, respectively. 
\item {\tt P()} calculates the value of a propagator and fills it into
the {\tt String\_Generator}. First of all, the number of the given
propagator will be determined via 
\\{\tt Basic\_Sfuncs::Get\_Mom\_Number()}. Then, the value is calculated
with \\{\tt Pcalc(Flavour fl,...)} and translated into a {\tt Kabbala}
value\\ ({\tt String\_Generator::Get\_Pnumber()}).
\item {\tt Pcalc(int fl,...)} translates the integer valued kf--code
into a proper\\ {\tt Flavour()} and calls {\tt Pcalc(Flavour fl,...)}.
\item {\tt Pcalc(Flavour fl,...)} calculates the value of a propagator
using\\ {\tt Propagator()}. 
\item {\tt Propagator()} calculates the value of a propagator for a
given four-momentum squared and the {\tt Flavour}. Note that
different types of the {\tt Flavour} gain different extra factors, 
i.e.\ fermionic and scalar propagators yield an extra factor of $i$ 
and vector boson propagators get a $-i$. 
\item {\tt Mass\_Term()} generates the extra mass term which occurs
during the decomposition of the fermionic propagator into spinor
products. The first step is to determine the {\tt Flavour} of the
propagator from the given number. Then, the mass term is
calculated with {\tt Mass\_Term\_Calc(..., Flavour fl)} and it is filled
into the {\tt String\_Generator} (via {\tt Get\_Massnumber()}).  
\item {\tt Mass\_Term\_Calc(..., int fl)} translates the given kf--code
into a proper\\ {\tt Flavour} and calls 
{\tt Mass\_Term\_Calc(..., Flavour fl)}. 
\item {\tt Mass\_Term\_Calc(..., Flavour fl)} calculates the mass term. 
\item {\tt M()} determines the mass for a given particle number in
case it is a propagator. This method is used for the calculation of
the mass terms in massive vector boson propagators. The term $1/M^2$
is then filled into the {\tt String\_Generator()} as a complex number
({\tt Get\_Enumber()}).  
\item {\tt Map()} translates a given propagator number ($>99$) into a
proper four--momentum number. Therefore the list of {\tt Pfunc}s is
searched for the given number. Now, the new number is determined from
the appropriate {\tt Pfunc} using 
\\{\tt Basic\_Sfuncs::Get\_Mom\_Number()}.
\end{enumerate}
We now turn our attention to the calculation of the basic functions
which are necessary for the determination of every $Z$ function. 

The structure {\tt Momfunc} represents a four--momentum of a
particle. Since in this structure not only incoming and outgoing
momenta, but also the four-momenta of the propagators are stored, a
list of arguments is available as well (compare {\tt Pfunc}).

The class {\tt Basic\_Sfuncs} generates and translates a list of
four-momenta, i.e.\ {\tt Momfunc}s, into the basic functions of the
helicity formalism, see Eq.~(\ref{Sfunctions}) for the $S$ functions
and Eq.~(\ref{muneta}) for the definition of the $\mu$ and $\eta$
functions. With these tools, all other $Z$ functions can be
calculated: 
\begin{enumerate}
\item The {\tt Constructor()} initializes the list of {\tt Momfunc}s
with the call to\\ {\tt Initialize\_Momlist()}.
\item {\tt Initialize()} constructs all necessary lists, i.e.\ for the
$\mu$--, $\eta$--, $S_0$-- and $S_1$--functions which correspond to the
$S(+,\cdot)$ and $S(-,\cdot)$ functions. Furthermore, a list of masses
for the different particles is created which is used later on for the
determination of the $\mu$ functions.
\item {\tt Initialize\_Momlist()} constructs the list of {\tt Momfunc}s
for the incoming and outgoing particles. Since at this stage no
propagators will be initialized, the list of arguments is
filled only with the number of the particle. At the end
the current number of {\tt Momfunc}s is returned.
\item {\tt Build\_Momlist()} adds to the list of {\tt Momfunc}s all
possible propagators. A loop over the given list of {\tt Pfunc}s
ensures that all propagators will be covered. First of all, a check
will be enforced, if the current propagator is already in the list
with {\tt Get\_Mom\_Number()}. This method returns the number of a given
{\tt Pfunc} in the list of {\tt Momfunc}s or returns a $-1$ in case
this propagator does not exist within the list. Now, a new {\tt Momfunc}
can be established, where the list of arguments will be taken over
from the {\tt Pfunc}. At the end, the new number of 
{\tt Momfunc}s will be returned.
\item {\tt Print\_Momlist()} prints the current list of 
{\tt Momfunc}s including the dependencies for every propagator type
momentum.
\item {\tt Get\_Mom\_Number()} returns for a given {\tt Pfunc} the
number of its propagator in the list of {\tt Momfunc}s. In
case this propagator does not exist, a $-1$ is returned. Accordingly,
a loop over all existing {\tt Momfunc}s is performed. Each list of
arguments for the current {\tt Momfunc} and the {\tt Pfunc} are now
compared. In case they are equal, the current number of the {\tt
Momfunc} is returned. Note that the comparison is sensitive to
different sequences within the list of arguments.   
\item {\tt Calc\_Momlist()} initializes the list of 
{\tt Momfunc}s. This means that the {\tt Momfunc}s for incoming and
outgoing particles will be simply set and all propagators are
calculated depending on the given four--momenta using the list of
arguments and the sign of each particle (see {\tt Sign()}). 
\item {\tt setS()} is the main method which is called from outside
for every new combination of incoming and outgoing four-momenta. Here,
all the different functions as well as all four-momenta for the
propagators will be calculated in the following steps:
\begin{itemize}
\item First of all, the four-momenta of all propagators will be
calculated with {\tt Calc\_Momlist()}.
\item Now, within a loop over all {\tt Momfunc}s all $\mu$-- and 
$\eta$--functions will be set. However, whereas the $\eta$--functions can be
calculated without any ambiguities, the sign of the $\mu$--functions
depends on the particle or anti-particle character. In our notation
a minus sign comes with:
\begin{itemize}
\item All anti-particles which do not correspond to the first incoming
particle and 
\item the first incoming particle which is not an anti-particle. 
\end{itemize}
All other particles yield no sign. Note that the sign
should have no influence for the propagator part of the 
{\tt Momfunc}s. 
\item In the last step the $S_0$-- and $S_1$--functions will be calculated
using an outer and an inner loop over the {\tt Momfunc}s, since these
functions depend on two four-momenta. The different relations between
the functions will be used in order to minimize their number to be evaluated.
\end{itemize}
\item {\tt N()} returns the normalization factor for two given
four-momenta. This method is used during the calculation of the
normalization for the massless vector boson polarizations.
\item {\tt Momentum()} returns the four-momentum for a given particle
number by searching the list of {\tt Momfunc}s. 
\item The methods {\tt S0()}, {\tt S1()}, {\tt mu()} and {\tt eta()}
return $S(+;p_1,p_2)$, $S(-;p_1,p_2)$, $\mu(p)$ and $\eta(p)$.
\item {\tt Sign()} returns the sign of a momentum, i.e.\ a $-1$ for
incoming and a $+1$ for outgoing particles.
\item {\tt Flav()} returns the {\tt Flavour} of the given particle.
\end{enumerate}
\subsubsection{\label{AmplCoCo}Calculating the Color matrix and the
Coulomb factor} 

The class {\tt Color} calculates all color factors from the list of
{\tt Cfunc}s and the appropriate strings of color matrices which have
been established in the class {\tt Color\_Generator()}. Therefore, the
list of {\tt Cfunc}s will be first shortened and then the calculation
is performed by manipulating the color matrices in form of a character
string, the different replacement rules are listed in
Tab.~\ref{ReplRule}.  
\begin{table}[t] 
\bc
\begin{tabular}{l|l} 
Method & Replacement\\ 
\hline
&\\ 
{\tt Single\_ReplaceFT()} &
$\displaystyle 
f^{abc} T^c = - i (T^a T^b-T^b T^a)$\\
&\\
{\tt Single\_ReplaceF()}&
$\displaystyle f^{abc} = 
2i T^a_{ij} (T^c_{jk}T^b_{ki}-T^b_{jk}T^c_{ki})$\\
&\\
{\tt ReplaceT()}&
$\displaystyle 
T^a_{ij} T^a_{kl} = \frac12\left[\delta_{il} \delta_{jk}-
\frac13 \delta_{ij} \delta_{kl}\right]$\\
&\\
&Special case: $\displaystyle 
T^a_{ij} T^a_{jl} = C_F \delta_{il}$\\
&\\
{\tt ReplaceD()}& $\delta_{ii} = 3$\\
&\\ 

\end{tabular}
\caption{\label{ReplRule} The replacement rules for the different
methods in the class {\tt Color}.}
\ec
\end{table}                                                                   
\begin{enumerate}
\item The {\tt Constructor()} calculates all elements of the color
matrix or reads them in from a saved file in the following steps:
\begin{itemize}
\item At first, all diagrams (the list of {\tt Single\_Amplitude}s)
will be switched on. Later on, this switch will be used to discard a
repeated calculation of the color factor for identical color structures.  
\item The data file {\tt Color.dat} will be probed. In case the
color factors have been already determined and saved into this file,
they will be read in and the {\tt Constructor()} can be
terminated. 
\item If the color factors have not been calculated before,
their determination starts with finding out all equal color
structures. This is mandatory in order to reduce the number of the
color factors to a minimum. It is performed in two steps:
\begin{enumerate}
\item All color matrices (i.e.\ {\tt Cfunc}s) will be brought into the
same form, i.e.\ the propagators will be unified. Therefore the numbers
of all incoming and outgoing particles will be searched in the list
of {\tt Cfuncs} for one diagram. If the appropriate number has been
found (every number has to emerge once), a propagator which stands in
the same {\tt Cfunc} will be changed to a number beginning from $120$
(these numbers do not occur in the usual numbering scheme for
propagators, see Tab.~\ref{PolNumb}). All other occurrences of this
propagator in the current list of {\tt Cfuncs} will be altered
accordingly. This procedure ensures that identical {\tt Cfunc}s in
different diagrams have the same numbering order for the propagators.     
\item Now, the list of {\tt Cfunc}s for all diagrams can be
compared. If two list of {\tt Cfunc}s exactly coincide, one of them
will be dropped by switching of this diagram. In a list of identities
the occurrence of the duplicate color structure will be saved in order
to build the full matrix of color factors at the very end.   
\end{enumerate}
\item Now, all products of the remaining color structures can be
evaluated. Loops over the graphs ensure that every diagram will be
connected with all other diagrams in order to fill the full matrix of
color factors. This is done by multiplying the string form of the color
structure for the current diagram with the string of the
conjugated color structure of the second diagram. Accordingly, the
value of this string has to be determined following these steps:
\begin{enumerate}
\item Both strings will be translated into a binary string tree with
the method {\tt String\_Tree::String2Tree()}. 
\item All structure constants $F_{ABC}$ will be replaced with a
combination of generators $T^A_{bc}$ with the method 
{\tt ReplaceF()}. This is done for both strings separately.   
\item Now, the strings will be multiplied by creating a new 
{\tt sknot} of the binary string tree, where the operation is a
multiplication and the string trees hang on the left and right hand
side of the new root {\tt sknot}.
\item All brackets will be expanded ({\tt  String\_Tree::Expand()}),
and the tree is brought into a linear form 
({\tt String\_Tree::Linear()}). Now, it will be sorted with 
{\tt String\_Tree::Sort()}.  
\item At this stage all generators $T^A_{bc}$ will be replaced or
combined to delta functions $D_{ab}$ with {\tt ReplaceT()}. 
\item Now, within the string only constants and delta functions
are left. Again, the string tree is expanded, linearized and the delta
functions will be replaced with the method {\tt ReplaceD()}. 
\item A string tree, where only constants exist, can be easily
evaluated with the method {\tt String\_Tree::Color\_Evaluate()} and the
color factor is now available.
\end{enumerate}  
\item Having filled the matrix of color factors for the irreducible cases,
the full matrix can be determined.  
\item At the end the matrix of color factors will be saved into a file
for later usage. Note, that the user might switch of individual
diagrams by means of this matrix, i.e. \ by setting the corresponding
colour factors to zero.
\end{itemize}
\item {\tt ReplaceF()} replaces all occurrences of the structure
constants $F_{ABC}$ with generators $T^A_{bc}$, see
Tab.~\ref{ReplRule}. A loop over the different steps of replacements
ensures that at the very end no structure constant is left over. At
first the method tries to find combinations of structure constants and
generators, i.e.\ with the same color octet indices and replaces them. This
step is performed with the method\\ {\tt Single\_ReplaceFT()} which is
repeated until no such combination exists any more. Prior to each call of
this routine, the string tree is expanded and linearized with the
methods {\tt String\_Tree::Expand()} and {\tt String\_Tree::Linear()},
respectively. Then, if any structure constants 
are still left, a single structure constant will be replaced
with the method {\tt Single\_ReplaceF()}. During the replacement new
generators appear which can be combined with any remaining structure
constants, i.e.\ the loop starts again.       
\item {\tt Single\_ReplaceFT()} changes a combination of a structure
constant $F_{ABC}$ and a generator $T^A_{bc}$ with one common color octet
index into a combination of generators, see Tab.~\ref{ReplRule}. 
Therefore, an expanded
and linearized list is searched for such combinations. This is done 
recursively in the following way: First of all a multiplication operator
is looked for, where a generator or structure constant is attached at one
side of the {\tt sknot}. Then, all the other factors of this
multiplication series are searched for the appropriate color index. If
a match occurs, the two factors will be replaced by the combination of
generators. Special care has to be taken for the sign of the product,
which depends on the place of the matched color index in the list of
arguments for the structure constant. Since the binary tree is
searched recursively for such a combination, two calls to 
{\tt Single\_ReplaceFT()} at the end ensure this procedure.    
\item {\tt Single\_ReplaceF()} replaces a structure constant $F_{ABC}$
by a combination of generators, see Tab.~\ref{ReplRule}. A simple find
and replace strategy in the binary tree is the underlying idea of this
method. Note that only the first structure constant will be
substituted, since in a global strategy it is faster to search for
combinations of structure constants and generators (see the proceeding
in {\tt ReplaceF()}). 
\item {\tt ReplaceT()} substitutes all the combination of two
generators with a number of delta functions, see Tab.~\ref{ReplRule}.
Since the product of all
generators has to yield a complex number, no generator will be left
after this method. The procedure is quite simple, since we have an
expanded, linearized and sorted binary string tree. Primarily this
means that generators with the same color index are immediate
neighbours. Therefore, only these two neighbours have to be
combined. A further simplification can be achieved, if one of the
matrix indices of the two generators are equal. However, in this way
all generators will be replaced recursively by calling {\tt
ReplaceT()} twice at the end of the method.   
\item {\tt ReplaceD()} applies in an expanded and linearized binary
string tree all delta functions. This means especially that at the
end of this method no delta function should remain (usually this method
is called, when no color matrices exist any more). First of all, a
delta function will be searched, where two different cases can occur:
\begin{enumerate}
\item Both arguments of the delta function are equal. Then this delta
function will be replaced by a factor of $3$ (according to the
underlying group structure). 
\item The arguments could differ. Then, all other delta functions
which are multiplied with the current one, are searched for the first
argument of the delta function which is then replaced by the second
argument. At the end, the current delta function was applied and will
be substituted by a factor of $1$. 
\end{enumerate}  
Two recursive calls to {\tt ReplaceD()} ensure that the whole binary
tree is examined. 
\item {\tt Mij()} returns the appropriate entry of the color matrix.
\end{enumerate}
{\tt coulomb} calculates the Coulomb factor of an amplitude with two
$W$ bosons. This is a first higher--order correction which takes into
account a Coulomb force between the two, slow moving, $W$ bosons:
\begin{enumerate}
\item The {\tt Constructor()} builds the matrix of Coulomb
factors. Note that only products of a graph and a complex conjugate
graph will be considered, where both graphs have two $W$ bosons. After
the matrix was constructed it will be filled with the method 
{\tt Build\_Matrix()}. 
\item {\tt Build\_Matrix()} searches for every graph which has two
$W$ bosons and marks them with the method 
{\tt Single\_Amplitude::Set\_coulomb()}. Then, at each place in the
Coulomb matrix, where two marked diagrams meet, the value $1$ will be
entered.   
\item {\tt Calculate()} evaluates the Coulomb factor for a given point
in the phase space, i.e.\ a list of four-momenta for the incoming and
outgoing particles with Eq.~(\ref{CoulEq}). For the calculation the
momenta of the two $W$ 
bosons are mandatory and will be determined with the help of the given
four--momenta. Now, all ingredients are at hand for the evaluation of
the Coulomb factor. 
\item {\tt Factor()} returns one plus the Coulomb factor multiplied
with the appropriate matrix element of Coulomb factors.
\end{enumerate}
In the long run, this class will be expanded to top--pairs etc..
  
\subsection{The Strings\label{Strings}}

\begin{figure}
\includegraphics[height=13cm]{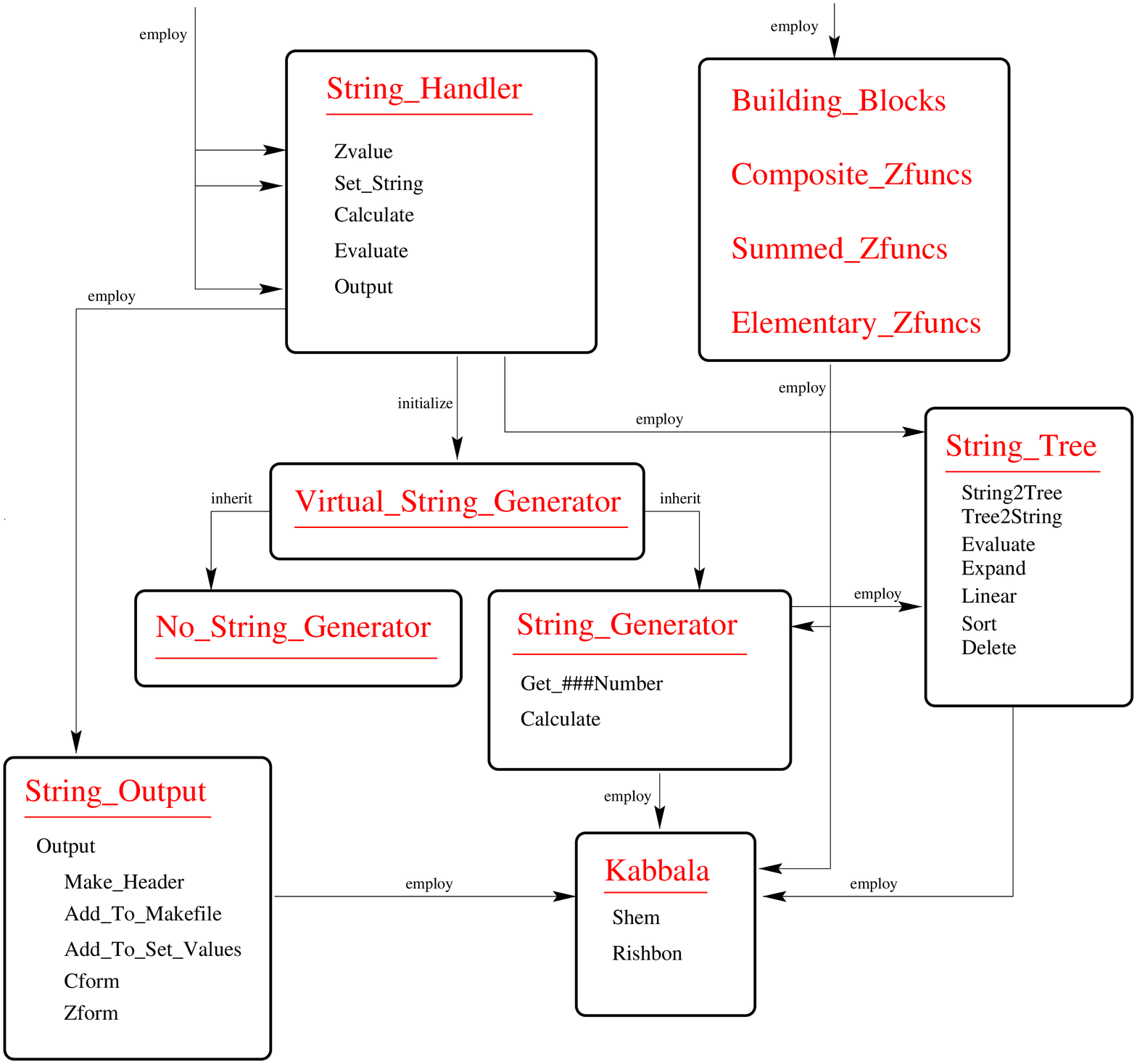}
\caption{\label{StringOrg} The String classes are responsible for writing the
helicity amplitudes into {\tt C++} files.}
\end{figure}

Within this section all basic methods to construct strings and save
them into libraries will be explained. The basic motivation is the
following: 
For every graph and helicity combination one expression
consisting of algebraic operations and $Z$ functions exists. Usually
the value of this part of the amplitude can be calculated directly,
but this has several large disadvantages: 
\begin{enumerate}
\item Every $Z$ function is calculated in the very moment it is
used. Hence some of the $Z$ functions will be calculated more than
once. 
\item The sequence of operations between the different $Z$ functions
has to be determined for every sample of external four-momenta again,
since it could not be stored.
\item A lot of helicity combinations can yield zero which is
usually reflected in a $Z$ function. Nevertheless, all $Z$ functions
belonging to this combination will be calculated because a direct
calculation is not very flexible. 
\end{enumerate}
These problems can be simply cured by the following steps, respectively:
\begin{enumerate}
\item Every $Z$ function will be stored with its value and its list of
arguments. Now, every $Z$ function has to be calculated only once at
the beginning. This task is performed by the {\tt
String\_Generator()}. Accordingly, every $Z$ function which was
calculated in the class {\tt Elementary\_Zfuncs} will now be stored
into a list.
\item The whole expression for the calculation can be casted into a
character string. Accordingly, for every sample of four-momenta only this
string has to be calculated. However, several problems occur during
the generation and calculation of this string:
\begin{itemize}
\item The string has to be built during the evaluation of the
amplitude which results in a quite complicated program code, since every
algebraic operation has to be stored. A simple way out is the usage of
a class called {\tt Kabbala} which allows a simultaneous generation
of a string during the calculation of complex numbers. 
\item The calculation of the string is very time consuming, since this
string has to be interpreted, i.e.\ the algebraic expression has to be
examined and evaluated accordingly. The way out is the translation of
this string into a binary string tree ({\tt String\_Tree}). This tree
has an operation saved at each knot and the $Z$ functions at the
leafs, i.e.\ end points of the tree. Now, during the calculation this
tree has to be followed only and every operation has to be performed
using the values of the left and the right hand branch of the
appropriate knot. This usually saves a lot of time during the
evaluation. However, the by far fastest interpreter of an algebraic
expression is the {\tt C++} compiler itself. Accordingly, all strings
will be written out into {\tt C++} files, translated and linked for
the evaluation. This will be performed within the class 
{\tt String\_Output} and is the optimal way of calculating an
expression within \AME. 
\end{itemize}
\item The last problem belongs to the unnecessary calculation of zero
parts of the amplitude. Again, the translation into a character string can
help to omit these parts. Since all zero $Z$ functions are
well known during the first evaluation of the amplitude, all connected
parts can be deleted within the character string. Of course, having the
string in form of a binary tree relieves this task a lot.   
\end{enumerate}
In our description of the single classes we now start with the 
{\tt String\_Handler} which is responsible for the whole
organization. Then, we proceed with the\\ {\tt String\_Generator}
family which takes care for the proper storage of the $Z$
functions. In order to manipulate a string (for instance to
delete zero parts), a representation of a string, namely a 
{\tt String\_Tree}, was chosen. This form allows an easy modification
and evaluation of the strings. If the strings are at hand, they have
to be saved into a library form, i.e.\ a {\tt C++} file which later
can be translated and linked. The classes {\tt Value} and {\tt
String\_Output} are responsible for this task. Last but not least two
helper classes are introduced, one for the handling of strings ({\tt
MyString}) and one for tracking algebraic expressions ({\tt
Kabbala}). A short description of all these classes can be found in
Tab.~\ref{StringClasses} and their relations are exhibited in
Fig.~\ref{StringOrg}. 
\begin{table}[h] 
\begin{tabular}{l|l} 
Class/Struct & Purpose\\
\hline 
&\\ 
String\_Handler            & Handles all string manipulations.\\
ZXlist                     & This list will be built from the 
			     {\tt String\_Generator}\\ 
                           & and all values necessary for the
                             calculation of a string\\ 
                           & are stored here.\\
Virtual\_String\_Generator & The mother class of all {\tt
                             String\_Generator}s.\\ 
No\_String\_Generator      & Calculating an amplitude without generating\\
                           & any string requires this class.\\
String\_Generator          & Translates all $Z$ functions into strings,\\
                           & stores and calculates them later on.\\
String\_Tree               & Translates an algebraic equation into a
                             binary tree.\\ 
sknot                      & Is a string knot of this binary tree.\\ 
sknotlink                  & Is a list of {\tt sknot}s.\\ 
Values                     & Is the mother class of all connected libraries.\\
String\_Output             & Maintains the output into {\tt C++} files.\\
MyString                   & A class for handling strings, i.e.\ adding
                             or searching\\ 
                           & for strings etc..\\                
Kabbala                    & Strings ({\tt shem}) and complex numbers \\
                           & ({\tt rishpon}) in one term.
\end{tabular}
\caption{\label{StringClasses}A short description of the string classes.}
\end{table}                                                                   

\subsubsection{The Handling of strings}

The {\tt String\_Handler} organizes the whole string generation, the
output of the strings into libraries and their calculation. Therefore
it not only handles the\\ {\tt String\_Generator} for the storage of the
$Z$ function but also saves the character strings which have been
generated during the evaluation of the amplitudes. These strings can
now be translated into a binary tree form. In order to calculate them,
a connection between the leafs or end points of the string and the real
$Z$ functions has to be drawn which is done as well in this class. At
the end an output into {\tt C++} files is performed. 
\begin{enumerate}
\item The {\tt Constructor} of this class initializes the 
{\tt String\_Generator}, where three different cases could occur:
\begin{enumerate} 
\item No string should be generated at all and therefore a 
{\tt No\_String\_Generator} which is a dummy class, will be constructed. 
\item A string should be generated, i.e.\ a regular 
{\tt String\_Generator} is constructed, but no library should be used.  
\item An already produced string library shall be used for the
evaluation. In this case no string will be generated either. However,
if no string library exists, it will be created. 
\end{enumerate}
In the latter two cases an identity, i.e. \ a name, has to be produced
in order to label and identify the appropriate library. 
\item {\tt Get\_Generator()} returns a pointer to the current
{\tt String\_Generator}.
\item {\tt Init()} initializes a matrix of {\tt sknot}s, where every
knot represents the root of a {\tt String\_Tree}, i.e.\ a tree
representation of a string. The different elements are labeled by the
graph number and the appropriate helicity combination. In this way,
all character strings for the different parts of the amplitude can be
stored. 
\item {\tt Set\_String()} performs the translation of a given string
into a binary string tree, deletes all zero parts and stores the
result into the matrix of {\tt sknot}s. For this purpose a {\tt
String\_Tree} object is constructed and filled via the translation method 
{\tt String\_Tree::String2Tree()}. Now, a full binary tree of strings
is available, where every knot belongs to an algebraic operation
($+$,$-$,$*$,$/$) and every end point can be identified with a $Z$
function. However, since all $Z$ functions with the string identity
$Z[0]$ yield zero, these parts can now be deleted with 
{\tt String\_Tree::Delete()}. Note that the current 
{\tt String\_Tree()} is a local object only and a translation
into a global {\tt String\_Tree} object is mandatory which is achieved
via {\tt String\_Tree::Copy()}. The root of this tree can now be
stored into the matrix of {\tt sknot}s, where the place is determined
through the current graph and helicity number.
\item {\tt Complete()} is performed when all character strings are already
stored in the matrix of {\tt sknot}s. Here, all $Z$ functions which
are attached at the end points of the different binary trees will be
connected with the $Z$ functions stored in the 
{\tt String\_Generator}. This step is mandatory, since until now, the
list of $Z$ functions was generated independently of the character string
representing the algebraic expression for the calculation of one part
of the amplitude. In order to calculate the full
algebraic equation with the help of the binary tree, the values of
these $Z$ functions have to be known. On the other hand, all $Z$
functions are stored in the {\tt String\_Generator}. Therefore links
between these different representations have to be drawn. 

Within loops over the number of graphs and helicities the binary
trees are handled one by one. At first, all leafs are collected with the
method\\ {\tt String\_Tree::Get\_End()}, where a list of {\tt sknot}s
will be produced. Now, every leaf is connected with the appropriate $Z$
function via\\ {\tt String\_Generator::Get\_Kabbala()}. At the very end,
all $Z$ functions which do not appear in the strings any more
(i.e.\ they belonged to zero parts of the term) are deleted via 
{\tt Z\_Kill()}. 
\item {\tt Z\_Kill()} simply deletes all $Z$ functions which do not
appear in a string any more. Therefore, a loop over all $Z$ functions is
performed. Every string from all graphs and helicity combinations
is searched for the appropriate function. Of course, since composed
$Z$ functions consist of strings themselves, they have to be searched as
well. At the very end, if the $Z$ function could not be found in any of
the strings, it will be deleted via {\tt String\_Generator::Kill\_Z()}. 
\item {\tt Calculate()} determines the value of all stored $Z$
functions using the method {\tt String\_Generator::Calculate()}. This
is the first step of the evaluation of the whole amplitude.
\item {\tt Zvalue()} calculates the value of the string in the form
of a binary tree for a given combination of graph and helicity
employing the method\\ {\tt String\_Tree::Evaluate()}.
\item {\tt Output()} generates the libraries from the
strings. Therefore, a {\tt String\_Output} object is generated and
the output is performed using\\ {\tt String\_Output::Output()}.
\item {\tt Set\_Values()} chooses the library to be used from a given
identity.   
\item {\tt Is\_String()} returns $1$, if all strings are established. 
\end{enumerate}

\subsubsection{Generating the strings}

The structure {\tt ZXlist} contains all informations about a $Z$
function, i.e.\ all arguments, the value, the type of $Z$ function, a
pointer to a {\tt String\_Tree} for the composed $Z$ functions and an
{\tt on}--switch. Note that we use the label $Z$--function for a variety
of functions which differ by their type, i.e.\ a real $Z$
function, an $X$ function, a $Y$ function, a complex number, 
a scalar product of two four-momenta, a $S$ function, a coupling, a
propagator, a composed $Z$ function and a mass term. Any algebraic
equation for the determination of an amplitude can contain only these
possible pieces. 

The basic structure for all {\tt String\_Generator}s is {\tt
Virtual\_String\_Generator}  which is a purely virtual class. 
Therefore it contains only virtual methods which are overwritten in
the derived classes. Accordingly, these methods will be described in
these classes.  

The {\tt No\_String\_Generator} is a dummy class. Its methods only
return the given complex value and are used to keep the same program
structure for a pure calculation without the generation of any strings.

The {\tt String\_Generator} is used to generate and store all
different types of $Z$ functions. The idea is that every $Z$ function
depending on its list of arguments is stored only once. For a
representation within the algebraic expression in form of a character
string these $Z$ functions have to gain a number as well for later
recognition during the evaluation. Therefore, the main tasks are
comparing a new $Z$ function with all existing ones as well as
creating a new one and storing it in the list of {\tt ZXlist}s in case
it does not yet exist and finally returning an appropriate string
representation including the number of 
the $Z$ function. Accordingly, most methods are concerned with
comparing arguments of $Z$ functions, storing and calculating them:  
\begin{enumerate}
\item The {\tt Constructor()} generates a new list of couplings,
flavours and {\tt ZXlist}s.
\item {\tt Reset()} deletes the contents of all lists. 
\item {\tt \{Z,X,Y\}\_Number()} has the aim to give a well defined
number for any $\{Z,X,Y\}$ function. For this purpose, it compares the
list of arguments and couplings for the appropriate function with all
already stored $\{Z,X,Y\}$ functions. In case a complete match can be
achieved the number of the function is returned, otherwise a $-1$ will be
passed back. In this way a double counting of any function can be
omitted.     
\item {\tt ZX\_Count()} counts the number of all {\tt ZXlist}s.
\item {\tt \{Z,X,E\}\_Count()} counts the number of $\{Z,X,E\}$
functions, where the latter one represents a complex number.
\item {\tt Get\_Cnumber()} compares a given coupling with all other
couplings in the list. If the coupling could be found, the
number within the list is returned, otherwise the new coupling will be
attached to the list of couplings. 
\item {\tt Get\_Fnumber()} works in the same way as {\tt
Get\_Cnumber()}, but for the list of {\tt Flavour}s.
\item {\tt Number()} constructs a {\tt Kabbala} (i.e.\ a string and a
complex number in one class) out of a given complex value and a number
for the $Z$ function. The string part of the {\tt Kabbala} is
generated in the form of ``{\tt Z[Z\_number]}''.
\item {\tt Get\_Znumber()} is the method called from the class
{\tt Elementary\_Zfuncs}. The input is the list of arguments, the list of
couplings and the value of the $Z$ function. First of all, the list of
couplings is translated into a list of numbers which hint on a global
list via {\tt Get\_Cnumber()}. Then, the number of the $Z$ function in
the list of {\tt ZXlist}s is determined with the method 
{\tt Z\_Number()}. If this $Z$ function already exists, the {\tt Kabbala}
value is returned. Otherwise a new {\tt ZXlist} will be
created. Therefore the lists of arguments and couplings as well as the
{\tt Kabbala} value (via {\tt Number}) are filled into a {\tt ZXlist}. 
\item {\tt Get\_Xnumber()} works in the same way like {\tt
Get\_Znumber()} for $X$ functions. 
\item {\tt Get\_Ynumber()} works in the same way like {\tt
Get\_Znumber()}  for $Y$ functions. 
\item {\tt Get\_Enumber()} adds a complex number to the list of 
{\tt ZXlist} (this could be the mass of a vector boson for
instance). Therefore the existing list is searched for this complex
number and a new {\tt ZXlist} is attached in case it could not be found.
\item {\tt Get\_Pnumber()} adds a propagator term to the list of
{\tt ZXlist}s. Here, the complex values and the flavour of the
propagator are compared with the existing list.  
\item {\tt Get\_Massnumber()} adds a mass term to the list of {\tt
ZXlist}s. These terms occur during the transformation of fermion
propagators into spinor products. The complex value of the term as well
as the flavour are compared with the existing list.  
\item {\tt Get\_Snumber()} adds a scalar product of two four-momenta
to the list of {\tt ZXlist}s. The two arguments, i.e.\ the numbers of the
four-momenta, are compared with the existing list.  
\item {\tt Get\_CZnumber()} adds a composed $Z$ function to the list
of {\tt ZXlist}s. These functions occur, when the building blocks
for the calculation of a multiple boson vertex will be stored as an
extra $Z$ function. Therefore only the complex values are compared
with the already existing list. However, since a composed $Z$ function
consists mainly of an algebraic equation in form of a string, this
string has to be stored. In order to speed up the evaluation, the
string is first of all translated into a binary tree. Then the zero
parts are deleted (i.e.\ where a $Z[0]$ function occurs) and the end
points of the tree are connected with the real $Z$ functions. Within
the {\tt ZXlist} only the pointer to the root of this tree is stored
(a similar treatment can be found within the method {\tt
String\_Handler::Complete()}).   
\item {\tt Calculate()} evaluates all complex values for the different
$Z$ functions of the list of {\tt ZXlist}s. Accordingly, the different
methods of the class\\ {\tt Elementary\_Zfuncs} are employed, see
Tab.~\ref{ZXlist_type}.
In case a string library is linked this job is performed by 
{\tt Values::Calculate()}.   
\item {\tt Get\_Kabbala()} returns a pointer to the {\tt Kabbala}
value of a {\tt ZXlist} for a given string representation of a $Z$
function. If the string consists of a complex value only and could not
be found in the current list, a new {\tt ZXlist} is created and
attached to the list.    
\item {\tt Set\_EZ()} sets the pointer to the actual 
{\tt Elementary\_Zfunc} object. 
\item {\tt ZX\_Max\_Number()} returns the maximum number of 
{\tt ZXlist}s. 
\item {\tt Get\_ZXl()} returns a {\tt ZXlist} for a given list number.
\item {\tt Coupl\_Max\_Number()} returns the maximum number of
couplings stored in the list of couplings.
\item {\tt Get\_Coupl()} returns the coupling for a given list number.
\item {\tt Kill\_Z()} switches a {\tt ZXlist} off.
\end{enumerate}

\subsubsection{Translating an equation into a binary string tree}

The structure {\tt sknot} represents one algebraic operation and is a
knot of the tree for a whole algebraic expression. Therefore it
contains a left and a right pointer to the next {\tt sknot}s, a string
and a pointer to a {\tt Kabbala} value for the end points of the tree
and an operation for the knots of the tree.

The structure {\tt sknotlink} is built from a list of {\tt sknot}s
with a fixed number and contains a link to the next {\tt sknotlink}. 
The actual number is stored as well.

The {\tt String\_Tree} is responsible for the translation of an algebraic
expression in form of a character string into a binary tree form. Every
knot of the tree contains one operation and the leafs, i.e.\ end points
correspond to the appropriate variables, i.e.\ for instance $Z$
functions. Consequently, this class not only enables the generation of
this tree but also allows its proper manipulation:
\begin{enumerate}
\item {\tt Reset()} deletes all linked {\tt sknotlink}s and therefore
all {\tt sknot}s in the list. 
\item {\tt newsk()} provides a new {\tt sknot} from the actual 
{\tt sknotlink} which saves a fixed number of {\tt sknot}s, usually
around $200$. If all {\tt sknots} are already used up, a new 
{\tt sknotlink} will be created. In this way a hybrid memory
management is achieved, i.e.\ something between a static and a dynamic
memory allocation model.  
\item {\tt String2Tree()} translates a given algebraic expression
represented by a string into a binary string tree. The string itself
will be examined and recursively cut into pieces in the following
steps:
\begin{enumerate}
\item If the string is empty a zero pointer will be returned.
\item According to the priorities of the different operators ($+$,$*$,
etc.) the summations and subtractions are examined first. Therefore the string
will be searched for global plus and minus signs, where global means
that at the very point, where the sign occurs, the number of opening
and closing brackets must be equal. If a plus sign is found this
way, a new {\tt sknot} will be created with the plus operation. The
whole string will be cut into a left and a right hand part with
respect to the place of the plus sign. Consequently, the left and the right
pointer of the {\tt sknot} are set by recursively calling the method
{\tt String2Tree()} with the left and right hand part of the string. Now, the
new {\tt sknot} will be returned. The procedure for the minus sign is
the same, only the right hand side of the string obtains an overall
minus sign, and both parts are added.
\item Since at this stage all global plus and minus signs are already
treated, only global multiplications and divisions can occur. If a
global multiplication was found, a new {\tt sknot} is created in the
same manner as for the plus sign. The same holds true for global
divisions.   
\item At this stage no global sign should occur any more (expect
leading plus and minus signs). Now, an expression can be enclosed by
brackets and therefore global brackets which open at the first
character of the string and close at the last one are removed and the
examination of the strings continues with point (b).
\item If no global brackets can be removed, a leading sign could
appear, i.e.\ a plus or minus sign at the first character of the
string. In case it is a plus sign it could be simply removed, for a
minus sign a new {\tt sknot} will be built with the left side
pointing to a {\tt sknot} with a '0' string and the right hand side to
the {\tt sknot} which will be created through a recursive call to
{\tt String2Tree()} with the rest of the string as an argument.     
\item At the very end only a term with no internal structure can
remain. This rest string is stored into a new {\tt sknot} which has a
zero as operation. Thus it is marked as a leaf of the tree and
will be returned.
\end{enumerate}
In this way the whole string can be translated into a tree and the
method returns its root. Note that the different knots of the tree
are stored internally in this class.
\item {\tt Color\_Evaluate()} calculates the value of a string
originating from the determination of the color coefficients of the
amplitude. Accordingly, only a distinct number of different strings
could emerge at the end points (i.e.\ these are integer numbers, color
factors like $C_F$ and imaginary units). Therefore they are handled
within this method. The evaluation is performed recursively, i.e.\ the
operation of the  {\tt sknot} is examined and accomplished accordingly
by recursive calls of {\tt Color\_Evaluate()} with the knot of the
left and right hand side, respectively. If the knot is an end point,
the string of this {\tt sknot} is compared with the different
possibilities and the value is returned.
\item {\tt Evaluate()} calculates the complex value of a string in the
same way as\\ {\tt Color\_Evaluate()} with a slightly different
treatment of the end points. Since this method is used for the
evaluation of strings consisting of $Z$ functions, the end points are
connected to the real values of the $Z$ functions which are calculated
elsewhere (i.e.\ in {\tt String\_Generator}). Therefore the value of
every end point is obtained by using the pointer value of the {\tt sknot}.    
\item {\tt Get\_End()} is used for linking the end points of the
tree with the $Z$ functions. Consequently, a list of all end points is
mandatory in order to draw the connection. This list is constructed in
this method recursively.
\item {\tt Tree2String()} translates a binary tree back into a
string. Of course, a tree does not contain any bracket structure and
to reestablish these is the main task of this routine. A ``granny'' 
knot is used in order to avoid an overburden bracket
structure. Now, simple summations do not need an extra bracket, when
the granny option was a plus as well. The output is therefore easy to
read and often much better than the string which originally was
filled into the tree. Note that in its most simplest form the 
{\tt String\_Tree} can be used to simplify the bracket structure of a
string term.    
\item {\tt Tree2Tex()} works in the same way as {\tt Tree2String()},
with the difference that the output string is LaTeX conformable. This
means in particular that the multiplications are replaced by a space
and the division is substituted by LaTeX commands.

\item {\tt Delete()} deletes all parts which are connected with a
given zero string for a binary string tree. This method is used to
drop all zero parts of a helicity combination. Since this can not be
cast into a completely recursive procedure, a loop over calls of {\tt
Single\_Delete()} is performed until all zero parts have been eliminated.
\item {\tt Single\_Delete()} deletes recursively all parts of a tree
which belong to a zero string: 
\begin{enumerate}
\item If the given {\tt sknot} is zero or an end point the method is
terminated with a return.
\item If the left hand knot is an end point and its string equals the
given zero string it will be dropped according to the operation of the
actual knot:
\begin{itemize} 
\item In case it is a multiplication, the actual {\tt sknot} is
set equal to the left {\tt sknot}, i.e.\ the right hand part is deleted. 
\item If the operation is a plus, the actual {\tt sknot} is set to the
right hand side, i.e.\ the left part is dropped.
\item The most difficult case occurs, when the operation is a minus. Now,
the granny knot plays an important role. If no granny exists, i.e.\ the
actual knot is the root, the string of the left knot is set to '0'. In
case the granny operation is a multiplication, the string of the left
knot is set to '0' as well. If it is a plus and the actual knot is on
the left hand side of the granny the string of the left knot is again
set to zero, otherwise the granny operation is set to a minus sign and
the actual {\tt sknot} is set onto the right hand side. Last but not
least if the granny operation is a minus the same steps as in the plus
case will be performed with a slight difference in the changing of the
sign of the granny {\tt sknot} to a plus instead of a minus.
\end{itemize}
\item If the left hand side has not been changed, the right hand side
can be examined. If it is an end point and its string equals the zero
string the following manipulations are done:
\begin{itemize}
\item If the operation of the actual {\tt sknot} is a multiplication,
the {\tt sknot} is set equal to its right hand side, i.e.\ the
left part is dropped.
\item In all other cases the actual {\tt sknot} will be set onto the
{\tt sknot} on the right hand side, if the left string is equal '0',  
otherwise it is set to the {\tt sknot} of the left hand side.
\end{itemize}    
\end{enumerate}
At the very end two recursive calls to {\tt Single\_Delete()} are
performed using the {\tt sknot}s of the left and right hand side as
arguments, respectively.  
\item {\tt Expand()} expands an algebraic equation represented by a
string tree, i.e.\ all brackets will be resolved. Since this is not a
totally recursive method (as {\tt Delete()}), a loop over calls to 
{\tt Single\_Expand()} is performed until every term is expanded.     
\item {\tt Single\_Expand()} performs one single expansion, where two
different cases can be distinguished:
\begin{itemize}
\item The operator of the mother is a multiplication and the left or
right hand {\tt sknot} has a plus or minus sign as operation. This is
the typical bracket structure in the from of $a*(b\pm c)$. Since a
multiplication is a commutative operation the left and right hand side
will be changed, if the plus or minus sign occurs on the left hand
side. Now, only the right hand side has to be regarded. At this
stage, a special case has to be considered, where inside the brackets
only a minus sign emerges ($a*(-b)$). Then, the {\tt sknots} will be
rearranged in order to get the form $-a*b$. In the usual case the
equation will be simply transformed to $a*b\pm a*c$.
\item The second case is connected with the break-up of extra minus
signs, i.e.\ a piece of equation looks like $a\pm(-b)$. Of course, then
the expression will be simplified to gain the form $a\mp b$.     
\end{itemize} 
However, if one of the two cases occur, a switch will be set which
ensures that no more manipulations in this step of 
{\tt Single\_Expand()} can be performed. This is necessary, because
one manipulation changes the structure of the tree in such a drastic
way that the tree has to be reexamined from the very beginning (see
{\tt Expand()}). Two recursive calls to {\tt Single\_Expand()} with
the left and right hand side {\tt sknot} as arguments, respectively
finalize this method.  
\item {\tt Linear()} linearizes the sequence of multiplications,
i.e.\ at the very end all multiplications are lined up. If one draws
brackets around every {\tt sknot} with a times operation the following
transformation will be done $(a*b)*(c*d)\to a*(b*(c*d))$. 
This means that in the first case a {\tt sknot} for multiplication with a
multiplication on both sides will be transformed to a line on the left
hand side with three subsequent multiplications. This transformation
becomes mandatory, if one easily wants to change the sequence of
multiplications or wants to examine the different factors in a
prescribed way. However, the first criterion is that the actual 
{\tt sknot} has a times operation. Now, two different cases can appear:
\begin{itemize}
\item The left and the right hand side have a multiplication
operation, then the transformation is done (by again drawing brackets
around every multiplication) in the form of 
$(a*b)*(c*d)\to c*(d*(a*b))$. Note that the difference to the
example above lies in the arbitrariness of the sequence of the times
operator. 
\item Only the right hand side has a times operator. Since
all multiplications should by construction lined up on the left hand
side, the two sides will be exchanged. In our example this will look
like $(b*c)*a\to a*(b*c)$.     
\end{itemize}
Since this routine is built fully recursive, two calls to {\tt Linear()} with
the typical left and right hand side {\tt sknot}, respectively, as
arguments conclude this method.  
\item The {\tt Sort()} method assumes that an already linearized and
expanded tree (see {\tt Linear()} and {\tt Expand()}, respectively)
exists and sorts the line of multiplications after their strings with
help of a standard bubble sort algorithm. 
This is an interesting option, if one wants to have similar
factors side by side. In one step (the method is of course
recursive) one line of multiplications will be ordered. An outer loop
takes care that the whole line will be perused from the very
beginning again and again as long as one exchange within the line
was performed. If no more changes occur, the line is considered as sorted. 
The inner loop compares step by step two subsequent factors, and
exchanges them if the first factor is smaller (in a string notation)
than the second one. At the very end of the method the classical
two recursive calls to {\tt Sort()} will be realized. 
\item {\tt Copy()} simply fills a given tree represented by its root
{\tt sknot} into the actual tree recursively.
\item {\tt Set\_Root()} sets the root of the actual tree.
\item {\tt Get\_Root()} returns the root of the actual tree.
\end{enumerate}
As part of further work we envision further simplifications of the
strings, like identification of common factors in different amplitudes
and their replacement.

\subsubsection{Casting strings into {\tt C++} files and the string libraries}

For every process which is to be linked to \AME a separate class (and 
{\tt C++} files) is generated. The class name is deduced from the
specific and unique process--id and the class itself is derived from a
purely virtual mother class {\tt Values}. This class has two essential
methods, namely {\tt Calculate()} and {\tt Evaluate()}. The former one
is responsible for the precalculation of the $Z$ functions and the
latter one is used to calculate the value for a specified graph and
helicity combination. All {\tt C++} files as well as the appropriate
header files are saved in a directory which is named after the process
as well. However, all the different files for the processes will be
packed into one library called {\tt Process}. Therefore, the {\tt
Makefile} for the production of this library has to be enhanced by the
different objects from every process in order to make the linkage as
easy as possible. The last step is the inclusion of the calling
sequence for this special process. These statements are added to the
method {\tt String\_Handler::Set\_Values()} which is saved into an
external file called {\tt Set\_Values.C}. Now, the user only has to
recompile the new objects via typing {\tt make install}.    

The {\tt String\_Output} produces the {\tt C++} and header files for
the output into libraries and manipulates the {\tt Makefile} for their
translation and the file {\tt Set\_Values.C} for calling the
libraries from within the program. 
\begin{enumerate}
\item The {\tt Constructor()} generates the different identities for
the process path and the different files. Especially, all plus and
minus signs from the given id will be erased, since these are
protected operations within C++. 
\item The method {\tt Output()} organizes the whole output to all 
{\tt C++} and header files as well as all manipulations. It is the
method which is called from outside (see {\tt String\_Handler}):
\begin{itemize}
\item First of all, a header and a {\tt C++} filename will be
generated with the help of the path and the id strings. 
\item With the method {\tt Is\_File()} a check is performed, if the
header file already exists. In this case the output will be canceled,
since it is assumed that the whole library already exists.
\item Now, the header file, strictly speaking the head of the header
file, will be generated via {\tt Make\_Header()}.
\item With the method {\tt Zform()} the list of $Z$ functions which
have to be precalculated at every step, is created, i.e.\ the method
{\tt Values::Calculate()} will be built. 
\item The calculation of the different helicity combinations for the
different amplitudes is maintained from {\tt Values::Evaluate()}. 
Furthermore, for every amplitude a new method, named {\tt M\#}, 
where $\#$ stands for the graph number, is generated. These {\tt C++}
files will be written out with the help of the method {\tt Cform()}. 
\item Now, all {\tt C++} files are established and the header file can be
finished. Note that during the output of the {\tt C++} files every
new method has to be added to the header file for the
prescription of the class. 
\item At the very end the manipulations to the {\tt Set\_Values.C}
file will be performed via calling {\tt Add\_To\_Set\_Values()}.
\end{itemize} 
\item {\tt Is\_File()} returns $1$, if the given filename already
exists. 
\item {\tt Cform()} produces the {\tt C++} files for the evaluation of
all helicity combinations for all Feynman graphs:
\begin{itemize} 
\item First of all, a new {\tt C++} file will be generated and its
name is added to the list of objects in the {\tt Makefile} with 
{\tt Add\_To\_Makefile()}.  
\item The main routine which will be called from outside, is 
{\tt Values::Evaluate()} and will be produced at this stage. 
From here, the appropriate method for the given graph number will be 
called. Note that for every graph one method will be generated. 
\item Now, a loop over all graphs will be performed in order to
produce a new method for each graph. The only argument for these
methods is the number of the helicity combination. 
\item An equation for each of these will be written out using a loop
over all helicity combinations. First of all each equation will be 
translated from its binary tree form into a single character string. Then,
this string is written out, where special care is taken that each
line has a maximum number of characters by calling {\tt Line\_Form()} 
(this is mandatory, since these string expressions could be quite long)     
\footnote{Since the resulting {\tt C++} file can be rather big, an
additional complication arises due to a maximum line number for
different compiler and machine types, e.g.\ we had some problems
compiling the extra libraries of \AME with a {\tt Digital Unix} compiler. 
Therefore, the lines have to be
counted and the different methods have to be casted into different
{\tt C++} files. This is quite simple, but sometimes, already one 
single method is longer than the allowed maximum line number. Then,
this method has to be partitioned as well.}.
\end{itemize}
At the end a number of {\tt C++} files is generated and recorded in
the list of objects in the {\tt Makefile}. 
\item {\tt Zform()} produces the method {\tt Values::Calculate()}
which is responsible for the precalculation of the $Z$
functions. Therefore, the main task is the translation of the list of
{\tt ZXlist}s from the {\tt String\_Generator} into {\tt C++}
statements. First of all, the name of the {\tt C++} file is added to the
{\tt Makefile} via\\ {\tt Add\_To\_Makefile()}. Then, a loop over all
{\tt ZXlist}s is performed, where the calling sequence for the
calculation of the actual $Z$ function depends on its type, see 
Tab.~\ref{ZXlist_type}. 
\begin{table}[t]
\bc
\begin{tabular}{l|l|l} 
Type & Purpose & Calling Sequence\\ 
\hline
&&\\ 
0&$X$ function     & {\tt Elementary\_Zfuncs::Xcalc()}\\
1&$Z$ function     & {\tt Elementary\_Zfuncs::Zcalc()}\\
2&Constant         & The method tries to reinterpret the constant\\
 &                 & as a coupling or a $1/M^2$ of a vector boson.\\
 &                 & These are the only constants which appear\\
 &                 & during the calculation.\\
3&scalar product   & {\tt Elementary\_Zfuncs::Scalc()}\\
4&$Y$ function     & {\tt Elementary\_Zfuncs::Ycalc()}\\
5&Propagator       & {\tt Elementary\_Zfuncs::Pcalc()}\\
6&Composed function& A composed function will be written out in its\\
 &                 & string form using the method {\tt Line\_Form()}.\\
7&Mass term        & {\tt Elementary\_Zfuncs::Mass\_Term\_Calc()}\\
&&\\ 
\end{tabular}
\caption{\label{ZXlist_type}Calling sequence for the different $Z$
function types.}  
\ec
\end{table}                                                                   
Note that in the same way as during {\tt Cform()} additional care has
to be taken for the maximum number of lines. Therefore, new files will
be opened if necessary.  
\item {\tt Make\_Header()} produces the head of the header which
means that the class definition and the standard methods will be
enlisted. 
\item {\tt Line\_Form()} writes a given string into a file. Special
care is taken to a maximum number of line characters (usually around
$70$). If the maximum is reached, a line breaking is performed after
the next operation ($+$,$-$,$*$). 
\item {\tt Add\_To\_Set\_Values()} adds to the {\tt Set\_Values.C}
file the call to the newly generated libraries. Therefore, an {\tt include}
for the appropriate header file and a calling sequence within the method
{\tt String\_Handler::Set\_Values()} is subjoined.
\item {\tt Add\_To\_Makefile()} enters the name of a file into the
list of objects for a given Makefile. First of all, the given file
will be searched ({\tt Search()}) and in case it already exists, the
method terminates. Otherwise, the name of the file will be added at two
different places, namely for {\tt libProcess\_la\_SOURCES} and 
{\tt libProcess\_la\_OBJECTS}, since these object files should belong to
the library {\tt Process}. If the key word is found, the file name
will be included in the next line. Note that for this purpose a temporary
file will be used which is copied onto the appropriate {\tt Makefile}
at the very end ({\tt Copy()}).
\item {\tt Copy()} makes a copy of one file. 
\item {\tt Search()} searches for a given string in a file. 
\end{enumerate}

\subsubsection{Strings and Kabbala}

{\tt MyString} is the implementation of a typical string class
extended by some methods compared to the standard template libraries
(STL):  
\begin{enumerate}
\item The usual standard {\tt Constructor()} initializes a zero
string. A copy constructor version is available as well.
\item {\tt remove()} removes a number of characters out of the string
for a given position and the length of the piece to be deleted.  
\item {\tt c\_str()} returns a standard {\tt C} character string.
\item {\tt length()} returns the length of the string.
\item {\tt substr(long int)} returns a substring beginning at the
given position until the end of the string.
\item {\tt substr(long int,long int)} returns a substring beginning at
the given position and ending at the position plus the given length of
the substring.
\item {\tt find()} returns the position of a given substring within
the string. If no representation could be found, a $-1$ is returned. 
\item {\tt insert()} inserts a substring at the given position.
\item {\tt Convert(double)} converts a double value into a string
using {\tt Convert(int)}.
\item {\tt Convert(int)} converts an integer value into a string.
\item The {\tt operators $=,+=$,[],$+,==,!=$,<<} do exactly what one might
expect. Therefore a special description will be discarded at this point.
\end{enumerate}
The class {\tt Kabbala} is a representation for a string and a complex
number in one class. The name of the class originates from the kabbalistic
system which attributes to every character in the hebrew alphabet a certain
number and tries to interpret the resulting number of words. In our
case this class has the big advantage that one can calculate
with {\tt Kabbala}s instead of complex numbers and not only gains the
numerical result, but also the string representation of the whole
calculation. This is extensively used during the generation of the
string libraries. Technically, this class provides the user with all
possible operations of a complex number in the same syntax, but is
taking care for the correct representation in a string form with all
possible preferences of the operators and the appropriate use of
brackets. Therefore it can be easily used instead of complex numbers. 
Two extra methods {\tt Value()} and {\tt String()} return the complex
number or the string representation, respectively.

\subsection{The Phase Space\label{PhaseSpace}}

\begin{figure}
\includegraphics[height=15cm,angle=90]{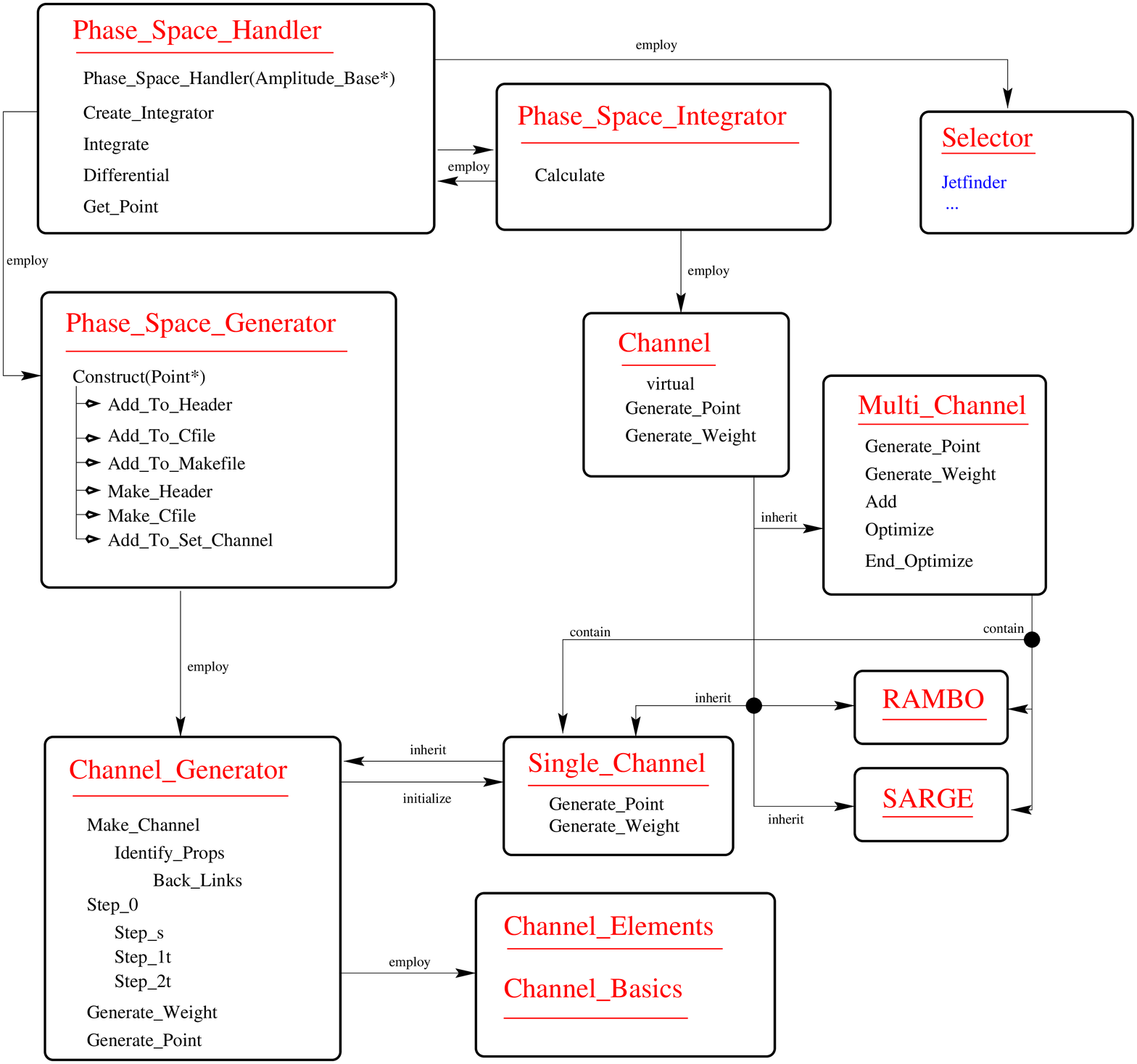}
\caption{\label{PSClassesFig}
The evaluation and integration of the phase space.}
\end{figure}

In this section we will describe the classes and methods used to
generate the various channels within \AME and to produce suitable sets 
of phase space points for the integration. The various classes are
summarized in Tab.~\ref{PSClasses}, their connections and
interrelations are depicted in  Fig.~\ref{PSClassesFig}.
\begin{table}[h] 
\bc 
\begin{tabular}{l|l} 
Class/Struct & Purpose\\ 
\hline
&\\ 
Phase\_Space\_Handler    & Handles the generation of integrators and\\ 
                         & the integration over phase space.\\     
Phase\_Space\_Integrator & Integrates the phase space.\\  
Phase\_Space\_Generator  & Generates the different channels for\\
                         & multi--channel integration.\\   
\hline
Channel\_Generator       & Generates one channel.\\    
Channel\_Basics          & Basic functions for channels.\\       
Channel\_Elements        & Contains all structural elements of channels,\\
                         & e.g. propagators.\\   
\hline
Selector                 & Is the mother class of all selectors\\
	                 & which perform cuts in the phase space.\\
Single\_Selector         & One single selector.\\
All\_Selector            & A list of selectors\\
JetFinder                & A selector which cuts via jet-measures.\\               
\hline
Channel                  & Is the mother of all channels.\\ 
Single\_Channel          & Is one single channel.\\
Multi\_Channel           & Is a list of channels and \\
	                 & manages the optimization.\\
Rambo                    & Is one single channel (flat distribution).\\
Sarge                    & Is one single channel (antenna distribution).\\
\end{tabular}
\caption{\label{PSClasses} Classes used for the generation of
integration channels and the phase space integration.}
\ec
\end{table}                                                                   
\subsubsection{Organization : The {\tt Phase\_Space\_Handler}}
The class {\tt Phase\_Space\_Handler} organizes both the 
generation of integration channels and the integration by steering the 
corresponding classes {\tt Phase\_Space\_Generator} and 
{\tt Phase\_Space\_Integrator}, respectively. The methods available
within the\\ {\tt Phase\_Space\_Handler} are:
\begin{enumerate}
\item The {\tt Constructor} reads in the parameters relevant for the 
      generation of phase space points via {\tt Read\_Parameter()}.
      Then it initializes all possible selectors for phase space
      points. The only selector implemented in the
      moment is a jet measure, i.e.\ the requirement that the final
      state particles form distinguishable jets.
\item {\tt Read\_Parameter()} reads in parameters relevant for the
      phase space integration. Currently they consist of the 
      jet scheme plus its $y_{\rm cut}$, the relative error, at which
      the evaluation of cross section terminates and the strategy used
      for this calculation, i.e.\ {\tt Rambo} only, or {\tt Rambo +
      Sarge}, or multi--channel method with generic, process dependent
      channels in addition to {\tt Rambo + Sarge}.
\item {\tt Create\_Integrator()} organizes the number of channels
      according to the choices given in the parameters.
      In case, the user choses to use multi--channel methods with
      process dependent channels, they are constructed amplitude--wise
      via the method {\tt Construct()} of the class 
      {\tt Phase\_Space\_Generator}. For this purpose, the
      list of {\tt Point}s of the corresponding {\tt Amplitude} and the
      process--identifying string are handed over to 
      {\tt Phase\_Space\_Generator::Construct()}
      as well as the external flavours and the amplitude--number. 
\item {\tt Drop\_Channel()} in general is used to eliminate channels
      generated in\\ {\tt Create\_Integrator()} before the actual 
      calculation starts. Channels are drop\-ped, if
      \begin{itemize}
      \item they yield the same result under permutation of -- at
            least -- two particle momenta. The idea is the following:
            Any channel is correlated to one amplitude and produces a
            set of momenta accordingly. Thus, any two amplitudes
            differing by just permuting two identical particles
            give rise to a new channel which basically yields the
            same result as the original one. To avoid double counting
            and to keep the number of channels small, one of the two
            channels should be removed.\newline
            The difference of permutations of momenta and particles
            is tested via {\tt Compare()}.
      \item they do not have the maximum number of resonating
            propagators within the phase space allowed. So, for every
            channel, the number of potentially resonating propagators
            is counted, non--maximal number of such propagators
            lead to the channel being switched off.
      \end{itemize}
\item {\tt Compare()} checks, if two sets of outgoing momenta and
      particles are just permutations of each other.
\item {\tt Integrate()} calls a suitable method of the 
      {\tt Phase\_Space\_Integrator} to perform the actual integration. 
      Depending on the number of incoming particles, this is either
      {\tt Calculate()} or {\tt Calculate\_Decay()}. 
\item {\tt Differential()} creates one phase space point, i.e.\ one set
      of outgoing momenta fulfilling the criteria given by the
      selectors and returns the corresponding weight. 
\item {\tt Get\_Point()} returns one phase space point,
      i.e.\ one set of four--momenta for the incoming and outgoing
      particles. 
\end{enumerate}

\subsubsection{Channels}

Since all of the integration within \AME is performed by invoking
channels, we start the presentation of the various classes utilized in 
the construction of the phase space integrator and in the calculation
by presenting the virtual class {\tt Channel}. It consists of quite a
number of virtual methods, used either to generate a phase space
point, i.e.\ a set of four-momenta for the outgoing particles, and the
weight associated with it ({\tt Generate\_Point()} and 
{\tt Generate\_Weight()}), or for the actual phase space integration
({\tt Add\_Point()},{\tt Variance()}), or used for the running of the
multi--channel method. 

\begin{enumerate}
\item {\tt Generate\_Weight()} is a method to generate the phase space 
      weight of a channel for a given set of momenta. 
\item {\tt Generate\_Point()} is a variety of methods generating phase
      space points, i.e.\ sets of momenta.
\item {\tt Add()} adds a channel to the multi--channel integrator.
\item {\tt Drop\_Channel()} drops a channel of the ensemble employed
      during the multi--channel evaluation of the phase space integral.
\item {\tt Optimize()} optimizes the a priori weights $\alpha_i$ of
      the multi--channel integrator.
\item {\tt End\_Optimize()} performs the last optimization step, selecting the
      set of a priori weights yielding the smallest overall variance in
      all optimization steps so far.
\item {\tt Name()} returns the name of the channel.
\item {\tt Set\_Name()} sets the name of the channel.
\item {\tt Add\_Point()} adds the result for a phase space point to the 
       overall result and increments the total number of points by
       one.
\item {\tt Variance()} returns the variance accumulated in the channel.
\item {\tt Reset()} resets the result etc. accumulated so far.
\item {\tt Reset\_Opt()} resets all internal results, variances
      etc. used for the optimization procedure of the multi--channel
      strategy. 
\item {\tt Number()} returns the number of channels.
\item {\tt Count\_Resonances()} counts and returns the maximal number of
      potentially resonant propagators within one channel.
\end{enumerate}
Directly derived from it is the class {\tt RAMBO}, generating phase
space points according to a flat distribution with help of the
following methods in addition to the ones outlined above:

\begin{enumerate}
\item The {\tt Constructor} initializes an array of helper functions
      used for the determination of the phase space weight and
      calculates them. Furthermore, arrays for outgoing momenta squared,
      their masses and energies are initialized.
\item {\tt Massive\_Point()} rescales massless vectors in order to
      bring them on their mass shells.
\item {\tt Massive\_Weight()} calculates the weight related to the
      rescaling above.
\end{enumerate}
Similarly, the other default channel of {\tt AMEGIC++}, 
{\tt SARGE} is derived from {\tt Channel} with the additional methods: 
\begin{enumerate}
\item {\tt Qcd\_Antenna()} actually generates a phase space point
      according to the QCD antenna pattern, once after some first and
      last -- massless -- trial vector has been generated. For this
      purpose {\tt Basic\_Antenna()} is employed. 
\item {\tt Basic\_Antenna()} generates one antenna, i.e.\
      one four vector emitted by a dipole.
\item {\tt Perm\_P()} generates a random permutation of a number of
      integers with help of the P--algorithm in 
      \cite{Knuth:TheArtofComputerProgramming}.
\item {\tt Polytope()} produces a uniform random distribution $\xi_i$
      inside a polytope with $|\xi_k|<1$, $|\xi_k-\xi_l|<1$ along
      the lines of \cite{vanHameren:2000rm}.
\end{enumerate}
Last but not least, the integration by the multi--channel method is
realized with help of the -- again derived -- class 
{\tt Multi\_Channel}. The methods employed here have already
been described above. However, some of the algorithms change a little
bit when having a multi--channel instead of just a single channel.
These are:
\begin{enumerate}
\item In {\tt Generate\_Point()} one channel $i$ is selected randomly
      according to the a priori weights $\alpha_i$ to generate the
      point with its corresponding method.
\item The evaluation of the multi--channel weight $w$ for a phase space 
      point is then performed in {\tt Generate\_Weight()}. It is given
      in terms of the individual weights $w_i$ as 
      $1/w = \sum\alpha_i/w_i$.
\item In {\tt Add\_Point()} the accumulative result, the result
      squared and the weight per channel are increased and the
      number of generated points is incremented.
\end{enumerate}

\subsubsection{Generating channels}
The generation of the individual channels is steered by the class
\\{\tt Phase\_Space\_Generator} and actually performed by appropriate
methods of the class {\tt Channel\_Generator}. The basic idea is to
provide for every occurring amplitude one channel which has the
advantage that all different peaking structures of a matrix element
will be covered. The generated channels are then saved into {\tt C++}
files and managed from a newly created class. This object has a
similar name like the one of the process and is derived from the class
{\tt Single\_Channel}. Consequently, the three different methods 
{\tt Generate\_Point()}, {\tt Generate\_Weight()} and 
{\tt Count\_Resonance()} have to decide about the current channel and
call an appropriate method with the name {\tt C\#\_Momenta()}, 
{\tt C\#\_Weight()} or {\tt C\#\_Resonances()} (\# is the number
of the channel) accordingly. Note that these methods will be created
within the class {\tt Channel\_Generator}. However, we are going
to start with the presentation of the {\tt Phase\_Space\_Generator},
which is responsible for generating one new channel. It mainly
consists of methods to initialize and manipulate files: 
\begin{enumerate}
\item {\tt Construct()} decides, if a channel belonging to the given
      amplitude has already been realized or should be
      created. However, the strategy is the following: 
      \begin{itemize}
      \item The full name of the channel is constructed, including its
            path. Similarly, names for the header and the {\tt C++} file
            are generated. 
      \item The methods {\tt Add\_To\_Header()} and 
	    {\tt Add\_To\_Cfile()} test first of all, if an
            appropriate header or steering {\tt C++} file already exists. If
            this is the case, these files will be searched for the
            current channel name, if it does not exist, the
            appropriate calling sequence and representation in the
            header file will be added. 
      \item Now, two different cases can emerge. If the channel was
            already generated, a new {\tt Channel} type object will be
            created and initialized with the method 
	    {\tt Set\_Channel()}. Then, the channel is ready to use
            and will be returned.  
      \item In the second case, a new channel has to be
            generated. Accordingly, a new object 
	    {\tt Channel\_Generator} is produced and the channel is
            created via {\tt Channel\_Generator::Make\_Channel()}.
      \item Now, the new {\tt C++} files can be attached to the list
            of objects in the appropriate {\tt Makefile} via 
	    {\tt Add\_To\_Makefile()}.
      \item If no header file or steering {\tt C++} file exists, the
            methods {\tt Make\_Header()} and {\tt Make\_Cfile()}
            generate them.
      \item At the end, the calling sequence for the newly generated
            channel will be attached to the file {\tt Set\_Channel.C} with 
	    the method {\tt Add\_To\_Set\_Channel()}.   	
      \end{itemize}
\item {\tt Add\_To\_Header()} adds the three methods 
      {\tt C\#\_Momenta()}, {\tt C\#\_Weight()}, and {\tt
      C\#\_Resonances()},  where {\tt \#} stands for the number of
      the current channel, to a header file.
      This manipulation is only performed (by copying the old file to
      a temporary one, adding lines there and copying back) after it
      has been checked that the header file exist at all, and that these
      methods have not been implemented yet.
\item {\tt Add\_To\_Cfile()} similarly adds the calling sequence for
      these methods to a {\tt C++} file.
\item {\tt Make\_Header()} constructs a new header file for this
      process, where all channels will be stored. Accordingly, all
      standard methods will be attached.
\item {\tt Make\_Cfile()} generates all steering routines for this
      process which call the different channels appropriately.
\item {\tt Add\_To\_Makefile()} adds a newly generated {\tt C++} file
      to the list of objects in the corresponding {\tt Makefile}.
\item {\tt Add\_To\_Set\_Channel()} creates the calling sequence for
      the newly generated library in the file {\tt Set\_Channel.C}.
\item {\tt Is\_File()} checks, whether a specific file already exists.
\item {\tt Search()} checks, whether a specific string is within a file. 
\item {\tt Copy()} copies one file to another. 
\end{enumerate}
Having set up the file system the actual generation of one 
channel is performed using methods of the class 
{\tt Channel\_Generator} which is derived from 
{\tt Single\_Channel}. All three methods for handling a channel,
i.e.\ one for generating the four-momenta of the outgoing particles, one
for calculating the appropriate weight for a given sample of
four-momenta and one for specifying the resonant propagators will be
created in this class. Accordingly, in every method {\tt C++} commands
and calling sequences are written out into a {\tt C++} file resulting
in the appropriate methods. Finally these methods look like the
example in Appendix \ref{SampleCode}.   
\begin{enumerate}
\item The {\tt Constructor} initializes the individual channel by
      setting the number of external particles, incoming and outgoing.
      It copies the point list and then calls {\tt Identify\_Props()}
      in order to mark all $t$--channel particles. 
\item {\tt Identify\_Props()}: Starting from the first point (which
      carries the first incoming particle) the points of the list
      are connected recursively to their previous ones via 
      {\tt Back\_Links()}, until, finally the other point related to an
      incoming particle with the flag ${\tt b}=-1$ is found. Starting from this
      endpoint, the backward links are followed and intermediate lines 
      which are considered as $t$--channel propagators, are counted
      and marked on the way. 
\item {\tt Back\_Links()} recursively establishes backward links 
      {\tt prev} between the {\tt left} and {\tt right} offsprings
      of a point and the point itself. If a final point, i.e.\ a point
      having no left and right links has ${\tt b}=-1$, i.e.\ if it is 
      related to an incoming particle, this point is stored as
      the end point of the incoming line.
\item {\tt Make\_Channel()} maintains the generation of all three
      methods. Therefore it creates the {\tt C++} file and fills in
      the method named {\tt C\#\_Momenta()}, {\tt C\#\_Weight()} and 
      {\tt C\#\_Resonances()}. After the first two a call to 
      {\tt Step\_0()} with accordingly two different options generates
      the body of the functions. The last method is simply filled with
      a list of all resonating propagators.
\item {\tt Step\_0()} calls either {\tt Step\_s()}, {\tt Step\_1t()},
      or {\tt Step\_2t()}, depending on whether zero, one, or two
      $t$--channel propagators have been found. So, to some extent
      {\tt Step\_0()} decides about the basic topology of the
      channel. Note that two different modes are available for the
      creation of a method, i.e.\ generating four-momenta and 
      calculating the appropriate weight. In case, there is no
      $t$--channel propagator, $s$ is already set as the square of the
      sum of incoming momenta.\\ 
      
      The scheme to distinguish between various vectors and their
      squares within the channels, is to label them as ${\tt p12}$
      and ${\tt s12}$, respectively, where {\tt 1} and {\tt 2} are the 
      numbers of the external particles connected to the vectors.
      These numbers are determined using {\tt Linked\_Masses()}.
\item {\tt Step\_s()} creates the complete decay sequence of an
      $s$--channel propagator. Consequently, this method is built up
      recursively and one recursion step consists of the decay of one
      propagator:  
      \begin{itemize}
      \item If the point under consideration is an outgoing particle,
            nothing will be done.
      \item By calling {\tt Linked\_Masses()} for both the {\tt left}
            and the {\tt right} leg, the indices for their vectors and
            masses are determined. The names for the momenta are
            constructed accordingly.
      \item With the help of {\tt Generate\_Masses()} the 
            invariant masses squared for both the {\tt left} and the
            {\tt right} leg are determined. Note that for propagators
            these masses have to be generated. 
      \item Then the current propagator decays isotropically into the
            corresponding {\tt left} and {\tt right} vectors, an
            appropriate calling sequence will be attached to the 
	    {\tt C++} file.  
      \end{itemize}
      Finally, {\tt Step\_s()} is called for both the {\tt left} and
      the {\tt right} leg. 
\item {\tt Step\_1t()} is the root for a topology with one
      $t$--channel propagator. Similarly to {\tt Step\_s()} the
      masses squared for the two outgoing legs are determined calling
      {\tt Generate\_Masses()}. Then their momenta are fixed
      by invoking the corresponding {\tt T\_Channel\_X} building
      blocks, where {\tt X} is replaced by {\tt Weight} or {\tt
      Momenta} accordingly. 
      Finally, the two outgoing legs can be treated as $s$--channel
      propagators by calling {\tt Step\_s()}. Strictly speaking this
      method is quite similar to {\tt Step\_s()} but the isotropic
      decay replaced by a $t$--channel decay.
\item {\tt Step\_2t()} is the key stone for a topology with two
      $t$--channel propagators. The only differences to  
      {\tt Step\_1t()} are related to the fact that there are three
      instead of two outgoing particles, and that there are different
      kinematical regions for the two $t$--channel propagators,
      depending on whether their particles are massless or not.
      This last fact is reflected in invoking one out of a variety 
      of strategies. At this point it should be noted, however, that
      at the present stage the only strategy implemented is for
      two massive $t$--channel propagators fusing into a massive or
      massless state and the two other outgoing particles being
      massless. Obviously, further refinements here are left to
      further work. 
\item {\tt Linked\_Masses()} generates a string of numbers for a given
      {\tt Point} which consists of all particles attached to the same
      branch. Starting from this {\tt Point}, all {\tt left} and {\tt
      right} links are followed recursively and end points
      are added to the corresponding string.
\item {\tt Generate\_Masses()} basically produces masses squared for
      a list of propagators along the following algorithm for each:
      \begin{itemize}
      \item By using {\tt Linked\_Masses()} names for the squared
            masses of each point are generated. If the point
            corresponds to an outgoing particle, the mass squared is
            set directly, otherwise, its minimal value is given by
            the square root of the sum of squares of the outgoing
            particles connected to it,
            \bea
            s_{\rm min}^{ij\dots} = \sqrt{m_i^2+m_j^2+\dots}\,.
            \eea
      \item Now a loop over all points not treated so far starts.
            \begin{itemize}
            \item The most resonating propagator among the points
                  left is selected. The contribution of each single
                  propagator is estimated via $1/(M_f \Gamma_f)^2$, where
                  $M_f$ and $\Gamma_f$ are the mass and width of the
                  propagating flavour. 
            \item The maximal $s_{\rm max}^{ij\dots}$ for this propagator 
                  is evaluated which is the available $s$ minus all other 
                  $s$ of propagators already dealt with and minus all 
                  minimal $s_{\rm min}$ of the so far untreated
                  propagators.
            \item Now the actual $s$ of this
                  propagator is chosen according to either a
                  Breit--Wigner or a simple pole distribution for a
                  massive or massless particle, respectively.                  
            \end{itemize}
      \end{itemize}
\item {\tt Generate\_Weight()} and {\tt Generate\_Point()} are dummy
      methods, i.e.\ they are only used, when new {\tt Channel}s have
      been created and not linked properly. Then these methods give an
      error message. 
\item {\tt Get\_Pointlist()} returns the list of {\tt Point}s.
\item {\tt Init\_T()} resets all $t$--flags in a {\tt Point} list.
\end{enumerate}

\subsubsection{Integration}
The phase space integration, i.e.\ the sampling over Monte Carlo
generated sets of four-momenta for the outgoing particles is organized 
by the {\tt Phase\_Space\_Integrator} with help of the following 
methods

\begin{enumerate}
\item Within the {\tt Constructor()}, the number of optimization
      steps, the number of phase space points per step as well as a
      maximum number of phase space points to be generated are set.  
\item {\tt Calculate()} actually performs the calculation of a 
      $2\to n$ cross section depending on a jet measure.
      Then the channel responsible for the integration is reset and
      during a loop points, or better the values related to
      them are added via {\tt Channel::Add\_Point()}. The necessary
      set of four-momenta as well as the corresponding value of the
      Feynman amplitude are obtained via the method
      {\tt Differential()} of the {\tt Phase\_Space\_Handler}.
      Note that due to the structure outlined above, the channel used
      can be a multi--channel or just one channel.
\item Up to some different value for the incoming flux, the method 
      {\tt Calculate\_Decay()} is fairly similar and calculates
      $1\to n$ decay widths.
\end{enumerate}

\subsubsection{Building blocks for the channels, selectors}

The basic building blocks for the construction of specific channels
have been listed already in Tab.~\ref{PSBB}. The corresponding
methods are organized in the class \\{\tt Channel\_Elements}
(where {\tt X} stands for {\tt Momenta} or {\tt Weight}): 

\begin{enumerate}
\item {\tt Isotropic2\_X()} generates momenta or the weight for an
      isotropic two--body decay.
\item {\tt Isotropic3\_X()} generates momenta or the weight for an
      isotropic three--body decay.
\item {\tt Anisotropic2\_X()} generates momenta or the weight for an
      anisotropic two--body decay.
\item {\tt Massless\_Prop\_X()} generates a mass squared or the weight
      for a massless propagator, i.e.\ according to a simple pole 
      distribution.
\item {\tt Massive\_Prop\_X()} generates a mass squared or the weight
      for a massive propagator, i.e.\ according to a Breit--Wigner 
      distribution.
\item {\tt T\_Channel\_X()} generates momenta or the weight for a
      $t$--channel propagator.
\end{enumerate}
Within these building blocks, some elementary functions are widely
used, they are organized in the class {\tt Channel\_Basics}:

\begin{enumerate}
\item {\tt rotat()} is used to set up $3\times 3$ matrices for 
      spatial rotations or to rotate a vector with such a matrix
      already set up.
\item {\tt boost()} is used to define a boost along an arbitrary axis
      or to transform a vector with such a boost.
\item {\tt sqlam(a,b,c)} returns $\sqrt{((a-b-c)^2-4bc)}/a$.
\item {\tt tj()} generates a number $s$ according to a simple pole
      $1/(a\pm s)^\eta$ with the corresponding arguments passed in the 
      call.
\item {\tt hj()} yields the normalization for {\tt tj()}.
\item {\tt tj1()} specializes {\tt tj()} for the case of a pole in the
      form of $1/(a-s)^\eta$.
\item {\tt hj1()} yields the weight for {\tt tj1()}.
\item {\tt Pseudo\_Angle\_Cut()} calculates and returns an angular cut
      for the $t$--channel methods {\tt T\_Channel\_X()} above. So it basically
      constraints the $t$ in the propagator and avoids potential
      singularities which might occur for massless particles.
\end{enumerate}
To impose additional cuts on the phase space, the virtual 
class {\tt Selector} is used. It consists of one method only, 
{\tt Trigger()} yielding a 0 or a 1 depending, on whether the cut was
passed or not. This is then used in the sampling to decide, of whether 
a specific point should be added to the result, or whether a 0 result
should be added, in case the point generated failed to pass the cuts.
 
Derived from this class is a class {\tt All\_Selector} which might be
used for a non--trivial combination of constraints.

The only selector actually implemented so far is a cut on jets which
is passed if all outgoing particles form different jets according to
some jet--measure. This selector is organized in the class
{\tt Jet\_Finder} and consists -- in addition to the trigger -- of the
methods: 
\begin{enumerate}
\item {\tt Constructor()} initializes the jet finder, i.e.\ the
      clustering scheme, the number of vectors etc..
\item {\tt y\_jettest()} determines the minimal jet--measure 
      $y_{\rm min}$ for a number of four-momenta with {\tt ymin()}. 
\item {\tt ymin()} gives the minimal jet--measure for the two
      four-momenta out of a set, having the smallest $y_{ij}$
      (determined by calling {\tt jet()}) according to the scheme
      selected.  
\item {\tt durham()} returns the argument in the Durham--algorithm to
      be compared with $y_{\rm cut}E_{\rm c.m.}^2$.
\item {\tt jade()} returns the argument in the Jade--algorithm to
      be compared with $y_{\rm cut}E_{\rm c.m.}^2$.
\item {\tt geneva()} returns the argument in the Geneva--algorithm to
      be compared with $y_{\rm cut}E_{\rm c.m.}^2$.
\item {\tt jet()} is the wrapper of the three methods above. Depending 
      on the jet--scheme selected, for two vectors the argument of the 
      methods above is returned. 
\item {\tt recomb()} recombines two specific vectors out of a set
      according to the $E$--scheme.
\end{enumerate}

\subsection{Parameters and Switches\label{ParaSwitch}}

\begin{table}[h] 
\bc 
\begin{tabular}{l|l} 
Class/Struct & Purpose\\ 
\hline
&\\ 
Switch         & Can be ON or OFF.\\  
Model\_Type    & Gives the type of the model.\\  
Output         & Gives the output level.\\
Data\_Pointer  & Is a purely virtual class for all {\tt Data}.\\ 
Data           & Is a template for reading in data.\\
Run\_Parameter & Steers the reading in of parameters and switches.\\  
&\\ 
\end{tabular}
\caption{\label{ParSWTab}A short description of the classes connected with the
handling of parameters and switches.} 
\ec
\end{table}                                                                   

Within the different files for the steering of the program via
parameters and switches, a number of different data types can
occur. Not only the usual {\tt C++} types {\tt int}, {\tt double} and
{\tt string} can be used during the input, but special types were
created. Among them, the classes {\tt Switch} and {\tt Model\_Type}
are the most important. The first one has the two settings {\tt On}
and {\tt Off} and is the standard type for all simple switches. With the
second one, the model can be specified, where {\tt pure QCD}, 
{\tt QCD}, {\tt EW} and {\tt SM} are the possible options for a pure
QCD model, a QCD model with the particles $e^+$, $e^-$ and all
their interactions with the QCD particles added, a pure
electroweak sector and the whole Standard Model. All
these different types have to be handled in one list of parameters and
switches. Therefore, an abstract and purely virtual class has to
be constructed, from which all other {\tt Data} types can be
derived, i.e.\ a {\tt Data\_Pointer}. Now, an abstract template class,
{\tt Data}, is used for all different data types which play the role
of a template parameter:
\begin{enumerate}
\item{\tt Set\_Name()} is used to set the name of the variable.
\item{\tt Get\_Name()} returns the name of the variable.
\item{\tt Get\_Value()} returns the value of the variable.
\item{\tt Set\_Value\_Direct()} sets the value of the variable
directly. 
\item{\tt Set\_Value()} converts a given string into the current
variable type. This template method is specified for the types 
{\tt Switch}, {\tt Model\_Type} and {\tt string}. 
\end{enumerate}
The class {\tt Run\_Parameter()} includes the main routines for
reading in parameter files. Supplementary to this, the class reads in
the parameter file {\tt Run.dat} and provides the program with all
parameters and switches necessary for one run. Note that the level of
output will be steered as well, one can choose between silent, normal
and noisy output.  
\begin{enumerate}
\item {\tt Init()} defines all parameters and switches (i.e.\ type and
name) which can be read in from the file {\tt Run.dat}. Accordingly,
a list of {\tt Data\_Pointer}s is built. Then, all the
data will be filled with the method {\tt Read()}. 
\item {\tt Read()} reads in a list of parameters and switches from a
given data file. This list includes all possible data, where every
variable has its own string name. Now, the data file will be searched
line by line for every given string name. If a name could be found,
the  pertaining value is filled into the list accordingly. 
\item {\tt Shorten()} deletes all initial and final spaces in a string.
\end{enumerate}
The rest of the methods are used to set or return the different
parameters and switches already read in from the file {\tt Run.dat}: 
\begin{enumerate}
\item {\tt CMS\_E()} returns the center of mass system energy. 
\item {\tt Set\_CMS\_E()} is used, to set the CM-energy.
\item {\tt Model\_File()} returns the name of the model file (in most
cases {\tt Const.dat}).
\item {\tt Model()} returns the {\tt Model\_Type}.
\item {\tt Model\_Mass()} returns a {\tt Switch::On}, if the masses
should be generated by the model.
\item {\tt Masses()} returns a {\tt Switch::On}, if all masses should
be taken into account.
\item {\tt Run\_Mass()} returns a {\tt Switch::On}, if the masses
should be regarded as running.
\item {\tt Run\_Width()} returns the number of the running width scheme
to be used, $0$ means no running. 
\item {\tt Run\_Aqed()} returns a {\tt Switch::On}, if 
$\alpha_{\rm QED}$ should be regarded as running.
\item {\tt Coulomb()} returns a {\tt Switch::On}, if Coulomb effects
should be taken into account.
\item {\tt Mass(Flavour,double)} returns the mass for a given 
{\tt Flavour} of a particle at a given scale. Accordingly, a possible
running of the masses will be taken into account in this method as well.
\item {\tt Mass(Flavour)} uses {\tt Mass(Flavour,double)} to yield the
mass of a particle at the CM-energy.
\item {\tt Width(Flavour,double)} works like  {\tt Mass(Flavour,double)},
but for the width.
\item {\tt Width(Flavour)} works like {\tt Mass(Flavour)}, but for the width.
\item {\tt Picobarn()} returns the conversion factor between $1/$GeV$^2$
and $pb$. 
\item {\tt Get\_Path()} returns the current path for all input data files.
\item {\tt Set\_Path()} sets the current path for all input data files.
\item {\tt Output()} returns the output level.
\item {\tt Set\_Output()} sets the output level.
\end{enumerate}
  
\subsection{Helpers\label{Help}}
A number of helper classes are always necessary for certain
purposes. Primarily, a random number generator is mandatory in every
Monte Carlo simulation. A tool to measure the elapsed time for the
different methods can be used for an optimization and of course a
three-- and a four--vector should be available as well. Last but not
least a matrix class can be used. A short description of the different
classes can be found in Tab.~\ref{HelpClass}.   
\begin{table}[h] 
\bc 
\begin{tabular}{l|l} 
Class/Struct & Purpose\\ 
\hline
&\\ 
Random   & Some random number generators.\\
MyTiming & A timer for measuring run times.\\                
vec3d    & A three-vector.\\  
vec4d    & A four-vector in Minkowski space.\\ 
Matrix   & A matrix with arbitrary rank.\\ 
&\\ 
\end{tabular}
\caption{\label{HelpClass}An overview of the helper classes.}
\ec
\end{table}                                                                   

\noindent The class {\tt Random} provides a number of different random number
generators. An additional option is the possibility, to store the
actual state of the appropriate generator. This ensures that after
every event a status can be saved and restored later on (for the
examination of a special event for instance). Note that all algorithms
for the different random number generators are taken from the
Numerical Recipes \cite{RECIPES}.
\begin{enumerate}
\item The {\tt Constructor()} initializes the random number generator
with a given seed via {\tt Set\_Seed()}.
\item {\tt ran1()} returns a random number after an algorithm of Park
and Miller with Bays-Durham shuffle and added safeguards, see \cite{RECIPES}. 
\item {\tt ran2()} returns a random number according to a long period random
number generator of L'Ecuyer with Bays-Durham shuffle and added
safeguards, see \cite{RECIPES}.
\item {\tt Ran3()} is a standard random number generator, see \cite{RECIPES}. 
\item {\tt Init\_Ran3()} initializes the random number generator.
\item {\tt get()} returns a random number calling {\tt Ran3()}.
\item {\tt getNZ()} returns a random number (with {\tt get()}) excluding zero.
\item {\tt Get\_Seed()} returns the actual seed.
\item {\tt Set\_Seed()} sets a given seed with the method {\tt Init\_Ran3()}.
\item {\tt theta()} returns an angle theta which is uniformly
distributed in the cosine of this angle.
\item {\tt WriteOutStatus()} writes out every status register of the
random number generator.  
\item {\tt ReadInStatus()} reads in every status register of the
random number generator.  
\end{enumerate}
The class {\tt MyTiming} can be used for internal time
measurements. The methods of this class are self-explanatory, i.e.\ 
{\tt Start()}, {\tt Stop()} and {\tt PrintTime()}.

The class {\tt vec3d} represents an Euclidean three-vector. Typical
operators, like $+,-,*$ and the cross product between three-vectors
and $*,/$ of three-vectors with scalars are defined accordingly. An
operator $<<$ ensures a proper output of a three-vector, other methods
are: 
\begin{enumerate}
\item The {\tt Constructor()} is available in the form of a standard
and a copy constructor as well as one constructor which can be
assigned with a four--vector ({\tt vec4d}). 
\item {\tt abs()} returns the absolute value (length) of the vector.
\item {\tt sqr()} returns the absolute value squared.
\item {\tt operator[]()} returns the element of the vector for a given
place number ($1\dots 3$). 
\end{enumerate}
The class {\tt vec4d} is the implementation of a four--vector in
Minkowski space. All the usual operators, i.e.\ $+,-,*$ between
four-vectors and $*,/$ of four-vectors with scalars are
implemented. Additional operators are $<<$ for the output, $==$ and
$!=$ for comparing two four-vectors.
\begin{enumerate}
\item Three different {\tt Constructor()} are available for the
four--vector, i.e.\ a standard, a copy and a constructor which can be
assigned with the energy and a three-vector.
\item {\tt operator[]()} returns the entry of the four-vector for a
given index ($0\dots 3$).
\item {\tt abs2()} returns the absolute value squared of the
four-vector. 
\item {\tt operator+=()} adds another {\tt vec4d}.
\item {\tt operator-=()} subtracts another {\tt vec4d}.
\item {\tt operator*=()} multiplies the current {\tt vec4d} with a scalar.
\end{enumerate}
The template class {\tt Matrix} represents a square matrix with arbitrary
rank, where the rank plays the role of the template parameter. The
usual multiplications with a matrix, a scalar or a four--vector (if
the rank is equal four) are mandatory. Note that not all possible
operations of a matrix have been implemented, since not all operations
are necessary for our purposes.
\begin{enumerate}
\item Two different types of {\tt Constructor()} exist in this
class, i.e.\ a standard and a copy constructor.
\item {\tt operator=()} is the copy operation.
\item {\tt operator[]()} returns a row of the matrix for a given index. 
\item {\tt matrix\_out()} makes a structured output of the matrix.
\item {\tt rank()} returns the rank of the matrix.
\item {\tt Num\_Recipes\_Notation()} translates the internal structure
of the matrix into a numerical recipes structure. The main difference
lies in the starting number for counting arrays, i.e.\ zero for \AME
and one for the numerical recipes.
\item {\tt Amegic\_Notation()} translates the numerical recipes
structure back into the internal structure.
\item {\tt Diagonalize()} diagonalizes a matrix, i.e.\ calculates the
eigenvalue and the eigenvectors with {\tt jacobi()}. For this purpose a
translation into the notation of the numerical recipes 
beforehand and a retranslation back into the original notation
afterwards are performed. 
\item {\tt Diagonalize\_Sort()} uses {\tt Diagonalize()} in order to
determine the eigenvalues and eigenvectors of a matrix. Afterwards
they will be sorted according to their eigenvalues.
\item {\tt jacobi()} diagonalizes a matrix, i.e.\ calculates the
eigenvalues and eigenvectors. This method has been borrowed from the
numerical recipes \cite{RECIPES}. 
\item {\tt Dagger()} returns the transposed matrix.
\end{enumerate}

\clearpage\newpage
\section{\label{Install}Installation guide}

\subsection{Installation}

The installation of \AME is quite simple, since a combination of 
{\tt automake} and {\tt autoconf} was used to generate the 
{\tt Makefile}'s of the program. The first script translates a number
of {\tt Makefile.am}'s, where a rough description of the object files
is included, into a {\tt Makefile.in}.  The second step is the
generation of a script called {\tt configure} from a basic file 
{\tt configure.in} which is achieved using {\tt autoconf}. This
script is able to translate an abstract {\tt Makefile.in} into a
proper {\tt Makefile} by including the actual path configuration of
the system. Note that these steps have been already performed by the
authors of this program. Accordingly, the last step which has to be
done by the user itself, is the translation of the 
{\tt Makefile.in}'s with the script {\tt configure}.

The steps to install \AME are the following:
\begin{enumerate} 
\item The program can be downloaded in form of the file
{\tt AMEGIC++-1.0.tar.gz}. It has to be unpacked with 
{\tt gzip} and  {\tt tar}. 
\item By calling the script {\tt configure} all 
{\tt Makefile}'s will be generated. Note that they are now adjusted to
the appropriate directory structure. 
\item With the command {\tt make install} all necessary libraries will
be built and the executable will be placed into the directory 
{\tt Amegic}, where it is ready to use.      
\item During the run of the program new {\tt C++} files might be
generated by {\tt AMEGIC++}. This is the case for the creation of
new integration channels or for saving the different helicity
combinations in form of a string. However, if this happens, a new
translation with {\tt make install} is all the user has to
do. Accordingly, all these files will be packed into a library called
{\tt Process} and linked as well. 
\end{enumerate}

\subsection{Running}

The program is executed by the script {\tt Amegic} with the directory
of the parameter files as an argument. However, the standard directory
is {\tt Testrun}, where the files {\tt Run.dat}, 
{\tt Const.dat}, {\tt Integration.dat}, {\tt Particle.dat} as well as
{\tt Processes.dat} should be available. The different possible
options are explained in the appropriate Tabs.~\ref{RunDat},
\ref{ConstDat} and \ref{IntDat}, whereas the properties of the Standard
model particles can be set according to Tab.~\ref{PartDat}. Note that
the particles which are marked as unstable will be provided with
calculated decay width in {\tt AMEGIC++}, the specified width in 
{\tt Particle.dat} will be ignored. 

The last step is to generate a list of processes within the data-file
{\tt Processes.dat}. Every process is built in the following way:
First the list of incoming particles is specified using their
kf--codes and an extra minus for anti-particles. The string ``{\tt ->}''
now parts the list of incoming and outgoing particles, where the
latter is created in the same way like the incoming list. A simple
example for the process $e^-\,e^+\to u\bar u$ could be 
``{\tt 11 -11 -> 2 -2}''. Note that it is also possible to indicate a list
of processes which will be handled one by one. Since now all
parameters and switches are determined, the program file
could be executed. 

However, a typical sample main file in {\tt C++} which is included in
the distribution as {\tt Amegic/main.C}, should look like: 
\begin{verbatim}
#include "Amegic.H"
#include "MyTiming.H"
#include "Run_Parameter.H"

using namespace AMEGIC;

Run_Parameter rpa;

int main(int argc,char* argv[]) 
{    

  MyTiming testtimer;
  testtimer.Start();
  string name("Testrun");
  if (argc==2) name = string(argv[1]);

  rpa.Init(name);

  particle_init(name);

  testtimer.PrintTime();
  Amegic Test(name);

  Test.Run();

  testtimer.Stop();
  testtimer.PrintTime();

}
\end{verbatim}
Note that an extra time measurement is performed as well.

\begin{table}[h] 
\bc
\begin{tabular}{l|c|l} 
Variable & Default & Purpose\\ 
\hline
&&\\
CMSENERGY &  91.       & CM-energy in GeV.\\
MODELFILE &  Const.dat & Model data-file.\\
MODEL     &  QCD       & Model used : {\tt pure QCD}, {\tt QCD}, {\tt EW}\\
          &            & and {\tt SM}.\\
MODELMASS &  Off       & Take the masses of the SM-particles from\\ 
          &            & the model data-file.\\
\hline
MASSES    &  On        & Quark masses.\\ 
RUNMASS   &  Off       & Running quark masses (LO).\\
RUNWIDTH  &  0         & Running width scheme.\\
RUNAQED   &  Off       & Running electroweak coupling $\alpha_{\rm QED}$.\\
COULOMB   &  Off       & Coulomb corrections.\\
OUTPUT    &  NORMAL    & Output level: {\tt SILENT}, {\tt NORMAL} or {\tt NOISY}.\\ 
&&\\
\end{tabular}
\caption{\label{RunDat}The parameters in {\tt Run.dat}.}
\ec
\end{table}                                                                   
\begin{table}[h] 
\bc
\begin{tabular}{l|c|l} 
Variable & Default & Purpose\\ 
\hline
&&\\
m\_up          & .005	& Mass of up-quark.\\
m\_down        & .01	& Mass of down-quark.\\
m\_e-          & .000511& Mass of electron.\\
m\_charm       & 1.3	& Mass of charm-quark.\\
m\_strange     & .170	& Mass of strange-quark.\\
m\_mu          & .105658& Mass of muon.\\
m\_top         & 174.	& Mass of top-quark.\\
m\_bottom      & 4.4	& Mass of bottom-quark.\\
m\_tau         & 1.77705& Mass of tauon.\\	
\hline
alphaS(M\_Z)   & .118	& $\alpha_S(M_Z)$\\                                   
v              & 246.0  & VEV of the SM Higgs-field.\\
m\_H\_SM       & 100.0  & SM Higgs mass.\\
alpha\_QED(MZ) & 128.   & $1/\alpha_{\rm QED}(M_Z)$\\
SinTW\^2       & .23124 & $\sin^2(\Theta_{\rm Weinberg})$\\
\hline
lambda         & .0     & $\lambda$ of Wolfenstein's CKM.\\
A              & .0     & $A$ of Wolfenstein's CKM.\\
rho            & .0     & $\rho$ of Wolfenstein's CKM.\\
eta            & .0     & $\eta$ of Wolfenstein's CKM.\\
&&\\
\end{tabular}
\caption{\label{ConstDat}The Standard Model parameters in {\tt Const.dat}.}
\ec
\end{table}                                                                   
\begin{table}[h] 
\bc
\begin{tabular}{l|c|l} 
Variable & Default & Purpose\\ 
\hline
&&\\
YCUT      &  0.01  & $y_{\rm cut}$ for jet-finders.\\
ERROR     &  0.01  & Allowed error when calculating matrix-elements.\\
\hline
INTEGRATOR &  0    & Phase space : Rambo=0, Rambo+Sarge=1,\\
           &       & Multichannel+Rambo=2, pure Multichannel=3.\\
JETFINDER  &  1    & Jet-finder:  DURHAM=1,JADE=2,GENEVA=3.\\
&&\\
\end{tabular}
\caption{\label{IntDat}The parameters in {\tt Integration.dat}.}
\ec
\end{table}                                                                   
\begin{table}[h] 
\bc
\begin{tabular}{l|r|r|r|r|c|c|c|c|l} 
kf-code & Mass & Width&3*e& Y&SU(3)&2*Spin&On&Stab.&Name\\
\hline
&&&&&&&&&\\
1	& .01  & .0   &	-1&-1&  1  &    1 &1 &  1   & d\_quark\\
2	& .005 & .0   &  2& 1&  1  &    1 &1 &  1   & u\_quark\\
3	& .170 & .0   &	-1&-1&  1  &    1 &1 &  1   & s\_quark\\
4	& 1.3  & .0   &	2 & 1&  1  &    1 &1 &  1   & c\_quark\\
5	& 4.4  & .0   &	-1&-1&  1  &    1 &1 &  1   & b\_quark\\
6       &174.0 & .0   &	2 & 1&  1  &    1 &1 &  1   & t\_quark\\
11	&.000511& .0  & -3&-1&  0  &    1 &1 &  1   & e-\\
12      &.0    & .0   &  0& 1&  0  &    1 &1 &  1   & nu\_e\\
13      &.10565& .0   & -3&-1&  0  &    1 &1 &  1   & mu-\\
14      &.0    & .0   &  0& 1&  0  &    1 &1 &  1   & nu\_mu\\
15      &1.777 & .0   & -3&-1&  0  &    1 &1 &  1   & tau-\\
16      &.0    & .0   &  0& 1&  0  &    1 &1 &  1   & nu\_tau\\
21	&.0    & .0   &	 0& 0&  1  &    2 &1 &  1   & gluon\\
22	&.0    & .0   &  0& 0&  0  &    2 &1 &  1   & photon\\	 
23      &80.356&2.07  & -3& 0&  0  &    2 &1 &  1   & W-\\
24      &91.188&2.49  &	 0& 0&  0  &    2 &1 &  1   & Z\\
25      &150.0 & .0   &  0& 0&  0  &    0 &1 &  1   & h\\
&&&&&&&&&\\
\end{tabular}
\caption{\label{PartDat}All particles included in {\tt Particle.dat}.}
\ec
\end{table}                                                                   

\clearpage\newpage
\section{\label{Summary}Summary}

In this paper we presented the newly developed matrix element
generator \AME which is capable of calculating Feynman amplitudes
as well as cross sections at tree level for the processes:
\begin{enumerate}
\item Electron positron annihilations into jets (up to a number of six
jets) and 
\item the scattering of QCD particles up to three jets.
\end{enumerate}
Two different kinds of problems occur during the calculation of the
Feynman amplitude and their integration to yield a cross
section. First of all, since the number of outgoing particles is quite
large, the resulting number of Feynman diagrams is enormous (for
example $e^+\,e^-\to q\bar q gggg$ yields $384$ diagrams in our
notation). Therefore, the usual method of summing and squaring the
amplitudes ``by hand'' is not sufficient enough. A way out provides
the helicity amplitude method which allows the decomposition of spinor
products into their helicity states. Now, only all other parts of the
Feynman diagram have to be translated into spinor products
accordingly, i.e.\ for instance propagators and polarization
vectors. Having at hand all helicity amplitudes for a given process,
the evaluation of the cross section could start. Here, the next
problem comes into play. Since the highly dimensional phase space
can contain a lot of differently strong peaks (i.e.\ resonant
propagators or soft and collinear gluons with respect to the quarks),
the usual method of a Monte Carlo integration with a uniformly
distributed phase space has to be abandoned. Accordingly, re-adjusted
phase space integrators have to be produced which take care for the
whole peak structure of a specified process. One possibility is the 
multi--channel integration, where every channel can be assigned to a Feynman
diagram of the process under consideration. Now, every channel is
built to cover the peak structure of a specific Feynman diagram and
therefore the whole peak structure is taken into account. 

The underlying ideas of the helicity formalism and the multi--channel
integration have been presented throughout this paper in different
forms, i.e.\ from a theoretical, algorithmical and technical point of
view. Primarily the detailed description of all classes and methods
within \AME should enable a possible user to enhance or simply to use
the program.  

Of course, \AME is far from being perfect, therfore a number of
extensions are planned for the near future:
\begin{itemize}
\item The list of Feynman rules and therefore {\tt Model}s will be
extended to the Two Higgs Doublet and to the Minimal Supersymmetric
Standard Model.  
\item The program will be enabled to calculate polarized cross
sections in order to make predictions for a possible next linear
collider with polarized electrons and positrons.
\item Initial state radiation will be possible for electron
positron and QCD parton scatterings. The former one will include the
Yennie--Frautschi--Suura approach and the latter one a structure
function scheme which makes it necessary to link parton distribution
functions (pdf).
\item Further technical refinements are in order, for instance
parallelization, further simplifications of the strings, more building 
blocks for the integration channels, more phase space selectors.
\end{itemize}

\acknowledgments{F.K. and R.K. would like to thank Mike Bisset,
Andreas Sch{\"a}licke, Steffen Schumann and Jan Winter for extensively
testing parts of the program and useful comments on both the program and
the manual. We appreciate gratefully the careful reading of the
manuscript by James Hetherington, Chris Harris and Peter Richardson
and their suggestions for its improvement.
F.K. would like to thank the Physics Department of the Technion,
Israel, where parts of the work were done, for friendly hospitality. 
R.K. is grateful for the kind hospitality of the Cavendish
Laboratory, where large parts of this work have been finalized. 
We acknowledge financial support by DAAD, BMBF and GSI.
}

\clearpage\newpage
\appendix
\section{\label{SampleCode}Sample channel for the phase space integration}

This is the {\tt C++} file which was produced in order to generate
the channel given in Fig.~\ref{st3}. Three different methods will be
created, the first one for the generation of the four-momenta, the
second one for the calculation of the appropriate weight and the last
one for finding resonant propagators within the channel. The last
option is used to drop irrelevant channels. 
\begin{verbatim}
#include "P2_5_e_e___s_cb_tau_nu_tau_P.H"
using namespace AMEGIC;

void P2_5_e_e___s_cb_tau_nu_tau_P::C1_Momenta(vec4d* pin,vec4d* pout,
	                                      double* ms_out,double* ran)
{
  vec4d  p32456 = pin[0] + pin[1];
  double s32456 = p32456.abs2();
  double s32_min = Max(pa.ycut_ini()*sqr(rpa.CMS_E()),
                       sqr(sqrt(ms_out[3]) + sqrt(ms_out[2])));
  double s456_min = Max(pa.ycut_ini()*sqr(rpa.CMS_E()),
                        sqr(sqrt(ms_out[4]) + sqrt(ms_out[5]) 
                                            + sqrt(ms_out[6])));
  double s32_max = sqr(sqrt(s32456)-sqrt(s456_min));
  vec4d  p32;
  double s32;
  s32 = CE.Massless_Prop_Momenta(1.,s32_min,s32_max,ran[1]);
  double s456_max = sqr(sqrt(s32456)-sqrt(s32));
  vec4d  p456;
  double s456;
  s456 = CE.Massless_Prop_Momenta(1.,s456_min,s456_max,ran[2]);
  double amct  = 1.;
  double alpha = 0.5;
  double ctmax = 0.;
  double ctmin = 2.;
  double tmass = Flavour(kf::code(12)).mass();
  CE.T_Channel_Momenta(pin[0],pin[1],p32,p456,s32,s456,tmass,alpha,
                       ctmax,ctmin,amct,0,ran[3],ran[4]);
  double s3 = ms_out[3];
  double s2 = ms_out[2];
  CE.Isotropic2_Momenta(p32,s3,s2,pout[3],pout[2],ran[5],ran[6]);
  double s45_min = Max(pa.ycut_ini()*sqr(rpa.CMS_E()),
	               sqr(sqrt(ms_out[4]) + sqrt(ms_out[5])));
  double s6 = ms_out[6];
  double s45_max = sqr(sqrt(s456)-sqrt(s6));
  vec4d  p45;
  double s45;
  s45 = CE.Massless_Prop_Momenta(1.,s45_min,s45_max,ran[7]);
  CE.Isotropic2_Momenta(p456,s45,s6,p45,pout[6],ran[8],ran[9]);
  double s4 = ms_out[4];
  double s5 = ms_out[5];
  CE.Isotropic2_Momenta(p45,s4,s5,pout[4],pout[5],ran[10],ran[11]);
}

double P2_5_e_e___s_cb_tau_nu_tau_P::C1_Weight(vec4d* pin,vec4d* pout,
	                                       double* ms_out)
{
  double wt = 1.;
  vec4d  p32456 = pin[0] + pin[1];
  double s32456 = p32456.abs2();
  double s32_min = Max(pa.ycut_ini()*sqr(rpa.CMS_E()),
                       sqr(sqrt(ms_out[3]) + sqrt(ms_out[2])));
  double s456_min = Max(pa.ycut_ini()*sqr(rpa.CMS_E()),
                        sqr(sqrt(ms_out[4]) + sqrt(ms_out[5]) 
                                            + sqrt(ms_out[6])));
  double s32_max = sqr(sqrt(s32456)-sqrt(s456_min));
  vec4d  p32 = pout[3] + pout[2];
  double s32 = p32.abs2();
  wt *= CE.Massless_Prop_Weight(1.,s32_min,s32_max,s32);
  double s456_max = sqr(sqrt(s32456)-sqrt(s32));
  vec4d  p456 = pout[4] + pout[5] + pout[6];
  double s456 = p456.abs2();
  wt *= CE.Massless_Prop_Weight(1.,s456_min,s456_max,s456);
  double amct  = 1.;
  double alpha = 0.5;
  double ctmax = 0.;
  double ctmin = 2.;
  double tmass = Flavour(kf::code(12)).mass();
  wt *= CE.T_Channel_Weight(pin[0],pin[1],p32,p456,tmass,
                            alpha,ctmax,ctmin,amct,0);
  double s3 = ms_out[3];
  double s2 = ms_out[2];
  wt *= CE.Isotropic2_Weight(pout[3],pout[2]);
  double s45_min = Max(pa.ycut_ini()*sqr(rpa.CMS_E()),
                       sqr(sqrt(ms_out[4]) + sqrt(ms_out[5])));
  double s6 = ms_out[6];
  double s45_max = sqr(sqrt(s456)-sqrt(s6));
  vec4d  p45 = pout[4] + pout[5];
  double s45 = p45.abs2();
  wt *= CE.Massless_Prop_Weight(1.,s45_min,s45_max,s45);
  wt *= CE.Isotropic2_Weight(p45,pout[6]);
  double s4 = ms_out[4];
  double s5 = ms_out[5];
  wt *= CE.Isotropic2_Weight(pout[4],pout[5]);
  if (!IsZero(wt)) wt = 1./wt/pow(2.*M_PI,5*3.-4.);

  return wt;
}

int P2_5_e_e___s_cb_tau_nu_tau_P::C1_Resonances(Flavour*& res_fl)
{
  res_fl = new Flavour[3];
  res_fl[0] = Flavour(kf::code(23));
  res_fl[1] = Flavour(kf::code(23));
  res_fl[2] = Flavour(kf::code(23));
  return 3;
}
\end{verbatim}
Note that only the methods of the class {\tt Channel\_Elements} will
be used, for details see Sec.~\ref{PhaseSpace}.

\section{\label{SampleMath}A sample for a Mathematica interfaced function}

This is a sample file generated by importing {\tt Mathematica}
output for the calculation of a three gluon vertex: 
\begin{verbatim}
#include "Mathematica_Interface.H"

using namespace AMEGIC;

Kabbala Mathematica_Interface::vGGG()
{
  return Z(1,0)*(X(2,0)-X(2,1))+Z(2,0)*(X(1,2)-X(1,0))
    +Z(2,1)*(X(0,1)-X(0,2));
}
\end{verbatim}

\section{\label{Loop}The loop over loops technique}

Sometimes it is necessary to have loops within loops to a depth which
is not known before run time. Then, a technique comes into play which
allows an arbitrary number of inner loops with different starting
and ending values. Therefore three different arrays are required, the
first one is filled with all beginning values, the
second with all ending values and the last one is used to store the
current combination of loop variables. Then a loop is performed
until the first variable reaches its endpoint. In every step the last
variable will be increased (or decreased accordingly). If it gets
larger than its endpoint, the variable is reset to its beginning
value and the variable at the previous position is increased (or
decreased). Consequently, every time a variable reaches its endpoint,
the previous one will be increased (or decreased). 
This procedure can be repeated until the first variable is counted
behind its endpoint and the loop ends.   

\section{\label{TRO}Test Run Output}

In this section we present a test run output for the given example
options in Sec.~\ref{Install} and the process 
$e^-\,e^+\to u\bar u g$:   
\begin{verbatim}
Starting Timer
Open File: Testrun/Run.dat
Open File: Testrun/Particle.dat
Time: 0 s   (clocks=0)
 (User: 0.01 s ,System: 0 s ,Children User: 0 s ,Children System: 0)
Open File: Testrun/Const.dat
Open File: Testrun/Const.dat
No FFS Vertex included in this Model
No SSV Vertex included in this Model
No VVS Vertex included in this Model
No SSS Vertex included in this Model
Number of Vertices: 21
1 process(es) !
No Value-Library available !
Building Topology...
Matching of topologies...
****File Process/2_3_e-_e+___u_ub_G/Color.dat not found.
Finding diagrams with same color structure...
 
2 different color structures left
++
+
File Process/2_3_e-_e+___u_ub_G/Color.dat saved.
2 Graphs found
Open File: Testrun/Integration.dat
 
using RAMBO for phase space integration
Starting the calculation. Lean back and enjoy ... .
1:********************************
2:********************************
Gauge(1): 0.0552225
Gauge(2): 0.0552225
Gauge test: 6.28295e-14%
3:********************************
String test: 0%
5000. LO-3-Jet: 1906.09 pb +- 2.32851%
10000. LO-3-Jet: 1946.15 pb +- 1.69762%
15000. LO-3-Jet: 1962.73 pb +- 1.39437%
20000. LO-3-Jet: 1986.16 pb +- 1.20933%
25000. LO-3-Jet: 1966.02 pb +- 1.08427%
30000. LO-3-Jet: 1965.21 pb +- 0.988323%
result: 1965.21
Stoping Timer
Time: 85.88 s   (clocks=8588)
 (User: 71.65 s ,System: 0.07 s ,Children User: 0 s ,Children System: 0)
\end{verbatim}
Note that this calculation was performed using internal strings for
the helicity amplitudes, if the expressions have been saved into a
library, the two last lines would look like follows:

\begin{verbatim}
Time: 16.22 s   (clocks=1622)
 (User: 16.22 s ,System: 0 s ,Children User: 0 s ,Children System: 0)
\end{verbatim}
One can see that roughly a factor of $4$ can be gained by using the
saved libraries.

\end{document}